\documentclass[fleqn,usenatbib]{mnras}

\usepackage{newtxtext,newtxmath}

\usepackage[T1]{fontenc}
\usepackage{ae,aecompl}

\DeclareRobustCommand{\VAN}[3]{#2}
\let\VANthebibliography\thebibliography
\def\thebibliography{\DeclareRobustCommand{\VAN}[3]{##3}\VANthebibliography}

\usepackage{graphicx}	
\usepackage{amsmath}	
\usepackage{adjustbox}
\usepackage{xcolor}
\usepackage{subfigure}
\usepackage{bm}

\title[$\omega$ Cen and its binary population]{Studying Binary Systems in Omega Centauri with MUSE: II. Observational constraints on the orbital period distribution}

\author[S. Saracino et al.]{S. Saracino$^{1,2}$\thanks{E-mail: sara.saracino@inaf.it},
S. Kamann$^{1}$, F. Wragg$^{1}$, S. Dreizler$^{3}$, K. Kremer$^{4}$, M. Latour$^{3}$, \and J. M\"uller-Horn$^{5}$, N. Neumayer$^{5}$, A. C. Seth$^{6}$, G. van de Ven$^{7}$, M. H\"aberle$^{5}$\\
\\
$^{1}$Astrophysics Research Institute, Liverpool John Moores University, 146 Brownlow Hill, Liverpool L3 5RF, UK\\
$^{2}$INAF – Osservatorio Astrofisico di Arcetri, Largo E. Fermi 5, 50125 Firenze, Italy\\
$^{3}$ Institute for Astrophysics, Georg-August-University G\"ottingen, Friedrich-Hund-Platz 1, D-37077 G\"ottingen, Germany\\
$^{4}$ Department of Astronomy \& Astrophysics, University of California, San Diego; La Jolla, CA 92093, USA\\
$^{5}$ Max-Planck-Institut f\"ur Astronomie, K\"onigstuhl 17, D-69117 Heidelberg, Germany \\
$^{6}$ Department of Physics and Astronomy, University of Utah, Salt Lake City, UT 84112, USA \\
$^{7}$ Department of Astrophysics, University of Vienna, T\"urkenschanzstrasse 17, 1180 Wien, Austria
\\
}
\date{Accepted 2025 March 13. Received 2025 March 11; in original form 2024 October 18}
\pubyear{2025}

\begin{document}
\label{firstpage}
\pagerange{\pageref{firstpage}--\pageref{lastpage}}
\maketitle
\begin{abstract}
Omega Centauri ($\omega$ Cen) is one of the most complex star clusters in the Milky Way, and likely the stripped nucleus of an accreted dwarf galaxy. Being the subject of debate between it hosting an intermediate-mass black hole (IMBH) or a collection of stellar-mass black holes (BHs) in its center, $\omega$ Cen has been intensively studied over the past decades. Our work focuses on characterizing the properties of binary systems in $\omega$ Cen via multi-epoch MUSE spectroscopic observations spanning over eight years and covering much of its central regions (i.e. core radius). We did not detect any stellar-mass BHs candidates orbiting luminous stars, although mock samples indicate a high sensitivity of our survey to such systems. This suggests that BHs orbiting stars may be rare in $\omega$ Cen or in wide orbits around low-mass companions (where our survey is 50\% complete) or that the periods of such systems are longer than expected from cluster dynamics. Additionally, we constrained the orbital properties of 19 binary systems in the cluster, with periods ranging from fractions of a day up to several hundred days. We observe an excess of binaries with P $\geq$ 10 d and find evidence that the intrinsic period distribution of binaries in $\omega$ Cen differs from those predicted by cluster evolutionary models.

\end{abstract}

\section{Introduction}
\label{sec:intro}
Omega Centauri ($\omega$ Cen) challenges the conventional categorization of Galactic globular clusters (GCs) thanks to its unique characteristics with up to 15 different stellar populations \citep{Bellini_2017} and complex nature such as a metallicity spread of up to 2 dex \citep{Johnson_2020,Nitschai2024} and evidence for a central stellar disk and tangential velocity anisotropy \citep{van_de_Ven_2006}. These findings are consistent with $\omega$ Cen being the stripped core of a disrupted dwarf galaxy that merged with the Milky Way early in its formation \citep{Lee1999,Lee2009}. In this picture, $\omega$ Cen would represent the former nuclear star cluster of such a galaxy (e.g. Gaia Enceladus/Sausage, \citealt{Pfeffer_2021,Limberg2022}). One of the most intriguing aspects of this stellar system is the ongoing debate over whether it hosts a massive central black hole (BH), so-called intermediate-mass black hole (IMBH, with a mass range $10^{2}$-$10^{5}M_{\odot}$, \citealt{Greene_2020}). \citet{Noyola_2008} first put forward this hypothesis and estimated a mass for this central dark object ($\approx$40 000 $M_{\odot}$), by comparing the surface brightness and line-of-sight velocity dispersion profiles of the cluster with early dynamical models. A massive dark central source became considerably less necessary later, after a new center \citep{Anderson_2010} and dynamical modeling of proper motions using the Hubble Space Telescope (HST) \citep{VanDerMarel_2010} were obtained for $\omega$ Cen, imposing an upper limit of 12 000 $M_{\odot}$ to the mass of the putative BH. While the idea of an IMBH in the cluster core is not surprising in principle, given the evidence that nuclear star clusters host massive BHs at their centers \citep{Neumayer_2020}, it does not rule out alternative hypotheses. In fact, a collection of 10s of thousands of stellar mass BHs (corresponding to $\sim$5\% of the mass of the cluster) could be present instead, as they would produce a similar signal \citep{Zocchi_2019}. Interestingly, a recent theoretical study by \citet{Sharma_Rodriguez_2024} have suggested that if a cluster hosts a central IMBH, the expected amount of stellar-mass BHs in its core is significantly reduced, by an amount that depends on how centrally-dense the cluster is, so that there cannot be a peaceful coexistence of these two entities. Along the same lines is the result by \citet{Leigh2014}, who suggested that the detection of one or more stellar-mass BHs strongly indicates against the presence of an IMBH more massive than $10^{3}$ $M_{\odot}$ in 80\% of their simulated clusters. 

The possibility that $\omega$ Cen contains an IMBH has sparked significant interest and debate within the astrophysical community. \citet{Baumgardt_2019} contributed to the discussion by presenting evidence suggesting the prevalence of a collection of stellar-mass BHs within $\omega$ Cen. This idea was based on tailored N-body simulations used to fit the velocity dispersion profile of the cluster, along with the absence of high-velocity stars in its central regions detected by observations. The search for an IMBH in the cluster was then suspended, leaving the question of the true nature of $\omega$ Cen's central dark component unresolved and triggering more analysis and investigations.

In fact, convincingly addressing this controversy requires a multifaceted observational approach. On one side, the search for high-velocity stars within the cluster's core offers a potential means of distinguishing between the IMBH and stellar-mass BH scenarios. On the other side, conducting a comprehensive multi-epoch spectroscopic campaign to identify and characterize binary systems within $\omega$ Cen, particularly those hosting stellar-mass BHs as companions, is essential. In fact, although a direct spectroscopic detection of isolated BHs or BH-BH binaries is not possible (i.e. these systems do not emit light and cannot be detected), a non-negligible fraction of BHs are expected to still form binaries with luminous companions, providing an indirect means of detection. It is worth mentioning that $\omega$ Cen contains white dwarfs (WDs) and neutron stars (NSs) orbiting luminous stars, discovered as cataclysmic variables, millisecond pulsars and a possible quiescent low mass X-ray binary, thanks to deep radio and X-ray observations \citep{Henleywillis2018,Dai2020}, but no stellar-mass BHs have been detected so far.

The advent of advanced observational facilities, such as the integral field spectrograph MUSE (Multi Unit Spectroscopic Explorer, \citealt{Bacon2010}) mounted at the Very Large Telescope (VLT), and high resolution HST observations, have significantly enhanced our ability to explore both pathways with unprecedented resolution, depth and completeness. Recent observational efforts in this direction have provided intriguing insights into the nature of the central object of $\omega$ Cen. Exploiting proper motion measurements from over 20 yrs of consecutive HST monitoring \citep{Haberle2024PM}, the same authors were able to identify seven fast-moving stars in the central 3" of $\omega$ Cen, providing convincing evidence for the actual presence of an IMBH within the cluster, with a mass of at least 8 200 $M_{\odot}$ \citep{Haberle_2024}. On the other hand, \citet{Platais_2024} made use of up to 13 yrs of Hubble observations of $\omega$ Cen's central regions to carry out a detailed astrometric acceleration search in order to detect any stellar-mass BHs in the cluster. They found four new binaries with significant accelerations, of which three were consistent with a WD companion and one possibly with a NS companion, but again no evidence for BHs. As previously mentioned, if an IMBH as massive as 40 000 $M_{\odot}$ is present in $\omega$ Cen, there might be not enough room for a large fraction of stellar-mass BHs, be they single BHs, BH-BH pairs or BHs orbiting stars.  

According to theoretical simulations with the Cluster Monte Carlo (CMC) and the MOnte Carlo Cluster simulAtor (MOCCA) codes, the number of BHs in binaries with luminous companions does not scale with the total number of BHs. In fact, only a small percentage of them will interact with luminous stars, with all the remaining pairing up with other BHs or remaining isolated \citep{Askar_2018,Kremer_2020}. However, although they represent only a small fraction of the total, BHs orbiting stars are the best to detect, given the signal they produce in radial velocity (RV) studies. 

RV studies are time consuming, as they require to observe the same field for multiple times over the years, but provide a lot of information, e.g. they allow to identify and characterize the orbital properties of all binary stars in the cluster. They can shed light on the dynamical interactions that occur, their frequency and how many binary systems with ongoing mass transfer we should expect. Binary stars are also critical for understanding the evolution, structure and dynamics of clusters. Mass segregation causes binaries to accumulate in cluster centers, where they act as dynamic energy sources, stabilizing the clusters against core collapse and affecting velocity dispersion measurements, which are essential for accurately estimating cluster masses \citep{Bianchini2016}. For all these reasons, a comprehensive analysis of the binary population content of $\omega$ Cen is urgently needed but still lacking. 

A study by \citet{MullerHorn2024} has recently highlighted important discrepancies in the comparison between the observed and predicted orbital period distributions of binaries in the Galactic GC 47 Tucanae (47 Tuc, \citealt{Ye_2022}), namely a large population of unobserved short-period binaries (P < 2-3 d). Both the uncertain treatment of the common envelope phase and the initial binary properties adopted by the simulations could be responsible for this difference. Making similar observations in other clusters with different dynamical times and nuclear densities can help us understand the physics driving this mismatch between theory and observation.

A first important step forward in this direction has been recently taken by a spectroscopic study of $\omega$ Cen led by \citet{Wragg2024}, the first paper in this series, who provided valuable information on the global binary fraction of the system (2.1\% $\pm$ 0.4\%), well in agreement with previous photometric estimates \citep{Elson_1995,Bellini_2017}. The sample consisted of MUSE observations spanning a time baseline of more than eight years, covering a large portion of the central regions of the cluster. Leveraging the dataset, in the second paper we perform Keplerian orbital fits to all binary candidates with more than six single-epoch RVs, to investigate the presence of any stars orbiting massive companions, i.e. stellar-mass BH or NS candidates, as well as to study the global properties of the binary population of $\omega$ Cen, in terms of their period distribution. The latter information is unavailable or rare for clusters as old as $\omega$ Cen (with 47 Tuc, \citealt{MullerHorn2024} and NGC 3201, \citealt{Giesers2019} being the only exception) but extremely useful for providing insights into reliable initial conditions for building tailored dynamical simulations of these systems.

In Section \ref{sec:obs}, we briefly introduce the dataset and outline the key steps used to identify binary stars in $\omega$ Cen. Section \ref{sec:orbitalfitting} details the orbital fitting methods applied to the observed binary sample and analyzes the results for systems with constrained orbital parameters. In Section \ref{sec:simulations}, we assess the completeness and purity of the binary star sample in $\omega$ Cen using mock datasets. Sections \ref{sec:corrected} and \ref{sec:LPB} focus on comparisons with theoretical predictions, while Section \ref{sec:concl} presents our conclusions.

\section{Observations, data analysis and binary system identification}
\label{sec:obs}
Spectroscopic observations of GCs face challenges due to crowded fields, resulting in source blending and limited samples for studying binary populations. However, VLT/MUSE offers a solution, enabling simultaneous spectroscopy of numerous stars within the crowded central regions of star clusters. It provides spectra covering the range of 4750 to 9350 {\AA} with a spectral resolution of $R\sim1 800-3 500$ across a $1\times1$~arcmin$^{2}$ field of view. This study utilizes MUSE Guaranteed Time Observations (GTO) data of $\omega$ Cen from the survey of Galactic GCs presented in \citet{Kamann_2018}, comprising 10 wide field mode (WFM) and 6 narrow field mode (NFM) pointings repeatedly observed between 2015 and 2022 with varying cadences and exposure times. 

As this work is based on the dataset already presented in \citet{Wragg2024}, we do not provide extensive details on how the MUSE observations were analysed. We briefly mention the main steps of the data analysis here and we refer any interested reader to their paper (and the references therein) for a more exhaustive explanation.
The MUSE raw data underwent the standard ESO pipeline reduction \citep{pipeline}, followed by stellar spectra extraction using \textsc{PampelMuse} software \citep{Kamann2013}. Iterative improvement of the point spread function ensured accurate extraction even in the most crowded region, the cluster core. High-resolution photometric data from the HST ACS survey of Galactic GCs \citep{Sarajedini_2007,Anderson_2008} and the study of \cite{Anderson_2010} served as an astrometric reference for stellar positions and magnitudes, facilitating the extraction of individual spectra.

Spectral analysis utilized \textsc{Spexxy} \citep{Husser2016}, performing full-spectrum fitting against PHOENIX template spectra (G\"ottingen Library \textsc{GLib} \citealt{Husser_2013}) to measure stellar parameters and RVs. Initial parameter guesses (e.g. T\textsubscript{eff} or log g) were determined by comparing the HST photometry against a PARSEC isochrone \citep{Marigo_2017} of appropriate age, metallicity, extinction and distance for $\omega$ Cen. Parameters such as T\textsubscript{eff}, [M/H], and RVs were refined through least-squares optimization, with different procedures adopted for stars across various evolutionary stages. For example, extreme horizontal branch (HB) stars were too hot (T\textsubscript{eff}$>$15,000K) to be compared against the \textsc{GLib} template spectra, hence were treated separately, following \citealt{Latour_2024}. Single-epoch RVs were obtained for all the observations available, and to address underestimated velocity uncertainties, a correction factor was determined based on comparison samples, as outlined in previous MUSE studies (see e.g. \citealt{Kamann_2016,Nitschai_2023}).

The final MUSE sample was obtained after several quality cuts and cluster membership selections were applied, including: i) The spectra successfully fitted by \textsc{Spexxy}. ii) The contamination from nearby sources was estimated to less than 5\%. iii) The spectrum was extracted at least 2 spaxels away from the edge of the MUSE field of view. iv) The magnitude accuracy parameter determined by \textsc{PampelMuse} was above 0.6. v) The RV measurement reliability parameter introduced in \citet{Giesers2019} was over 80\%. vi) The T$_{\text{eff}}$, log g, and [M/H] values were consistent with those obtained for other epochs. vii) A membership probability cut of 0.8 was used to discard field stars. viii) Photometric variable stars were removed via a cross-match with the catalogs by \citet{Clements_2001}, \citet{LebzelterWood2016} and \citet{Braga_2020}. ix) The photometric variability parameter estimated from the MUSE data was less than 0.25. Such variability most likely points to problems during the extraction of the spectra, not to the detection of new variable sources, so it is safe to discard those objects.

By applying these criteria, a sample of 266 816 individual spectra from 28 979 stars remained. We have a median number of six valid RV measurements per star, and stars with at least six measurements are also the only stars for which we estimated a probability to be variables, i.e. to be part of a binary system. Stars with less than six single-epoch RV measurements (47\% of the total, corresponding to 10 170 stars) were not considered due to the limited information available. This choice is based on previous works of this type (see \citealt{Giesers2019}) and allows to limit misclassifications that could influence subsequent results.

We adopted the method by \citet{Giesers2019} to calculate the probability of velocity variability for each star in the sample. By weighing $\chi^2$ values against the likelihood of statistical noise, this approach minimizes false positive detections, allowing to get a cleaner sample of binaries to analyze. 
We adopted a probability threshold of $P_{var}>0.8$ to distinguish binaries from single stars, as suggested in \citet{Wragg2024}, because this threshold allows to minimize the number of spurious detections in the sample (a visual representation of this selection is shown in their Figure 3). 

\begin{figure}
    \centering
	\includegraphics[width=0.48\textwidth]{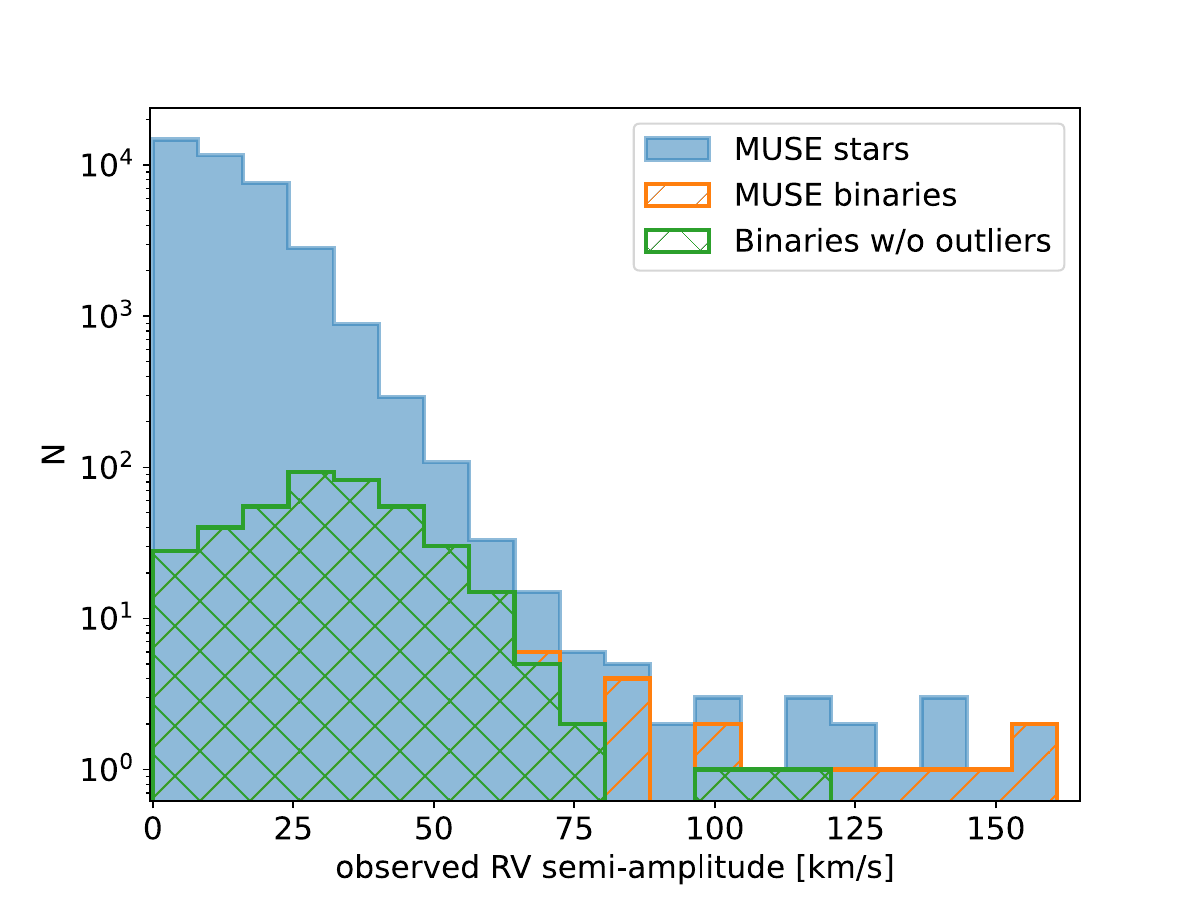}
    \caption{Observed RV semi-amplitude distribution of all stars in the MUSE observations of $\omega$ Cen (blue), compared to the distribution for the binaries detected (orange). A tail of high-amplitude binaries ($>$75 km/s) seems to be present. However, while plotting the distribution of binaries after removing those showing clear outliers in their observed RV curve (green), most of the tail previously observed disappears.}
    \label{fig:distribution}
\end{figure}

The sample thus obtained, composed of 222 binary stars for a total of 2 649 velocity measurements is used in the present work to attempt a Keplerian fit for each of the binaries. The catalog containing MUSE RVs of individual epochs for all stars in $\omega$ Cen, together with their $P_{var}$ values, is published as supplementary material to this paper and can be found on Vizier (the link to the page will be provided here). For binaries with fewer than six epochs, the $P_{var}$ column is empty. The distribution of the RV semi-amplitude (K) for all stars observed by MUSE in $\omega$ Cen is presented as a blue filled histogram in Figure \ref{fig:distribution}. The subsample of binary candidates with $P_{var}>0.8$ is instead shown in orange. For each star, K is measured as half of the observed peak-to-peak RVs variation. For single stars this quantity is close to 0 and increases for binaries, with highest values corresponding to binaries with high RV variations. The distribution shows a tail for K values of 75 km/s or above, suggesting the possible presence of an interesting population of high-amplitude binaries. However, after a careful inspection of the binaries in the tail we realized that the vast majority (12 binaries) might contain one or two outliers, artificially inflating the RV semi-amplitude. 

To try and discriminate between binaries with outliers and genuine variables, we fitted their RV curves before and after removing the RV measurement(s) responsible for the high-amplitude values (see details on the methodology in Section \ref{sec:orbitalfitting}). A good orbital solution was found for all systems when such epochs were included, but not when they were removed. This would support the idea that these measurements are outliers, probably caused by undetected blends or local minima in the $\chi^{2}$ space sampled by \textsc{Spexxy}.
On the other hand, if we instead assume that these are all genuine binaries, they must be very eccentric (with e$>$0.7-0.8). Given that only 1\% of all binaries with main sequence (MS) companions in the simulation presented in Section \ref{sec:testI} have e$>$0.7-0.8, it would look extremely unlikely that all these binary systems are genuine and highly eccentric. However, we cannot exclude that there is an issue with these stars, hence we make the conservative choice to discard these systems from the subsequent analysis. The new distribution, without these stars, is presented in green in Figure \ref{fig:distribution}. We note here that the inclusion of these systems is not expected to have any significant impact on the binary fraction of $\omega$ Cen derived by \citet{Wragg2024}. The small number of systems in fact produces differences that fall within the corresponding uncertainty.

\section{Orbital Fitting of individual binaries}
\label{sec:orbitalfitting}

The main aim of this study is to fit the observed RV curves of all binary candidates in $\omega$ Cen, to constrain their orbital properties (e.g. period, semi-amplitude, mass ratio etc.). The only assumption we make to perform the analysis is that all binaries consist of two stars, and one star dominates the light, simplifying the model to SB1 binaries. While the Generalised Lomb-Scargle periodogram (GLS, \citealt{Zechmeister2009}) method is widely used in astronomy for detecting periodic signals in irregularly sampled time-series data, it has been proven to be not ideal for sparse data like ours. In this study we have employed two different algorithms, able to infer binary orbital parameters in a Bayesian framework: \textsc{The Joker} \citep{Price-Whelan2020} and \textsc{Ultranest} \citep{buchner2021}. Both were proven to be rather effective with sparse and noisy RV measurements. 
\subsection{The Joker}
\label{sec:joker}
\textsc{The Joker} is a custom Monte-Carlo sampler \citep{Price-Whelan2017,Price-Whelan2020}. The software generates a library of possible orbits based on input parameters, scanning the parameter space to find orbits that match the observed RV curve. For our dataset, $2^{29}$ prior samples are generated log-uniformly within a period range of 0.1 d to 1000 d. We apply distributions for eccentricity, velocity semi-amplitude K, and systemic velocity consistent with previous studies \citep{Saracino_2023} and detailed in Table \ref{tab:prior_PDF}, left column. For each star we requested 512 posterior samples, discarding stars for which a significantly lower number of posterior samples were obtained. The results, both in terms of binary population and individual binary properties, are presented in subsequent Sections.
\subsection{Ultranest}
\label{sec:ultranest}
\textsc{Ultranest} is a nested sampling algorithm, originally introduced by \citet{buchner2021}. It is able to explore complex likelihood landscapes and compute posterior probability density functions (PDFs). For each binary in our sample we determined the PDFs for six orbital parameters (P, K, e, $\omega$, $v_{sys}$ and $M_{0}$), also including a jitter term (s), which takes into account the possible underestimation of observed RV uncertainties. The adopted prior distributions for all the parameters are detailed in Table \ref{tab:prior_PDF}, right column, and are in line with those suggested in similar studies (e.g. \citealt{MullerHorn2024}). The number of posterior samples produced by \textsc{Ultranest} is of the order of a few thousands, making the results for constrained binaries overall more robust and statistically reliable than those provided by \textsc{The Joker}.\\

\begin{table*}
\caption{Summary of the distributions adopted as a prior for the different parameters, both in \textsc{The Joker} (left-hand side) and in \textsc{Ultranest} (right-hand side).}\label{tab:prior_PDF}
\centering
\begin{tabular}{|lll|}
\hline
Parameter & \textsc{The Joker} & \textsc{Ultranest}\\
\hline
period, P [d] or frequency, $f$ [1/d] & $\ln P \sim \mathcal{U}(0.1,1000)$ & $\ln f \sim \mathcal{N}(\ln{0.1},2.3)$ \\
RV semi-amplitude, $K_1$ [km/s] & $\mathcal{N}(0,\sigma^{2}_{K})$ with $\sigma_{K}$ = 30 km/s & $\mathcal{N}(0,\sigma^{2}_{K})$ \\
mean anomaly, $M_0$ [rad] & $\mathcal{U}(0,2\pi)$ & $\mathcal{U}(0,2\pi)$\\
eccentricity, $e$ & $\beta(0.867,3.03)$ & $\beta(0.867,3.03)$ \\
argument of pericenter, $\omega$ [rad] & $\mathcal{U}(0,2\pi)$ & $\mathcal{U}(0,2\pi)$\\
jitter term, $s$ [km/s] & --- & $\ln s \sim \mathcal{N}(-4,2.3)$\\
system velocity, $v_{sys}$ [km/s] & $\mathcal{N}(233,20)$ & $\mathcal{N}(233,20)$ \\
\hline
\end{tabular}
\end{table*}

\subsection{Binaries with constrained orbits}
\label{sec:obsresults}
\begin{figure*}
    \centering
	\includegraphics[width=0.60\textwidth]{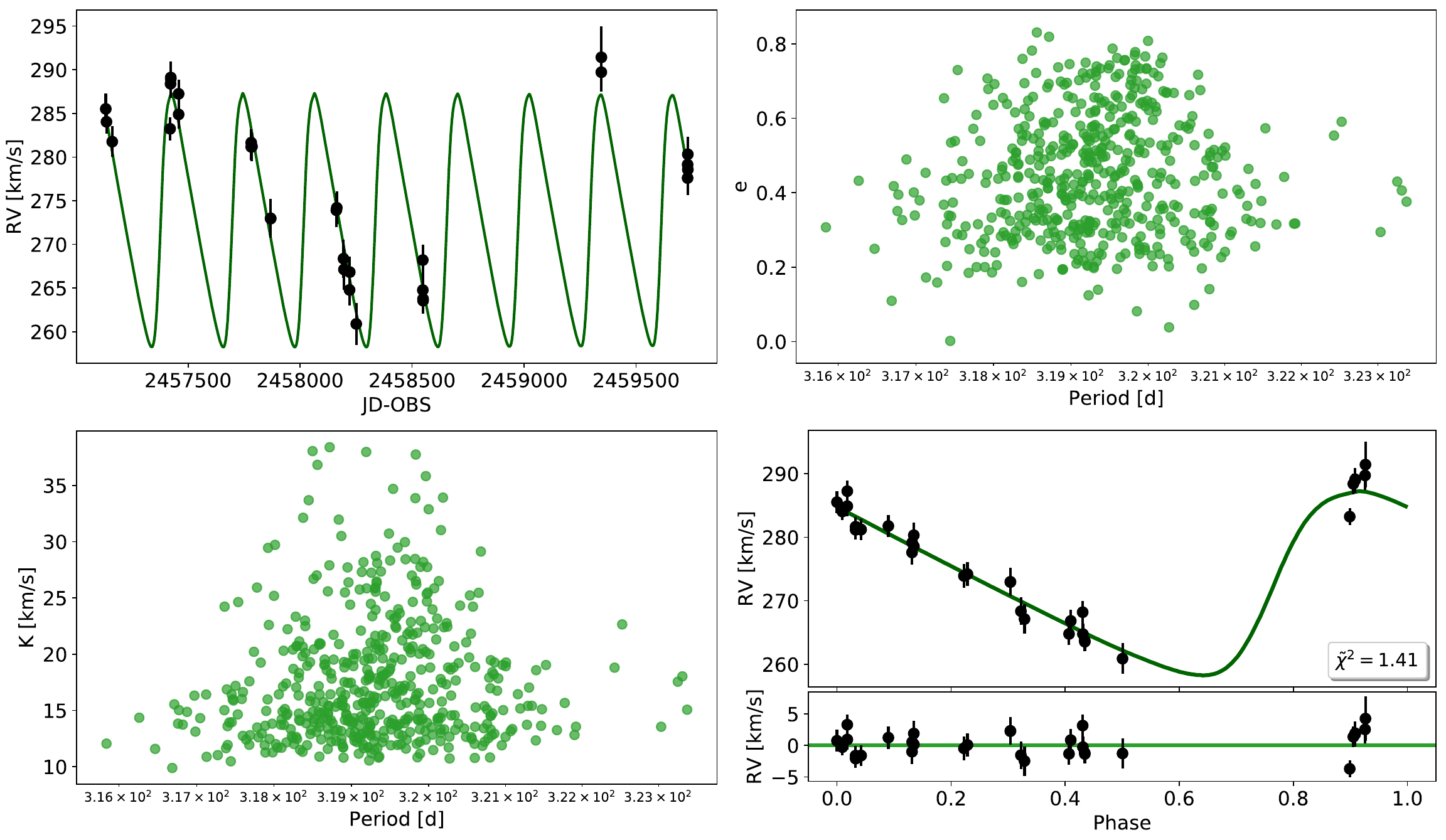}
    \caption{The best-fit orbital solution for the binary candidate ID \symbol{35}1665349 provided by \textsc{Ultranest}, to show what a unimodal orbital solution looks like. The green points in the top right and bottom left panels represent only a subsample of 512 posterior solutions, to avoid overcrowding the plot. The best-fit solution is shown as a green curve overplotted on the observed RV measurements, both in the time space (top left) and phase space (bottom right).}
    \label{fig:MUSE_fit_example}
\end{figure*}

The period distribution of binaries often remains multi-modal due to aliasing, data uncertainties or irregular time sampling with both \textsc{The Joker} and \textsc{Ultranest}. This reduces the number of constrained binaries sensibly relative to the original sample of candidates. The "golden" subset of binaries with well-constrained orbits is identified using clustering techniques (e.g., a Gaussian Mixture Model (GMM) algorithm), which lead to a distinction between stars with unimodal or bimodal solutions, and unconstrained solutions. Based on the classification adopted in previous binary studies (\citealt{Giesers2019}, \citealt{Saracino_2023}, \citealt{MullerHorn2024}), we define that a binary i) has a unimodal solution if $\sigma$(log P)$<$0.5, ii) has a bimodal solution if the period has a bimodal distribution and each of the two peaks satisfies the above criterion. The solution adopted is the one with the largest number of posterior samples associated with it. Finally, a binary has an unconstrained solution if it does not belong to any of the previous categories. An example of a well-constrained, unimodal, binary orbit is illustrated in Figure \ref{fig:MUSE_fit_example}, demonstrating a clear clustering around a single orbital solution.

To determine the final list of binaries with reliable solutions in $\omega$ Cen we adopted the following strategy: 1) we did include all binaries classified as unimodal or bimodal by both methods; 2) we did exclude all binaries constrained by \textsc{The Joker} but not by \textsc{Ultranest} if the number of posterior samples in the former case was below 512. In these cases we were not sure that the few posterior samples provided by \textsc{The Joker} indicated a very informative, but rather inconclusive, solution. 3) we included all binaries classified as constrained by \textsc{Ultranest} and not by \textsc{The Joker}, as this was often due to low number statistics of the latter. The total number of binaries with constrained solutions thus identified is 19, of which 5 are only constrained by \textsc{Ultranest}. The photometric and astrometric properties of this compilation of binaries are listed in Table \ref{tab:astrophoto}, while their orbital properties are presented in Table \ref{tab:constrained}, along with a comment specifying whether these are unimodal or bimodal solutions. Among the 19 constrained binaries there are 3 that have e = 0. This is the result of a test carried out to verify whether some binaries could be more easily and better constrained by adopting a fixed rather than variable eccentricity. The latter is indeed the most uncertain parameter among all we can retrieve but we expect some binary systems to have circular orbits, especially with periods P $<$ 2 d. The observed RV curves as well as the best-fit orbital solutions for the 19 constrained binaries in $\omega$ Cen, are presented in the Appendix, in Figures \ref{fig:plot}, \ref{fig:plot1}, \ref{fig:plot2} and \ref{fig:plot3}, both in time and phase space. On average, binaries with constrained solutions have a higher number of RV measurements. None of these systems were specifically targeted and this is simply the result of our observational setup.

\begin{figure}
    \centering
	\includegraphics[width=0.49\textwidth]{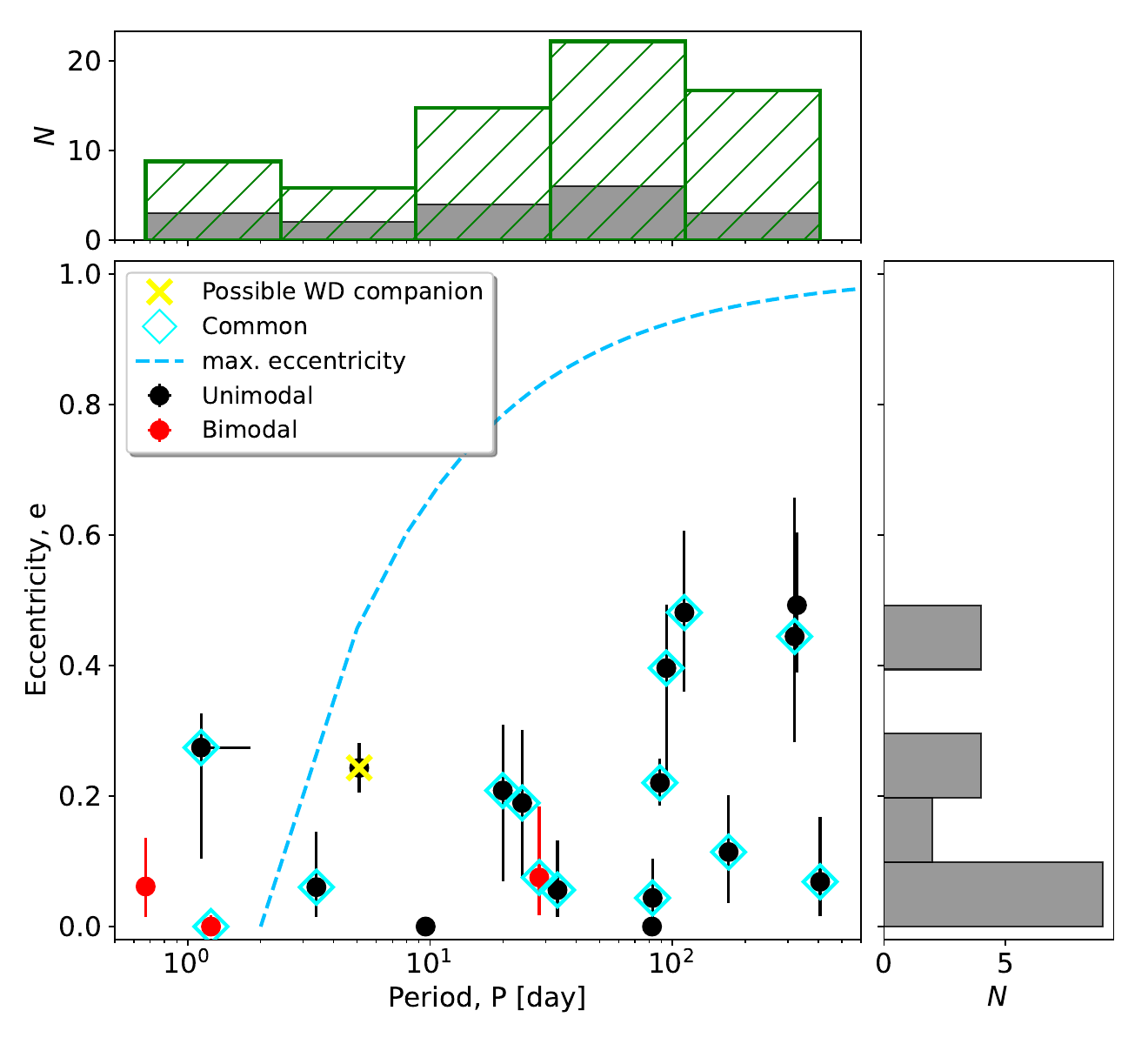}
    \caption{Eccentricity - Period plot of the well-constrained binaries in $\omega$~Cen. Binaries with unimodal and bimodal solutions in the posterior period sampling are shown as black and red dots, respectively. Cyan diamonds identify binaries constrained by both \textsc{The Joker} and \textsc{Ultranest}. The period distribution of the 19 binaries is shown in gray in logarithmic scale and spans the range between 1 and 500 days with multiple peaks. The green histogram shows the same period distribution, once corrected for the incompleteness derived in Section \ref{sec:testI}. The eccentricity distribution, on the other hand, varies only from 0 to 0.5, with a peak around 0.1/0.15, i.e. prefers low eccentricity orbits. The dashed cyan line defines the maximum expected eccentricity for a given period. Binaries with P$<$ 2 days are expected to have circular or close to circular orbits. The reported values are from \textsc{Ultranest}. The values from \textsc{The Joker} can be found in Table \ref{tab:constrained} in the Appendix.}
    \label{fig:Pe_plot}
\end{figure}

Among the 222 likely binary systems ($P_{var}$>0.8) of $\omega$ Cen, 19 have constrained orbital properties, representing 9\% of the entire sample. We present their properties in Figure \ref{fig:Pe_plot} using a linear-log plot of their eccentricity vs period distributions. Black and red dots refer to binaries with unimodal and bimodal solutions, respectively and large cyan diamonds highlight the 14 binaries constrained by both methods. Eccentricity and period distributions are also shown as gray histograms in the vertical and horizontal panels, to better visualize the results. The green histogram overlaid on the observed period distribution of the binaries shows the distribution corrected for incompleteness (see Section \ref{sec:testI} for details).

While binaries span a wide period range (from less than 1d to a few hundreds days), their distribution is not uniform due to observational sensitivity peaks. In other words, the sensitivity to different orbital periods changes over the entire range due to time sampling and cadence of our observations. The result is that, overall, we are significantly more sensitive to short-period binaries than long-period ones. In this context, it is interesting to note that we observe an overabundance of binaries with periods larger than 10-20 d compared to binaries with shorter periods. For instance, if we assume that the underlying period distribution of the binaries in $\omega$ Cen were uniform across the entire range, we would have expected to observe the opposite trend. The fact that this is not the case suggests an overabundance of moderately long periods in the cluster, a possibility that we will explore in the next sections. Eccentricity instead ranges from 0 to 0.5, with no highly eccentric binaries detected. This result is not surprising because we know that the eccentricity distribution is biased towards low values, i.e. fewer number of measurements are needed to constrain binary systems with low eccentricity values. On a similar note, Figure~\ref{fig:Pe_plot} also shows a dashed cyan curve representing the maximum eccentricity limit as defined by Equation 3 in \citet{MoeDiStefano2017}. The authors assume circular orbits for binaries with periods P $<$ 2 d due to tidal forces (consistent with both observations and tidal theory of early-type binaries, \citealt{Zahn1975,Abt1990,Sana2012}) and our small sample follows this trend, except for one binary with a rather large eccentricity uncertainty.

\begin{figure}
    \centering
	\includegraphics[width=0.49\textwidth]{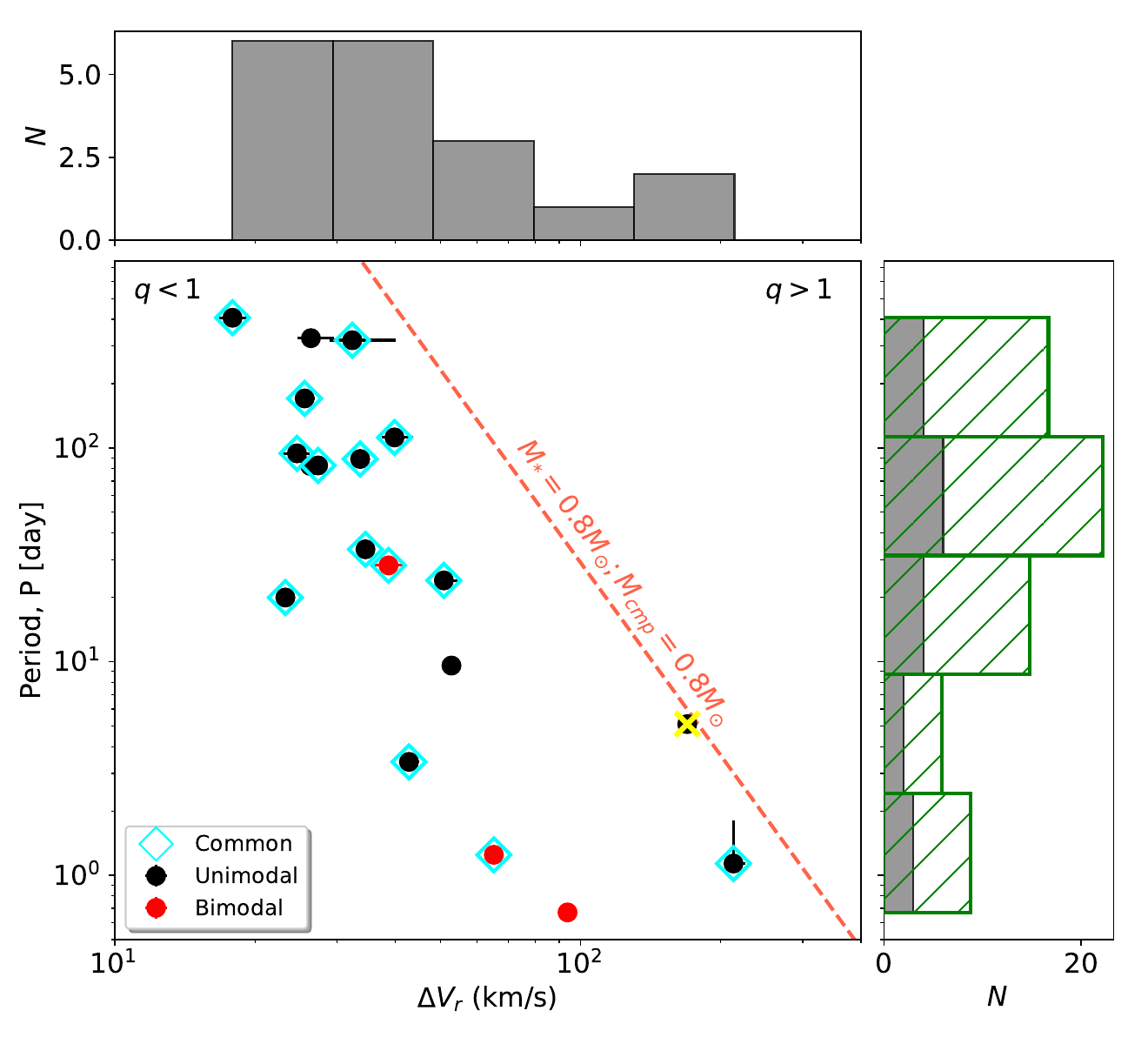}
    \caption{Period - Peak-to-Peak RV variation ($\Delta V_{r}$) plot of the 19 well constrained binaries in $\omega$ Cen. The colour code is the same as in Figure \ref{fig:Pe_plot}, also shown in the bottom-left legend. Stars with large orbital periods and/or high peak-to-peak RV variability occupy the upper right region of the plot. The dashed orange line defines the locus where equal-mass binaries composed of two stars with mass $0.8M_{\odot}$ -- the maximum stellar mass expected in $\omega$ Cen given its absolute age -- are located when observed edge-on. Binaries with q<1 are on the left of the orange line, while binaries with q>1 are on the right. The 1D period and $\Delta V_{r}$ distributions of the 19 binaries are also shown in the figure, in gray in logarithmic scale, spanning a large range of values. As in Fig. \ref{fig:Pe_plot}, the green histogram refers to the distribution of periods, once corrected for the results of Section \ref{sec:testI}. The reported values are from \textsc{Ultranest}. The values from \textsc{The Joker}, are provided in Table \ref{tab:constrained} in the Appendix.}
    \label{fig:PK_plot}
\end{figure}

In Figure \ref{fig:PK_plot} we used the same data to show the peak-to-peak RV distribution ($\Delta V_{r} = 2K$) of the 19 constrained binaries as a function of their period distribution. Binaries in $\omega$ Cen span a range of amplitude values, from tens to hundreds of km/s, peaking at around 30-40 km/s. Two systems even exceed 100 km/s. This plot is generally used to identify regions where binaries with massive companions may reside, based on predictions by \citet{Clavel2021}. The red shaded line identifies the locus of binaries with mass ratios q = $M_{2}$/$M_{1}$ = 1 and where each of the two components has a mass of 0.8 $M_{\odot}$, corresponding to a MS turn-off star in the cluster. The latter are the most massive stars we can find in a stellar system as old as $\omega$ Cen, if we do not take peculiar stars like blue straggler stars (BSSs, \citealt{Ferraro1999}) into consideration. Stars more massive than that have already evolved and died. Binaries to the left of the red dashed line have mass ratios q $<$ 1 (i.e. secondary components with masses lower than the 0.8 $M_{\odot}$ primary), while those to the right have mass ratios q $>$ 1 (i.e. more massive secondaries). None of the binary systems with constrained properties in $\omega$ Cen fall into the latter region, meaning there is no evidence of binary systems with potential NS and BH candidates within the sample. It is worth noticing, however, that binaries are generally shifted to the left of this plot, due to velocity damping and inclination effects. The term "velocity damping" refers to the phenomenon of intrinsic reduction of the velocity amplitude of binary systems made up of two stars with similar masses. More details are provided in Section \ref{sec:whitedwarf}.

\begin{figure*}
    \centering
	\includegraphics[width=0.8\textwidth]{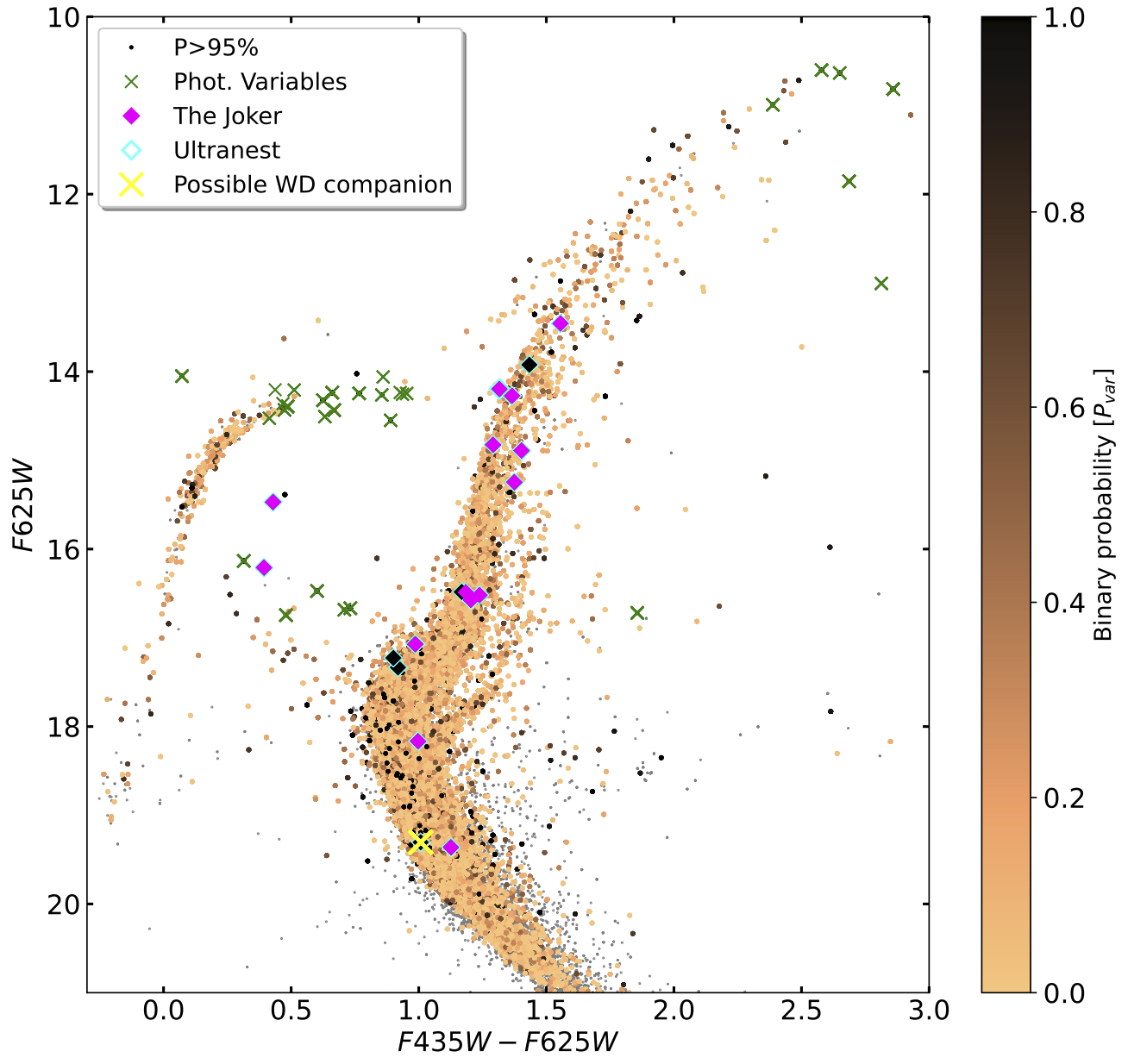}
    \caption{Color-magnitude diagram of $\omega$ Cen, where all stars with MUSE spectra are presented, colour-coded for their probability to be in binary systems. Light colors identify likely single stars while dark colors indicate binary candidates. Small green crosses are photometric variables identified thanks to the cross-correlation with the catalogs by \citet{Clements_2001} and \citet{Braga_2020} and were discarded from the subsequent analysis. Binary systems with constrained solutions are also shown: those constrained by \textsc{The Joker} are highlighted as large pink diamonds, while those constrained by \textsc{Ultranest} are highlighted as large cyan open diamonds. The vast majority of the binaries have been constrained by both methods. The large yellow cross identifies a candidate WD companion to one of the constrained binaries with MS primary.}
    \label{fig:cmd_constrained}
\end{figure*}

Figure \ref{fig:cmd_constrained} summarizes the results of this study from a photometric perspective. It presents the HST (F435W-F625W, F625W) colour-magnitude diagram (CMD) of $\omega$ Cen, where all stars observed in the MUSE field of view for multiple epochs are shown. Each star is color-coded for its probability to be a binary system ($P_{var}$) according to the investigation presented in Section \ref{sec:obs}. Dark colors indicate high probabilities of being binaries, while light orange suggests single stars. Green x-shaped symbols identify photometric variables from the literature \citep{Clements_2001,LebzelterWood2016,Braga_2020}, while large diamonds highlight the 19 binaries constrained in this work. Of those, the 14 coloured in violet identify those constrained by both \textsc{The Joker} and \textsc{Ultranest}, while the remaining 5 are those only constrained by \textsc{Ultranest}. It is worth noticing that the binary probability exceeds 99\% ($P_{var}>0.99$) for all constrained binaries, confirming the reliability of the method used to identify binaries in clusters. The 19 binaries with constrained orbital properties span the entire F625W range, from the red giant branch (RGB, F625W$\sim$13) down to the MS (F625W$\sim$19), with two systems also occupying the BSS region. We note here that most of the constrained systems are on the RGB (over 60\% of the total), despite MS stars dominating our sample in number. This is because RGB stars are significantly brighter and have lower RV uncertainties.

\subsection{A possible white dwarf candidate}
\label{sec:whitedwarf}
Based on the results in Figure \ref{fig:PK_plot}, there is no evidence for massive dark companions (BHs or NSs) in the sample of binaries we have constrained in $\omega$ Cen with the MUSE data available. However it is still important to investigate whether or not any of the constrained binaries contains a WD companion, which is not easy to tell given they share similar masses with stars in $\omega$ Cen.

Star clusters are characterized by a well-populated MS, as well as by a slightly redder sequence containing binaries composed of two MS stars. Since the two components of a binary system are too close to each other to be spatially resolved at the distance of $\omega$~Cen, such binaries appear as a single, yet brighter source. In particular, while the mass ratio increases, the system follows an arc that first gets redder and then returns back closer to the primary stars color. When the mass ratio is equal to 1, the MS binary has the same colour, but appears 0.75 mag brighter than either of its constituents. Photometry can then be used to find binaries and it has indeed been extensively used in the literature to estimate the binary fraction of old stellar systems such as Galactic GCs (see e.g. \citealt{Milone_2012}). The spectroscopic detection of binaries is instead influenced by the luminosity damping, i.e. the RV amplitude measured with MUSE is linearly damped by the flux ratio $f_{2}$/$f_{1}$ of the stars (see \citealt{Giesers2019}). In the extreme case of a binary made of two similarly bright stars, the spectral lines of the two components cannot be individually resolved so that the measured RV amplitude is 0. High mass ratio binaries are then the most difficult to detect spectroscopically but the easiest to identify photometrically. For stars on the MS, where this rule holds, such a different behavior between photometry and spectroscopy can be exploited to investigate the nature of the unseen companion.

Briefly, for a given orbital period, the maximum amplitude of a binary system with a MS companion can be predicted, assuming an edge-on configuration and considering the mass-ratio dependent damping factor. If the observed amplitude exceeds this value, a MS companion appears unlikely, suggesting that a WD may be present instead. Further, if the system's position on the CMD appears inconsistent with the one predicted by the mass ratio of its components, this lends further evidence to the idea that the companion is not a MS star but a WD. To test the method, we applied it to the three binaries in our sample located on the MS and found that the results for two of them were fully consistent with having a MS companion.

The binary system with ID \symbol{35}7634619 instead provided the most interesting result: the observed RV amplitude of the binary was too high compared to the predicted one and the derived mass ratio was not consistent with its position in the CMD of Figure \ref{fig:cmd_constrained} (yellow cross), exactly on top of the MS of the most populated stellar population of $\omega$ Cen (we refer the reader to \citealt{Bellini_2017} for a detailed discussion about the stellar populations of the cluster). This suggests that the secondary is a WD, because it does not contribute light to the system, nor does it produce the damping factor mentioned above. We estimated a mass for the WD candidate of about 0.97 $M_{\odot}$. The position of this binary in Figures \ref{fig:Pe_plot}, \ref{fig:PK_plot} and \ref{fig:cmd_constrained} is highlighted with a large yellow cross.
Further details on the method will be provided in an upcoming publication (Dreizler et al., in prep).

\section{The binary population of \texorpdfstring{$\omega$}{Omega} Cen}
\label{sec:simulations}
The primary goal of this study is to determine the orbital properties of as many binaries in $\omega$ Cen as possible, while investigating the overall characteristics of the cluster’s binary population. By inferring the period and mass ratio distributions, we aim to gain insights into the dynamical processes that have shaped the cluster’s evolution. This type of analysis has not been done for $\omega$ Cen and is rare even in more typical GCs (e.g., 47 Tuc; \citealt{MullerHorn2024}).

One key question is whether the apparent overabundance of binaries with P $>$ 10 d is a genuine feature or an artifact of incomplete sampling. 
$\omega$ Cen has no available binary population models based on dynamical evolutionary models (e.g. \citealt{Askar2018}, \citealt{Kremer_2020}), unlike the case of 47 Tuc \citep{Ye_2022}. To overcome this, we created our own distribution of binaries, with known binary populations and orbital parameters, following that of \citet{Wragg2024}. These mock samples, referred to as simulations, were processed identically to the real data (e.g., using the same probability methods and orbital fitting software) to assess completeness and purity of the sample and derive a corrected distribution for key binary properties, especially orbital periods, which strongly affect detectability. We generated two sets: one for binaries where the secondary is a MS star (simulation I), and another for binaries with dark remnants such as WDs, NSs, or stellar-mass BHs (simulation II). Both sets used the same time stamps and velocity errors as the real data to ensure identical observational limitations.

\subsection{Simulation I: MS companions}
\label{sec:testI}
\subsubsection{Setup}
To create the first mock sample, we followed the guidelines in \citet{Wragg2024} with some modifications. We matched the size of the simulated sample to the observed one (i.e. the full MUSE sample of stars, either single or binaries), assigning a random fraction of objects to binary systems. Although $\omega$ Cen has a low overall binary fraction ($2.1\pm0.4\%$; \citealt{Wragg2024}), we assumed a 50\% binary fraction for two reasons: 1) to ensure a large enough binary sample to reliably assess completeness and purity as a function of orbital properties, and 2) the binary fraction does not affect the recovery of individual orbital parameters.
The primary stars' properties were randomly selected from the observed sample, while the secondary components were assigned based on the following parameter distributions:
\begin{itemize}
    \item Mass ratio: uniform distribution, in agreement with \citet{Ivanova_2005}.
    \item Inclination: uniform distribution of $\cos{i}$ between 0 and 1.
    \item Cluster systemic velocity: normal distribution, with a mean value of 250 km/s and a velocity dispersion $\sigma$ of 20 km/s, as derived from the MUSE data.   
    \item Orbital period: log normal distribution, with a mean value of 10 d and a standard deviation of $10^{1.5}$ d.
    \item $t_{0}$: uniform distribution between 0 and the value of P for a given binary.
    \item Eccentricity: beta function, with $\alpha$=2 and $\beta$=5. Binaries with P $<$ 2 d have a fixed eccentricity of 0.
    \item Argument of periapsis $\omega$: uniform distribution between 0 and $2\pi$.
\end{itemize}
The magnitudes of the secondary components have been assigned by using an isochrone of appropriate parameters for $\omega$ Cen and by finding the closest match in terms of mass along the MS.
Of the initial 50\% of stars classified as binaries, only 34\% remained as such at the end of the simulation. The others were either "soft binaries", too weakly bound to survive in the cluster environment\footnote{This criterion will be explored in more detail in Section \ref{sec:LPB}, where a direct comparison will be made with CMC simulations of two types of GCs \citep{Kremer_2020}, one as dense as $\omega$ Cen and a denser one.}, or binaries where the more massive star overflowed its Roche lobe.

The final sample, comprising single stars and binaries with known orbital parameters, was processed identically to the observations. First, we derived the probability for each star being part of a binary, considering those with $P_{var} > 0.8$ as reliable. These were then processed using \textsc{Ultranest}\footnote{We chose to only use \textsc{Ultranest} in this part, for the orbital fitting of the simulated binaries. The latter method, in fact, proved to be the most reliable and robust, given the greater number of posterior samples it provides, compared to \textsc{The Joker}.} to determine their orbital properties by adopting the priors listed in Table~\ref{tab:prior_PDF}.

\subsubsection{Results}
\label{sec:simulated_pop}
In the following, we investigate our ability to recover the orbital properties of simulated binary stars. We discard the small fraction ($\sim$1\%, see also \citealt{Wragg2024}) of false positives, i.e. single stars whose single-epoch RV measurements show enough random dispersion to violate the $P_{var}>0.8$ threshold and be considered binaries.
We only use the sub-sample of objects that are actually true binaries and have $P_{var}>0.8$ so that we can make a one-to-one comparison between the orbital properties recovered by \textsc{Ultranest} and the simulated values. All binaries with six epochs or more, representing 33\% of the original amount of binaries in the simulation, were finally processed with \textsc{Ultranest}. Of them, 40\% could actually be constrained, with a unimodal or bimodal solution. The remaining 60\% were unconstrained, with the posteriors covering the full range of parameter space. Interestingly, the fraction of constrained binaries in the simulation is significantly higher than in the real sample (40\% vs 9\%). We discuss possible reasons for this discrepancy in Section \ref{sec:LPB}.

\begin{figure}
    \centering
	\includegraphics[width=0.4\textwidth]{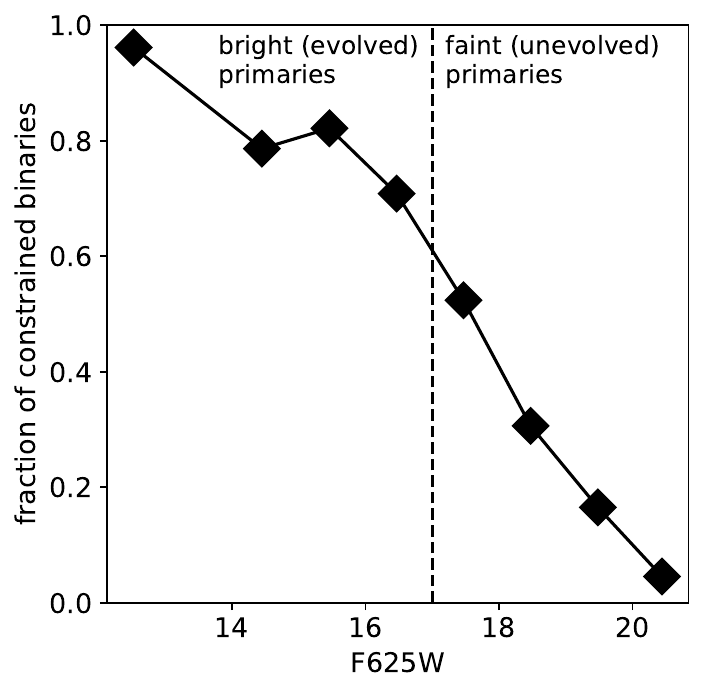}
    \caption{The fraction of constrained binaries in simulation I as a function of the F625W magnitude. Binaries with evolved (bright) primaries are more frequently constrained than binaries with unevolved (faint) primaries. The separation between the two categories is highlighted by the black dashed vertical line.}
    \label{fig:F625-comp}
\end{figure}

Figure \ref{fig:F625-comp} presents the fraction of constrained binaries as a function of F625W magnitude in the simulation. We observe that binaries with bright primaries have completeness levels beyond 70-80\%, significantly dropping when moving to binaries with fainter primaries. This is expected and mainly driven by the fact that the observed mean RV uncertainty of stars increases towards fainter magnitudes. In particular, if we arbitrarily use the base of the RGB at F625W = 17 to divide the sample of binaries into two categories -- those with evolved primaries (mainly belonging to RGB and HB) and those with unevolved primaries (e.g., sub-giant branch and the MS) -- we observe that the fraction of constrained binaries is very different, corresponding to 77\% and 33\%, respectively.

\begin{figure*}
    \centering
	\includegraphics[width=\textwidth]{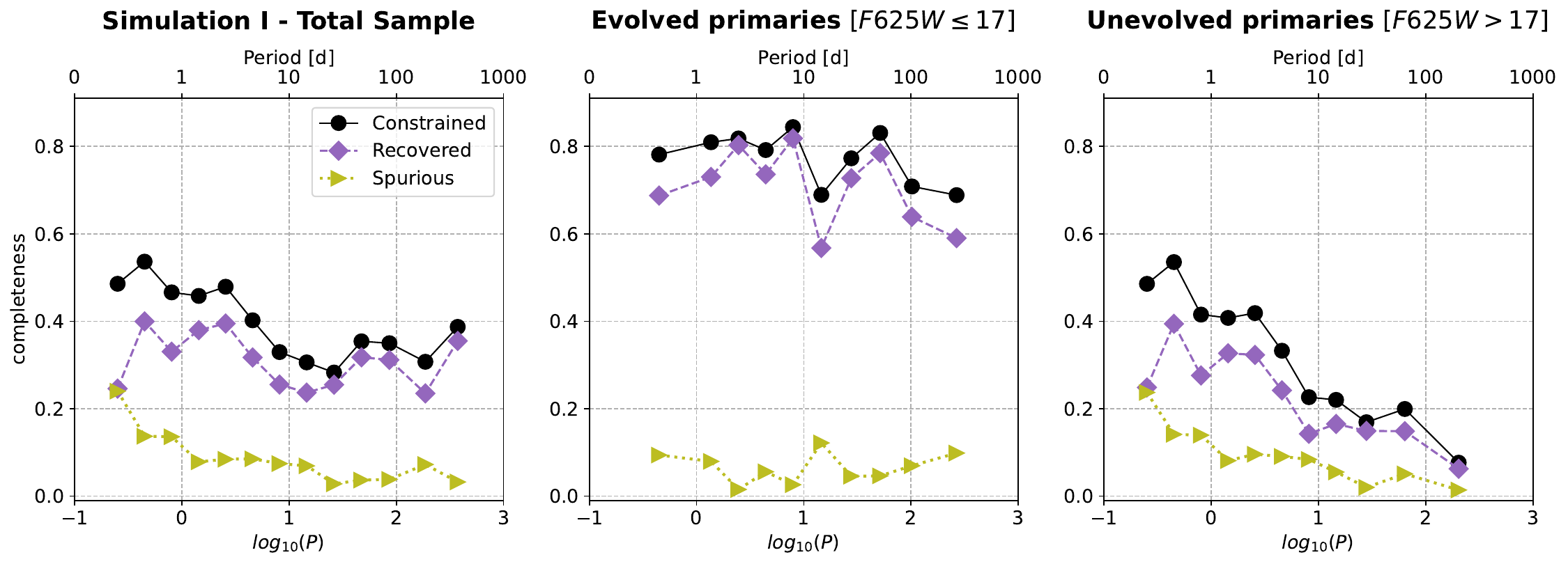}
    \caption{The completeness curves (or constrained fractions) of binaries in the first set of simulations (or simulation I), as a function of the orbital period~P. The left panel represents the total sample, the middle panel shows binaries with evolved primaries (F625W $\leq$ 17), and the right panel focuses on binaries with unevolved primaries (F625W $>$ 17). Constrained binaries are shown as black dots, binaries with well-recovered periods as violet diamonds, and spurious solutions as olive-green triangles. This plot highlights the dominance of binaries with MS primaries in the sample, although RGB or HB binaries exhibit significantly higher completeness values across all periods.}
    \label{fig:simulationI_plot}
\end{figure*}

In Figure \ref{fig:simulationI_plot}, we instead show the fraction of constrained binaries relative to the total number of detected binaries in simulation I, as a function of the orbital period P. The black dots represent the overall completeness, the violet diamonds show the fraction of binaries with well-recovered orbital periods (i.e. binaries whose period P is within 10\% of the simulated one) and the olive-green triangles indicate binaries with spurious solutions (i.e. binaries whose period P differs by more than 10\% from the simulated one). The left panel shows results for the entire sample, while the middle and right panels split the results into binaries with evolved primaries (F625W $\leq$ 17) and unevolved primaries (F625W $>$ 17), respectively. The grouping in each of the three panels was done in such a way as to guarantee a minimum number of 50 objects per bin (to avoid bins with low number statistics) and a minimum separation between them of log(P/1d) $>$ 0.25. We show the recovered values, since they are the only ones available for the observations.

In the left panel, we observe that the fraction of constrained binaries decreases from about 50\% for periods shorter than 1 d to around 30\% for periods between 10 and 100 d. While this decline is expected due to the dataset's sensitivity, it is notable that the decrease is not entirely smooth, showing small oscillations with slight increases in certain bins. Specifically, we recover the properties of approximately 50\% of binaries with MS companions for P$<$1 d, 42\% for 1 $\leq$ P $<$ 10 d, 32\% for 10 $\leq$ P $<$ 100 d, and 30\% for 100 $\leq$ P $<$ 500 d, indicating a higher recovery rate for shorter-period binaries. These completeness values exclude binaries with spurious solutions, as this information is not available for the observed dataset. However, simulation I shows that binaries with spurious solutions make up 22\% of the total, but their contribution is minimal (less than 10\%) for the period range of 1 $\leq$ P $<$ 500 d, with a significant increase only for P$<$1 d due to aliasing and sparse time sampling, which complicates orbital fitting. It is worth noting that this trend is robust and not sensitive to specific period or mass ratio distributions, as similar trends would be expected regardless of the underlying distributions.

As shown in Figure \ref{fig:F625-comp}, the completeness for binaries with evolved primaries (F625W $\leq$ 17, middle panel) is notably higher—up to 30\% more—than for binaries with unevolved primaries (F625W $>$ 17, right panel). Additionally, the contribution of spurious solutions remains low, never exceeding 10\% across all periods. Specifically, of the 22\% of binaries with spurious solutions, only 3\% have evolved primaries, while 19\% are from systems with unevolved primaries. This is reassuring, considering that over 60\% of the constrained binaries in $\omega$ Cen feature evolved primaries.

Although completeness for evolved primaries exceeds 80\%, their lower numbers mean that the overall trend is still driven by unevolved primaries, which make up 69\% of the total sample. Only for binaries with periods longer than log(P/1d) $>$ 1.5 (roughly 30 d) do evolved primaries significantly impact the sample, as these stars are generally found in wider orbits due to their evolutionary stage. We do not provide completeness values for the minimum secondary mass $M_{2,min}$ or K, nor track their behavior, as RV damping affects their distributions in a complex manner.

\subsection{Simulation II: WD, NS and BH companions}
\label{sec:darkremnants}
\subsubsection{Setup}
Simulation II aimed to investigate whether $\omega$ Cen hosts binary systems consisting of a star and a dark object, mainly a NS or a BH, which may have eluded detection due to the current observational setup. This simulation followed the same prescriptions as Simulation I, with one key change: the binary mass ratio was uniformly distributed between 1 and 5. We adopted the following classification: WDs had masses up to 1.4 $M_{\odot}$, NSs had masses between 1.4 $M_{\odot}$ and 2.5 $M_{\odot}$ and BHs had masses above 2.5 $M_{\odot}$.

\subsubsection{Results}
We generated as many sources as in Simulation I, of which 26\% were classified as binaries based on the binary hardness and Roche Lobe overflow criteria. The probability method identified 5 559 sources with $P_{var} > 0.8$ as potential binaries. After verification, 89\% were genuine binaries, while 11\% were false positives\footnote{Interestingly, the fraction of false binaries is 10 times higher than in simulation I. This supports the idea that the probability method used to identify binaries becomes less reliable if the binary fraction and the average K value of the binaries increase.}. As in simulation I, only the 4 933 real binaries were analyzed using \textsc{Ultranest} to compare simulated and recovered properties.

\begin{figure*}
    \centering
    \includegraphics[width=0.96\textwidth]{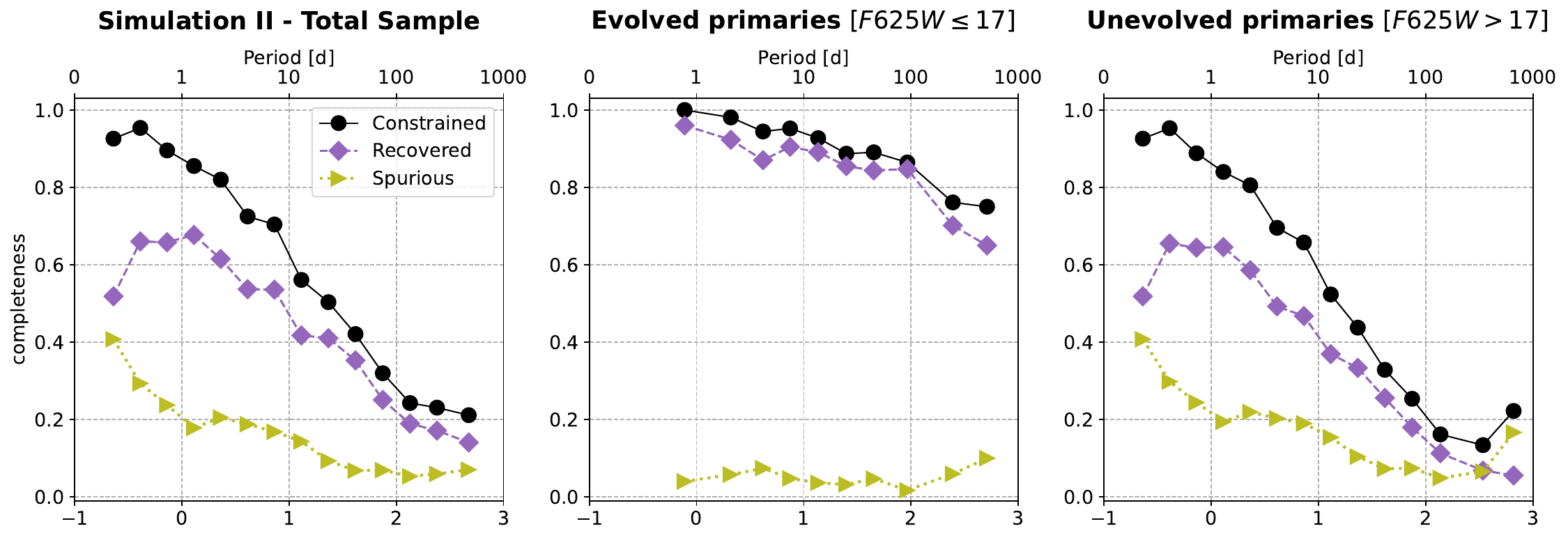}
    \includegraphics[width=0.96\textwidth]{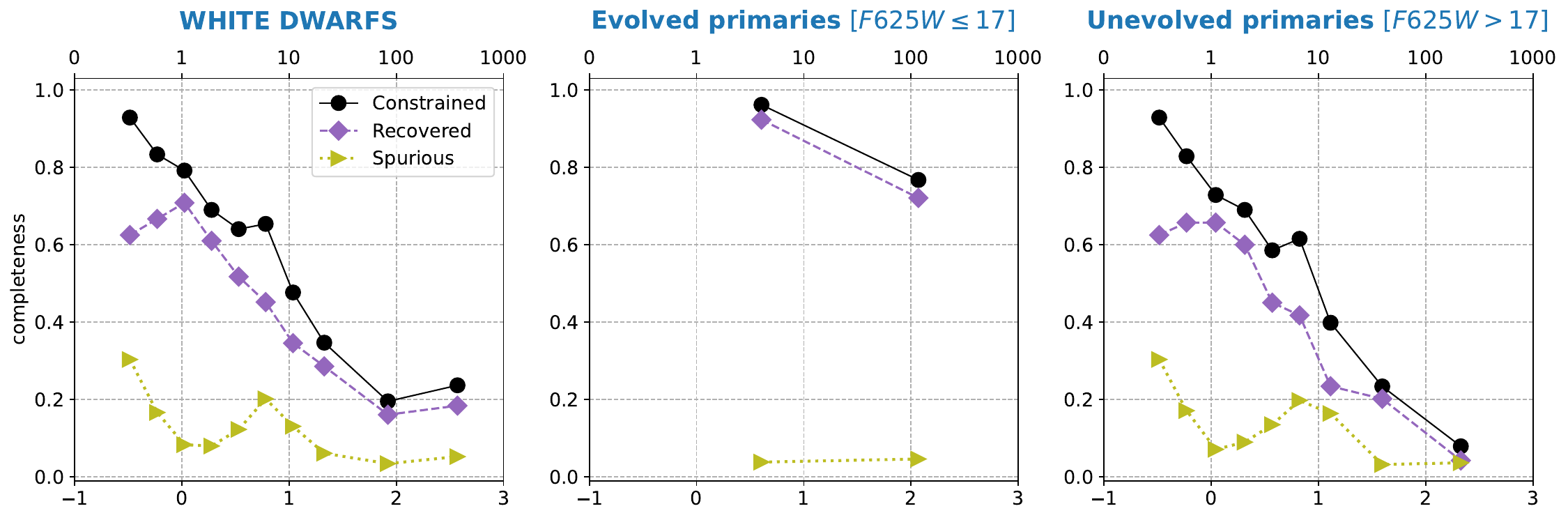}
    \includegraphics[width=0.96\textwidth]{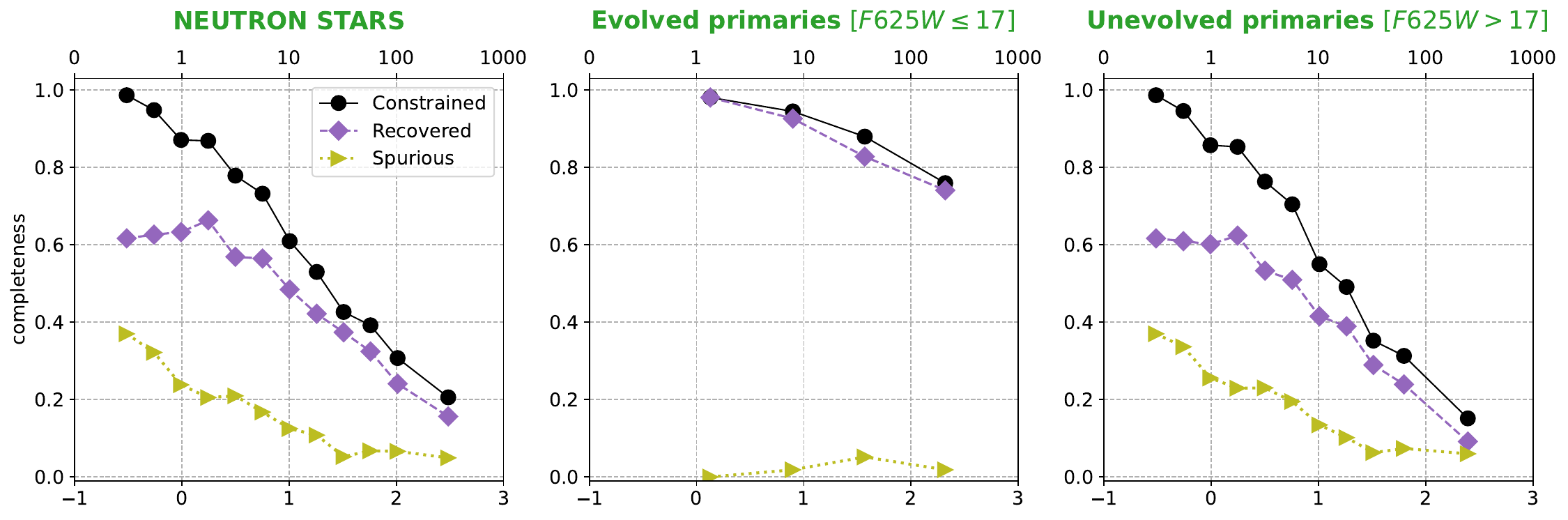}
    \includegraphics[width=0.96\textwidth]{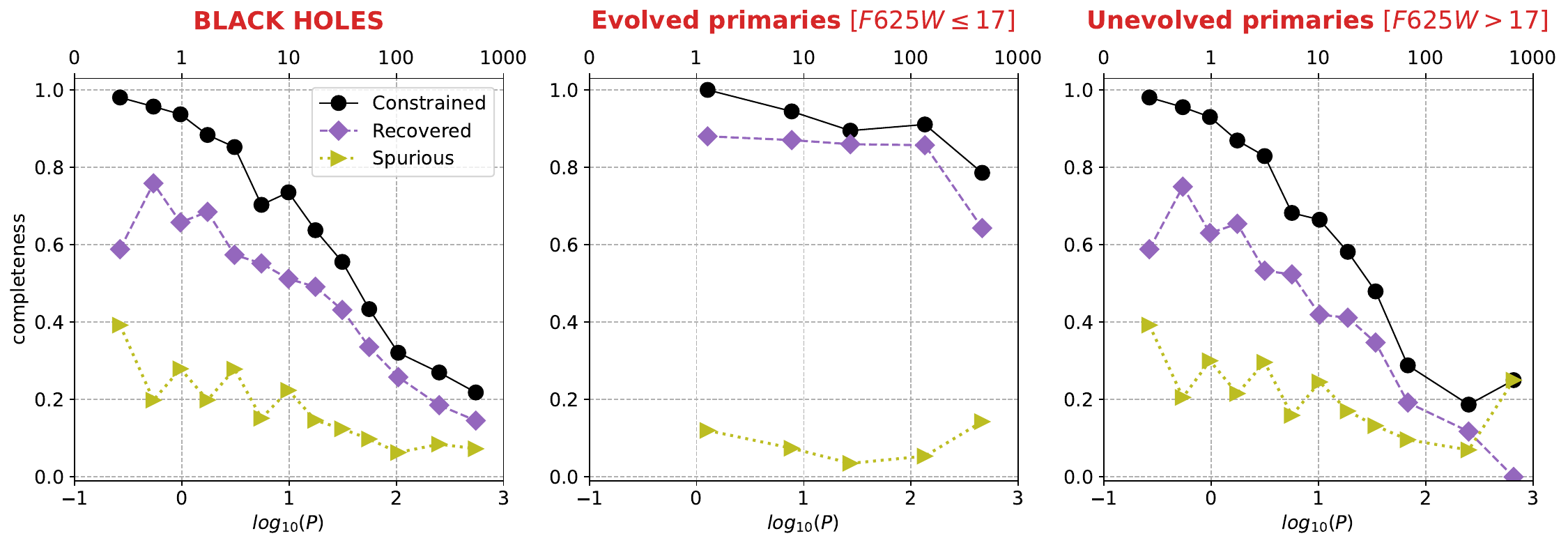}
\caption{Same as in Figure \ref{fig:simulationI_plot}, but for simulation II. The first row shows the total sample, while the subsequent rows correspond to specific classes of objects: WDs, NSs and BHs, respectively. The results are also presented for two subsets of binaries, those with F625W$\leq$17 (evolved primaries, middle panel) and those with F625W$>$17 (unevolved primaries, right panel).}
\label{fig:simulationII_plots}
\end{figure*}

In Figure \ref{fig:simulationII_plots} we present the completeness and purity of simulation II as a function of the orbital period distribution of the constrained binaries. Different rows correspond to a specific class of objects: the first row refers to the total sample, while the subsequent rows refer to binaries with WD, NS and BH secondaries. The columns correspond to different primary star categories: the whole sample of stars is shown in the first column, evolved stars (F625W $\leq$ 17,  e.g., RGB, HB) in the second, and unevolved stars (F625W $>$ 17, e.g., SGB, MS) in the third. As in Figure \ref{fig:simulationI_plot}, constrained binaries are shown in black, binaries with well-recovered periods in violet, and those with spurious solutions in olive-green. 

The trend observed in the distribution of orbital periods is interesting. It does not depend on the class of dark objects analyzed and simply shows that, if there are no additional biases that can alter the trend, the setup of our observations is such that we are very sensitive in identifying short-period binaries (with completeness close to 100\%), with sensitivity gradually decreasing towards longer periods. The fraction of binaries with spurious solutions rises sharply (up to over 40\%) for periods shorter than P=1 d, regardless of the secondary type, of which almost 20\% have unreliable \textsc{Ultranest} results due to sparse time sampling. Notably, completeness remains above 80\%, even for periods exceeding 100 d for binaries with evolved primaries (middle panel), with 90\% or more having well-recovered periods. In contrast, completeness drops to as low as 10\% for binaries with unevolved primaries (right panel) over the same period range.

These results suggest that if binaries with BH or NS companions to evolved stars were present in $\omega$ Cen with orbital periods of few hundred days, they would likely have been detected and their orbital properties accurately recovered, especially since more than 60\% of the constrained binaries in the observed sample have evolved primaries, for which completeness is nearly 100\%.

In Table \ref{tab:dark_fractions} we report the fractions of constrained binaries (or completeness) for different ranges of P and $M_{2,min}$, both for the overall sample of dark remnants and for the three categories (WDs, NSs and BHs) individually. The results show that more than half (50\%) of the binaries with dark remnant companions with periods shorter than 10 d have been successfully recovered by \textsc{Ultranest}, with this percentage decreasing toward long orbital periods, where the sensitivity of our observational setup drops. At the same time, if we observe the trend of the minimum companion mass for the different groups of remnants, we realize that binary systems containing BHs are the easiest to detect and best to constrain, compared to binary systems containing NSs and WDs, in that order. This is not surprising since binaries with BHs are usually the ones that produce the largest signal in terms of RV variation.

\begin{table*}
\caption{Quantitative estimate of the fraction of well-recovered binaries with dark companions (WDs, NSs and BHs companions) for different ranges of period and minimum secondary mass.}\label{tab:dark_fractions}
\centering
\begin{tabular}{|c|c|c|c|c|c|c|c|}
\hline
              & P$<$1d & 1d$\leq$P$<$10d & 10d$\leq$P$<$100d & 100d$\leq$P$<$500d & $M_{2,min}$$<$1.5$M_{\odot}$ & 1.5$M_{\odot}$$\leq$$M_{2,min}$$<$3$M_{\odot}$ & $M_{2,min}$$\geq$3$M_{\odot}$ \\ \hline
Dark Remnants & 92\%            & 76.9\%                           & 45.8\%                                & 23.4\%                              & 51.8\%                    & 63.2\%                                   & 74.8\%                        \\ \hline
WD companions & 84.2\%            & 67\%                             & 31.6\%                                & 15.3\%                                & 49.5\%                      & ---                                    & ---                           \\ \hline
NS companions & 94.5\%            & 77.5\%                             & 46.6\%                                & 22.7\%                              & 55.6\%                      & 63\%                                 & ---                           \\ \hline
BH companions & 95\%            & 82.4\%                             & 53.4\%                                & 28.5\%                                & 48.5\%                      & 63.6\%                                 & 74.8\%                        \\ \hline
\end{tabular}
\end{table*}

The analysis of simulation II provides a key insight: if binaries similar to those simulated exist in $\omega$ Cen, the probability of detecting them with the current data is high. However, since no such systems have been observed—specifically, no evidence of NS or BH secondaries among the constrained binaries in the $\omega$ Cen sample—we conclude that these systems are either rare, exist in configurations beyond our detection capabilities or have periods longer than expected from cluster dynamics. In this regard, the findings by \citet{Platais_2024} are particularly relevant. Of the four binaries with significant acceleration they identified in $\omega$ Cen, three likely contain a WD, and the fourth may host a NS. Their lack of stellar-mass BH detections aligns with our results. If confirmed, this would also support \citet{Haberle_2024}'s recent potential detection of an IMBH in $\omega$ Cen's core.

\section{Global binary properties}
\label{sec:corrected}
An important extension of the analysis presented so far would be to derive the overall orbital period distribution for the binary population in $\omega$ Cen. Unfortunately, due to the limited number of binaries (19) with constrained orbital solutions from \textsc{Ultranest} and \textsc{The Joker}, we do not have sufficient statistics to draw definitive conclusions about the global properties of the binary population in the cluster. However, the 1D histograms presented in Figures \ref{fig:Pe_plot} and \ref{fig:PK_plot} show an intriguing overabundance of binaries with P $>$ 10–20~d.

While the current dataset does not allow a complete determination of the period distribution for $\omega$ Cen binaries, valuable insights can be gained by comparing the observed sample to the completeness curves from Figures \ref{fig:simulationI_plot} and \ref{fig:simulationII_plots}. Simulations I and II have demonstrated that short-period binaries (P $<$ 2–3 d) are easier to detect and constrain than long-period binaries, with the detection completeness decreasing for systems with periods of tens of days or more. Notably, this incompleteness is independent of the assumed period distribution in the simulations, as the completeness curves primarily indicate the fraction of binary systems detectable within a specific orbital period range.

Given this, the observed "excess" of binaries with P $>$ 10–20 d, alongside the scarcity of binaries with shorter periods in $\omega$ Cen, appears counter-intuitive and unexpected. Importantly, this pattern cannot be attributed to incompleteness in the dataset. In fact, correcting the observed trend for incompleteness would only amplify the overabundance of long-period binaries. This is qualitatively shown in Figures \ref{fig:Pe_plot} and \ref{fig:PK_plot}, where the green histograms represent the completeness corrected distribution of periods in the observed sample. This suggests that the skew toward longer orbital periods is an intrinsic feature of the binary population in $\omega$ Cen. Specifically, the population appears to be dominated by long-period binaries, with only a few of them having periods shorter than 1–2 d.

This result overall aligns with the predictions of \citet{Ivanova_2005}, who simulated the binary fractions for clusters of varying densities. In their Figure 7, they show that the binary period distribution in dense clusters shifts from a flat distribution (in log P) for loose clusters to a sharply peaked distribution in denser clusters. For clusters with core densities similar to that of $\omega$ Cen—log $\rho_{c}$ = 3.23 $M_{\odot}/pc^{3}$ (1.7 × $10^{3}$ $M_{\odot}/pc^{3}$) as reported by \citet{Baumgardt_2018}—the predicted period distribution is skewed towards longer periods, with a peak between 1 and 5 d. When comparing our observed and corrected distributions (Figures \ref{fig:Pe_plot} and \ref{fig:PK_plot}) to the simulations of \citet{Ivanova_2005}, the shapes are broadly similar, with both showing an excess of long-period binaries and few systems with periods below 1d. A significant discrepancy, however, is that our observed distribution peaks at longer periods (P $>$ 10 d) than predicted (1–3 d), and the overall distribution does not look as flat as predicted.

\citet{MullerHorn2024} performed a similar analysis for the Galactic GC 47 Tuc. Through a MUSE multi-epoch spectroscopic study, they derived the orbital period distribution of constrained binaries in 47 Tuc and compared their results with a simulated population from CMC models tailored to its properties \citep{Ye_2022}. The core density of 47 Tuc (log $\rho_{c}$ = 4.72 $M_{\odot}/pc^{3}$, equivalent to 5.2 × $10^{4}$ $M_{\odot}/pc^{3}$) is much higher than that of $\omega$ Cen. Interestingly, the observed binary period distribution in 47 Tuc peaks at 20–30 d, consistent with our findings in $\omega$ Cen, with long periods being preferred despite selection effects enabling easier detection of shorter periods. This is in stark contrast to the theoretical simulation by \citet{Ye_2022}, who predict a strong peak around 1d in 47 Tuc, with an overabundance of binaries with very short periods (P $<$ 1d). The significant lack of short-period binaries in 47 Tuc, as observed by \citet{MullerHorn2024}, closely mirrors what we find in $\omega$ Cen.

\begin{figure}
    \centering
	\includegraphics[width=0.46\textwidth]{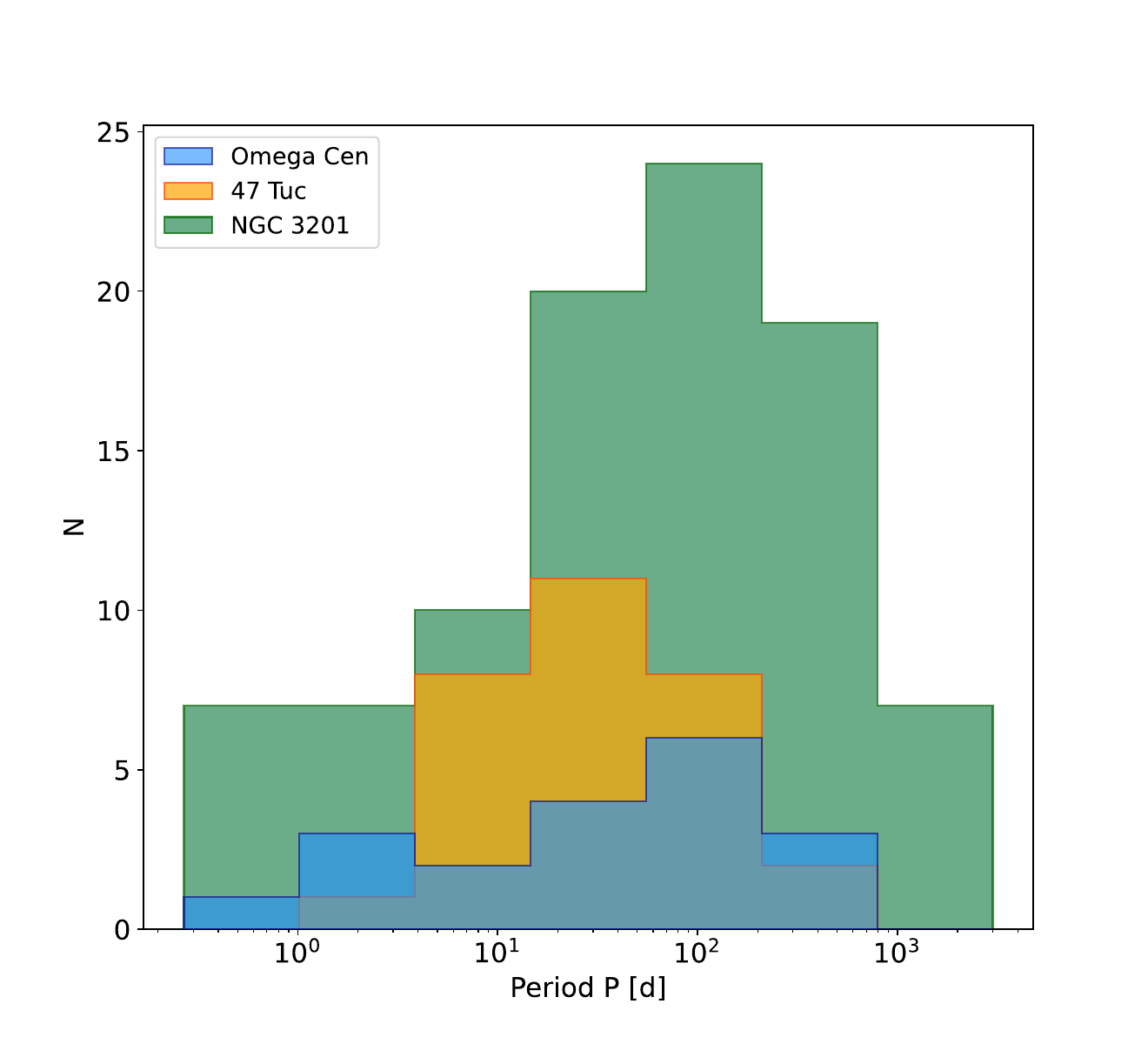}
    \caption{Qualitative comparison of the period distribution of binaries in $\omega$ Cen (blue), 47 Tuc (orange) and NGC 3201 (green). The sample is rather small for the first two clusters, but it seems that a larger number of long-period binaries (P$>$10 d) are observed, regardless of the cluster. This is unexpected since the MUSE observations used in these studies have the best sensitivity for short-period binaries. The histograms are not normalized, to give the reader an idea of the size of the different samples.}
    \label{fig:periods_obs}
\end{figure}

Figure \ref{fig:periods_obs} presents the observed distribution of periods for binaries in 47 Tuc\footnote{47 Tuc has a much higher central density compared to $\omega$ Cen and NGC 3201 by almost two orders of magnitude.} (in orange, \citealt{MullerHorn2024}), $\omega$ Cen (in blue, this work) and NGC 3201 (in green, \citealt{Giesers2019}), the only three clusters for which a detailed characterization of binaries has been performed so far. Interestingly, although the MUSE campaigns have been designed to be mostly sensitive to short-period binaries, we can clearly notice that the majority of the detected binaries have periods of at least 10d or more in all clusters. This is unexpected if compared with the results from \citet{Ivanova_2005} and/or \citet{Ye_2022}.

These findings suggest that the simulated binaries used in theoretical models may not fully capture the intrinsic properties of binary populations in dense clusters like NGC 3201, $\omega$ Cen and 47 Tuc, with a more important discrepancy for high-density environments such as the latter. It may indicate that the treatment of certain physical processes, particularly those occurring in dense stellar environments, requires further refinement.

To improve our understanding, more observational studies across a broader sample of star clusters with well-constrained binary populations are essential. These observations serve two important purposes: first, they can help determine whether the features seen in these stellar systems are common in other clusters with varying core densities. Second, they can provide crucial information to better inform future theoretical simulations.

\section{Observations and simulations compared}
\label{sec:LPB}
Although we observe some similarities between our results and the predictions of \citet{Ivanova_2005} for clusters with central densities similar to $\omega$ Cen, our dataset provides no information on binaries with orbital periods longer than a few hundred days. This is a significant gap, as \citet{Ivanova_2005} suggest that clusters with low central densities may host binaries with periods as long as $10^4$~d or more, contributing to the overall binary distribution.

Recent work by \citet{Platais_2024} using proper motions from multi-epoch HST observations identified four long-period binaries in $\omega$ Cen. These binaries (3 WDs and a potential NS) have orbital periods exceeding 10 years (3–4 × $10^3$ d). This finding, based on a technique with high sensitivity for long-period binaries, is particularly interesting because it targets regions farther from the cluster center, where dynamical interactions are less frequent. The results suggest that a substantial fraction of binaries in $\omega$ Cen may actually be in much wider orbits than previously thought, but survive because they are located on the outskirts of the cluster.

Unfortunately, the MUSE dataset is not sensitive to orbital periods on the order of $10^3$–$10^4$ d, making it unsurprising that we have not detected such binaries. However, an intriguing observation is that only about 9\% of the binaries in our observed sample are constrained by \textsc{Ultranest}, which is significantly lower than the more than 40\% constrained in our mock samples. Several factors could explain this discrepancy, including the fact that our simulations did not include binaries with periods longer than 500 d\footnote{We tested whether randomly changing the RV uncertainties by $\pm$20\% could have a significant impact on the number of constrained binaries in the simulation, but it had essentially no impact.}. Such binaries are treated as single stars in our simulations, based on the hard/soft binary boundary (\citealt{Heggie_1975}; see also \citealt{Ivanova_2005}). This approach assumes that all simulated binary systems with a hardness parameter $\eta$ $<$ 1 are soft and, therefore, are rapidly destroyed due to frequent dynamical interactions in the cluster environment. However, if $\eta$ = 1 proves to be too restrictive a threshold for distinguishing between soft and hard binaries, it could explain part of the observed discrepancy.

To assess the impact of the $\eta = 1$ threshold, we turned to CMC simulations of two other clusters: NGC 3201, which has a central density closest to $\omega$ Cen, and NGC 6752, which has a slightly higher density. It is worth noting that while $\omega$ Cen and NGC 3201 have similar core densities, their formation and evolutionary histories likely differ significantly. For example, recent evidence for a central IMBH in $\omega$ Cen \citep{Haberle_2024} could affect binary dynamics in ways that differ from NGC 3201.

We examined the upper end of the binary period distribution in both clusters and calculated $\eta$ for these systems to evaluate whether our assumption of $\eta = 1$ was appropriate. We tailored the CMC simulations to match our observed sample as closely as possible by selecting binaries with primary masses above 0.35 $M_{\odot}$—a limit imposed by our observational setup—and by limiting our sample to binaries within the same observational field. In both simulated clusters, only a small fraction of binaries had $\eta$ $<$ 1, supporting the appropriateness of our simulation assumptions.

The CMC simulations for NGC 3201 predict binaries with periods up to $10^4$ d, which are short-lived and likely formed dynamically in recent times. However, these systems represent only 8\% of the total binary population, making their absence from our simulations unlikely to be the primary reason for the large discrepancy between the observed and simulated constrained binary fractions (9\% vs. over 40\%). To verify this, we used the period and mass ratio distributions of the binaries from the NGC 3201 simulation to create a new mock sample, which resulted in only a 6\% reduction in the constrained binary fraction (from 41\% to 35\%), without resolving the discrepancy. The observed trends in magnitude and orbital period remained consistent, reinforcing the conclusions drawn in earlier sections.

\begin{figure}
    \centering
	\includegraphics[width=0.50\textwidth]{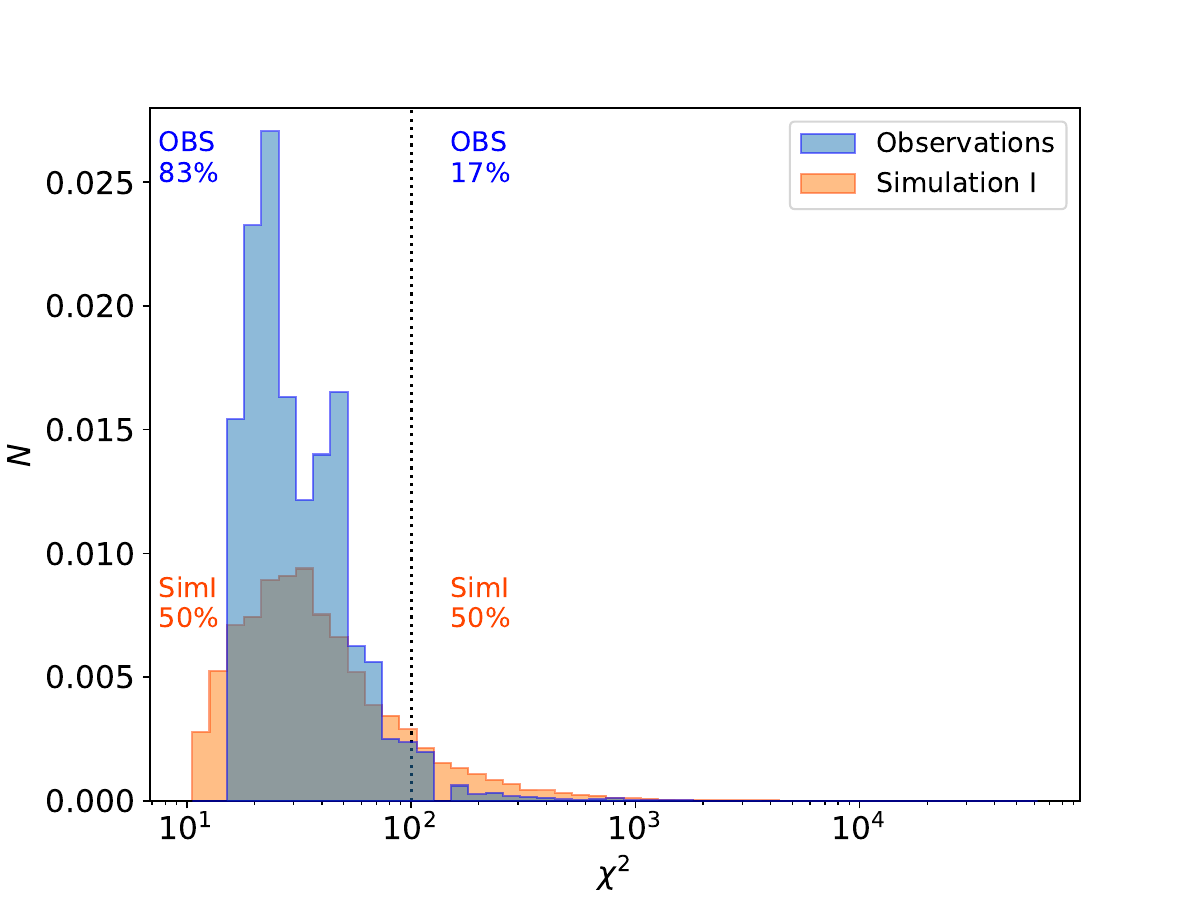}
    \caption{Distribution of $\chi^{2}$ values for binaries in the observations of $\omega$~Cen (in blue) and for the simulated binaries in Simulation I (in orange). The black vertical line arbitrarily divides each sample in two, and we report the fractions of binaries below and above this threshold in the figure.}
    \label{fig:chi-comp}
\end{figure}

We further investigated the discrepancy by comparing the distribution of $\chi^{2}$ values, calculated as the dispersion between a star's RV measurements (weighted by their uncertainties) and the assumption of a flat RV curve (i.e., a single star). As shown in Figure \ref{fig:chi-comp}, most observed binaries have $\chi^{2}$ values between 20 and 50, with very few exceeding 100. In contrast, simulation I shows a much more pronounced tail for high $\chi^{2}$ values, above 100—values typically associated with binaries likely to be constrained by \textsc{Ultranest}.

To quantify the difference between the two distributions, we measured the fraction of binaries with $\chi^{2}$ below and above 100 (threshold assumed arbitrarily), finding that 83\% of the observed binaries had $\chi^{2}$ values below 100, compared to only 50\% in simulation I. This difference of 33\% could explain the discrepancy between the 41\% constrained binaries in simulation I and the 9\% in the observations. Importantly, applying the same analysis to the mock sample with CMC priors yielded consistent results: the difference was reduced by 26\%, again consistent with reconciling the 35\% constrained binary fraction in the simulation with the 9\% in the observations.

This result strongly suggests that the binary period distribution in $\omega$ Cen differs from those in simulated clusters, implying that binary populations found in models of lower-mass clusters cannot be directly applied to $\omega$ Cen. To address this, future work should focus on developing tailor-made simulations of $\omega$ Cen, possibly incorporating a central IMBH, to see the impact of the latter on the binary fraction and its period distribution in the cluster itself. Exploratory work in this direction has recently been done by \citet{Aros_2021}, who examined the binary distributions in a sample of Milky Way globular clusters with and without IMBH, and found a depletion of binaries towards the cluster center, if an IMBH is indeed present.

\section{Conclusions}
\label{sec:concl} 
This paper, the second in the series, focused on the investigation of the orbital properties of binary systems in $\omega$ Cen, by using a collection of MUSE observations of the central regions of the cluster, acquired from 2015 to 2022, that have recently been used to measure an updated value for the fraction of binaries in the cluster (\citealt{Wragg2024}, Paper I). The observed RV curves of all stars with a probability $P_{var}>0.8$ of being binaries in the sample have been processed using a Bayesian approach with \textsc{The Joker} and \textsc{Ultranest} to find their best-fit orbital solutions. We were able to determine the orbital properties of 19 systems, for which we measured periods, velocity semi-amplitudes, minimum companion masses, and eccentricities. Our results offer a first glimpse into the central binary population of $\omega$ Cen. Relative to the parent sample of 222 binary candidates, the 19 systems correspond to a fraction of 9\%.

We made use of the time coverage and single-epoch RV uncertainties of the observed binaries in $\omega$ Cen to create two main sets of simulations, with the idea of investigating the incompleteness of our observational setup as well as to interpret the results obtained in Figures \ref{fig:Pe_plot} and \ref{fig:PK_plot}. The first simulation was created to contain only binaries with MS companions, while the second only contained dark remnants companions, such as WDs, NSs and BHs.
We observed the completeness of our sample vary as a function of the orbital period P, with short-period binaries being the easiest to detect. It also significantly changed with magnitude, moving from 77\% for binaries with evolved primaries down to 33\% for binaries with MS primaries.

In this study we have drawn four main conclusions regarding the population of binary systems in $\omega$ Cen, in light of the results obtained from the comparison to mock MUSE samples and CMC simulations:
\begin{itemize}
    \item The orbital period distribution of the binaries in the cluster shows an overabundance of systems with periods P $>$ 10-20 d. This feature persists even when the distribution is corrected for the incompleteness (in terms of sensitivity) of the current observational configuration. Also, a lack of binaries with P $<$ 1-2 d is observed. These findings are qualitatively in agreement with what predicted by \citet{Ivanova_2005} for a cluster as dense as $\omega$ Cen, but show significant differences if a detailed comparison is made.
    \item The binary period distribution of $\omega$ Cen likely extends up to periods of $10^{3-4}$d. This is supported by: i) the comparison with the CMC simulation of NGC 3201, which has similar core density to $\omega$ Cen; ii) the result recently published by \citet{Platais_2024}, of binary systems which are cluster members and have orbital periods of 10 years or more. We note that binaries in such wide orbits likely represent only a few percent of the total, and their absence in the simulations analyzed here has a negligible impact on the results.
    \item The intrinsic period distribution of binaries in $\omega$ Cen seems different from what is predicted by current theoretical simulations (e.g. CMC) after over 10 Gyrs of simulated evolution, potentially caused by the initial binary properties or the treatment of binary evolution. Indeed, the large discrepancy in the fraction of constrained binaries in the observations (9\%) and in the ad-hoc mock samples (also using CMC priors, 35-41\%) seems produced by the rather different distribution of $\chi^{2}$ values (by up to 26-33\%) between the observed and the simulated sample of binaries (i.e. the higher the value, the more easy to be detected as a binary).
    \item We do not see any evidence for stellar-mass BH or NS candidates orbiting stars in the observed dataset of binaries. We only find a possible WD companion. Moreover, the mock samples have shown that systems with such massive companions have the highest probability of being detected. As a result, this type of systems appear intrinsically rare in $\omega$ Cen, or may exist in wide orbits around low-mass companions (where our survey is 50\% complete) or even in wider orbits than expected from cluster dynamics which we are not sensitive to under the current observational setup.
    However, in order to draw firm conclusions on the total number of remnants residing in $\omega$ Cen will ultimately require estimates on the numbers of single remnants and remnant-remnant binaries, which our data are not sensitive to. Microlensing studies \citep{Zaris2020, Kiroglu_2022} or future gravitational wave observatories appear as promising avenues to provide additional observational constraints on the number of remnants in the cluster.
    Interestingly, a lack of stellar-mass BHs appears in line with \citet{Haberle_2024}'s recent discovery of seven fast-moving stars within the central 3" of the cluster, indicating the presence of an IMBH candidate in its core and also simulations of GCs hosting an IMBH \citep{Aros2021}.
\end{itemize}
Further investigations are needed to confirm the findings presented here about $\omega$ Cen, however the obtained results already represent important observational constraints that future evolutionary models of the cluster will have to satisfy.
 
\section*{Data Availability}
The single epoch MUSE radial velocity measurements are available on VizieR at the following link [the link to the catalog will be provided here] All other data underlying this article will be shared on reasonable request to the corresponding author.

\section*{Acknowledgements}
SS acknowledges funding from STFC under the grant no. R276234. SS also aknowledges funding from the European Union under the grant ERC-2022-AdG, {\em "StarDance: the non-canonical evolution of stars in clusters"}, Grant Agreement 101093572, PI: E. Pancino. SK acknowledges funding from UKRI in the form of a Future Leaders Fellowship (grant no. MR/T022868/1).
This study made use of Prospero high-performance computing facility at Liverpool John Moores University.


\bibliographystyle{mnras}
\bibliography{OmegaCen_binaries} 

\begin{thebibliography}{}
\makeatletter
\relax
\def\mn@urlcharsother{\let\do\@makeother \do\$\do\&\do\#\do\^\do\_\do\%\do\~}
\def\mn@doi{\begingroup\mn@urlcharsother \@ifnextchar [ {\mn@doi@} {\mn@doi@[]}}
\def\mn@doi@[#1]#2{\def\@tempa{#1}\ifx\@tempa\@empty \href {http://dx.doi.org/#2} {doi:#2}\else \href {http://dx.doi.org/#2} {#1}\fi \endgroup}
\def\mn@eprint#1#2{\mn@eprint@#1:#2::\@nil}
\def\mn@eprint@arXiv#1{\href {http://arxiv.org/abs/#1} {{\tt arXiv:#1}}}
\def\mn@eprint@dblp#1{\href {http://dblp.uni-trier.de/rec/bibtex/#1.xml} {dblp:#1}}
\def\mn@eprint@#1:#2:#3:#4\@nil{\def\@tempa {#1}\def\@tempb {#2}\def\@tempc {#3}\ifx \@tempc \@empty \let \@tempc \@tempb \let \@tempb \@tempa \fi \ifx \@tempb \@empty \def\@tempb {arXiv}\fi \@ifundefined {mn@eprint@\@tempb}{\@tempb:\@tempc}{\expandafter \expandafter \csname mn@eprint@\@tempb\endcsname \expandafter{\@tempc}}}

\bibitem[\protect\citeauthoryear{{Abt}, {Gomez}  \& {Levy}}{{Abt} et~al.}{1990}]{Abt1990}
{Abt} H.~A.,  {Gomez} A.~E.,   {Levy} S.~G.,  1990, \mn@doi [\apjs] {10.1086/191508}, \href {https://ui.adsabs.harvard.edu/abs/1990ApJS...74..551A} {74, 551}

\bibitem[\protect\citeauthoryear{{Anderson} \& {van der Marel}}{{Anderson} \& {van der Marel}}{2010}]{Anderson_2010}
{Anderson} J.,  {van der Marel} R.~P.,  2010, \mn@doi [\apj] {10.1088/0004-637X/710/2/1032}, \href {https://ui.adsabs.harvard.edu/abs/2010ApJ...710.1032A} {710, 1032}

\bibitem[\protect\citeauthoryear{{Anderson} et~al.,}{{Anderson} et~al.}{2008}]{Anderson_2008}
{Anderson} J.,  et~al., 2008, \mn@doi [\aj] {10.1088/0004-6256/135/6/2055}, \href {https://ui.adsabs.harvard.edu/abs/2008AJ....135.2055A} {135, 2055}

\bibitem[\protect\citeauthoryear{{Aros}, {Sippel}, {Mastrobuono-Battisti}, {Bianchini}, {Askar}  \& {van de Ven}}{{Aros} et~al.}{2021a}]{Aros_2021}
{Aros} F.~I.,  {Sippel} A.~C.,  {Mastrobuono-Battisti} A.,  {Bianchini} P.,  {Askar} A.,   {van de Ven} G.,  2021a, \mn@doi [\mnras] {10.1093/mnras/stab2872}, \href {https://ui.adsabs.harvard.edu/abs/2021MNRAS.508.4385A} {508, 4385}

\bibitem[\protect\citeauthoryear{{Aros}, {Sippel}, {Mastrobuono-Battisti}, {Bianchini}, {Askar}  \& {van de Ven}}{{Aros} et~al.}{2021b}]{Aros2021}
{Aros} F.~I.,  {Sippel} A.~C.,  {Mastrobuono-Battisti} A.,  {Bianchini} P.,  {Askar} A.,   {van de Ven} G.,  2021b, \mn@doi [\mnras] {10.1093/mnras/stab2872}, \href {https://ui.adsabs.harvard.edu/abs/2021MNRAS.508.4385A} {508, 4385}

\bibitem[\protect\citeauthoryear{{Askar}, {Arca Sedda}  \& {Giersz}}{{Askar} et~al.}{2018a}]{Askar_2018}
{Askar} A.,  {Arca Sedda} M.,   {Giersz} M.,  2018a, \mn@doi [\mnras] {10.1093/mnras/sty1186}, \href {https://ui.adsabs.harvard.edu/abs/2018MNRAS.478.1844A} {478, 1844}

\bibitem[\protect\citeauthoryear{{Askar}, {Arca Sedda}  \& {Giersz}}{{Askar} et~al.}{2018b}]{Askar2018}
{Askar} A.,  {Arca Sedda} M.,   {Giersz} M.,  2018b, \mn@doi [\mnras] {10.1093/mnras/sty1186}, \href {https://ui.adsabs.harvard.edu/abs/2018MNRAS.478.1844A} {478, 1844}

\bibitem[\protect\citeauthoryear{{Bacon} et~al.,}{{Bacon} et~al.}{2010}]{Bacon2010}
{Bacon} R.,  et~al., 2010, in {McLean} I.~S.,  {Ramsay} S.~K.,   {Takami} H.,  eds,  Society of Photo-Optical Instrumentation Engineers (SPIE) Conference Series Vol. 7735, Ground-based and Airborne Instrumentation for Astronomy III. p. 773508 (\mn@eprint {arXiv} {2211.16795}), \mn@doi{10.1117/12.856027}

\bibitem[\protect\citeauthoryear{{Baumgardt} \& {Hilker}}{{Baumgardt} \& {Hilker}}{2018}]{Baumgardt_2018}
{Baumgardt} H.,  {Hilker} M.,  2018, \mn@doi [\mnras] {10.1093/mnras/sty1057}, \href {https://ui.adsabs.harvard.edu/abs/2018MNRAS.478.1520B} {478, 1520}

\bibitem[\protect\citeauthoryear{{Baumgardt} et~al.,}{{Baumgardt} et~al.}{2019}]{Baumgardt_2019}
{Baumgardt} H.,  et~al., 2019, \mn@doi [\mnras] {10.1093/mnras/stz2060}, \href {https://ui.adsabs.harvard.edu/abs/2019MNRAS.488.5340B} {488, 5340}

\bibitem[\protect\citeauthoryear{{Bellini}, {Milone}, {Anderson}, {Marino}, {Piotto}, {van der Marel}, {Bedin}  \& {King}}{{Bellini} et~al.}{2017}]{Bellini_2017}
{Bellini} A.,  {Milone} A.~P.,  {Anderson} J.,  {Marino} A.~F.,  {Piotto} G.,  {van der Marel} R.~P.,  {Bedin} L.~R.,   {King} I.~R.,  2017, \mn@doi [\apj] {10.3847/1538-4357/aa7b7e}, \href {https://ui.adsabs.harvard.edu/abs/2017ApJ...844..164B} {844, 164}

\bibitem[\protect\citeauthoryear{{Bianchini}, {Norris}, {van de Ven}, {Schinnerer}, {Bellini}, {van der Marel}, {Watkins}  \& {Anderson}}{{Bianchini} et~al.}{2016}]{Bianchini2016}
{Bianchini} P.,  {Norris} M.~A.,  {van de Ven} G.,  {Schinnerer} E.,  {Bellini} A.,  {van der Marel} R.~P.,  {Watkins} L.~L.,   {Anderson} J.,  2016, \mn@doi [\apjl] {10.3847/2041-8205/820/1/L22}, \href {https://ui.adsabs.harvard.edu/abs/2016ApJ...820L..22B} {820, L22}

\bibitem[\protect\citeauthoryear{{Braga} et~al.,}{{Braga} et~al.}{2020}]{Braga_2020}
{Braga} V.~F.,  et~al., 2020, \mn@doi [\aap] {10.1051/0004-6361/20203914510.48550/arXiv.2010.06368}, \href {https://ui.adsabs.harvard.edu/abs/2020A&A...644A..95B} {644, A95}

\bibitem[\protect\citeauthoryear{{Buchner}}{{Buchner}}{2021}]{buchner2021}
{Buchner} J.,  2021, \mn@doi [The Journal of Open Source Software] {10.21105/joss.03001}, \href {https://ui.adsabs.harvard.edu/abs/2021JOSS....6.3001B} {6, 3001}

\bibitem[\protect\citeauthoryear{{Clavel}, {Dubus}, {Casares}  \& {Babusiaux}}{{Clavel} et~al.}{2021}]{Clavel2021}
{Clavel} M.,  {Dubus} G.,  {Casares} J.,   {Babusiaux} C.,  2021, \mn@doi [\aap] {10.1051/0004-6361/202039317}, \href {https://ui.adsabs.harvard.edu/abs/2021A&A...645A..72C} {645, A72}

\bibitem[\protect\citeauthoryear{{Clement} et~al.,}{{Clement} et~al.}{2001}]{Clements_2001}
{Clement} C.~M.,  et~al., 2001, \mn@doi [\aj] {10.1086/323719}, \href {https://ui.adsabs.harvard.edu/abs/2001AJ....122.2587C} {122, 2587}

\bibitem[\protect\citeauthoryear{{Dai}, {Johnston}, {Kerr}, {Camilo}, {Cameron}, {Toomey}  \& {Kumamoto}}{{Dai} et~al.}{2020}]{Dai2020}
{Dai} S.,  {Johnston} S.,  {Kerr} M.,  {Camilo} F.,  {Cameron} A.,  {Toomey} L.,   {Kumamoto} H.,  2020, \mn@doi [\apjl] {10.3847/2041-8213/ab621a}, \href {https://ui.adsabs.harvard.edu/abs/2020ApJ...888L..18D} {888, L18}

\bibitem[\protect\citeauthoryear{{Elson}, {Gilmore}, {Santiago}  \& {Casertano}}{{Elson} et~al.}{1995}]{Elson_1995}
{Elson} R. A.~W.,  {Gilmore} G.~F.,  {Santiago} B.~X.,   {Casertano} S.,  1995, \mn@doi [\aj] {10.1086/117553}, \href {https://ui.adsabs.harvard.edu/abs/1995AJ....110..682E} {110, 682}

\bibitem[\protect\citeauthoryear{{Ferraro}, {Paltrinieri}, {Rood}  \& {Dorman}}{{Ferraro} et~al.}{1999}]{Ferraro1999}
{Ferraro} F.~R.,  {Paltrinieri} B.,  {Rood} R.~T.,   {Dorman} B.,  1999, \mn@doi [\apj] {10.1086/307700}, \href {https://ui.adsabs.harvard.edu/abs/1999ApJ...522..983F} {522, 983}

\bibitem[\protect\citeauthoryear{{Giesers} et~al.,}{{Giesers} et~al.}{2019}]{Giesers2019}
{Giesers} B.,  et~al., 2019, \mn@doi [\aap] {10.1051/0004-6361/201936203}, \href {https://ui.adsabs.harvard.edu/abs/2019A&A...632A...3G} {632, A3}

\bibitem[\protect\citeauthoryear{{Greene}, {Strader}  \& {Ho}}{{Greene} et~al.}{2020}]{Greene_2020}
{Greene} J.~E.,  {Strader} J.,   {Ho} L.~C.,  2020, \mn@doi [\araa] {10.1146/annurev-astro-032620-021835}, \href {https://ui.adsabs.harvard.edu/abs/2020ARA&A..58..257G} {58, 257}

\bibitem[\protect\citeauthoryear{{H{\"a}berle} et~al.,}{{H{\"a}berle} et~al.}{2024a}]{Haberle_2024}
{H{\"a}berle} M.,  et~al., 2024a, \mn@doi [\nat] {10.1038/s41586-024-07511-z}, \href {https://ui.adsabs.harvard.edu/abs/2024Natur.631..285H} {631, 285}

\bibitem[\protect\citeauthoryear{{H{\"a}berle} et~al.,}{{H{\"a}berle} et~al.}{2024b}]{Haberle2024PM}
{H{\"a}berle} M.,  et~al., 2024b, \mn@doi [\apj] {10.3847/1538-4357/ad47f5}, \href {https://ui.adsabs.harvard.edu/abs/2024ApJ...970..192H} {970, 192}

\bibitem[\protect\citeauthoryear{{Heggie}}{{Heggie}}{1975}]{Heggie_1975}
{Heggie} D.~C.,  1975, \mn@doi [\mnras] {10.1093/mnras/173.3.729}, \href {https://ui.adsabs.harvard.edu/abs/1975MNRAS.173..729H} {173, 729}

\bibitem[\protect\citeauthoryear{{Henleywillis}, {Cool}, {Haggard}, {Heinke}, {Callanan}  \& {Zhao}}{{Henleywillis} et~al.}{2018}]{Henleywillis2018}
{Henleywillis} S.,  {Cool} A.~M.,  {Haggard} D.,  {Heinke} C.,  {Callanan} P.,   {Zhao} Y.,  2018, \mn@doi [\mnras] {10.1093/mnras/sty675}, \href {https://ui.adsabs.harvard.edu/abs/2018MNRAS.479.2834H} {479, 2834}

\bibitem[\protect\citeauthoryear{{Husser}, {Wende-von Berg}, {Dreizler}, {Homeier}, {Reiners}, {Barman}  \& {Hauschildt}}{{Husser} et~al.}{2013}]{Husser_2013}
{Husser} T.~O.,  {Wende-von Berg} S.,  {Dreizler} S.,  {Homeier} D.,  {Reiners} A.,  {Barman} T.,   {Hauschildt} P.~H.,  2013, \mn@doi [\aap] {10.1051/0004-6361/201219058}, \href {https://ui.adsabs.harvard.edu/abs/2013A&A...553A...6H} {553, A6}

\bibitem[\protect\citeauthoryear{{Husser} et~al.,}{{Husser} et~al.}{2016}]{Husser2016}
{Husser} T.-O.,  et~al., 2016, \mn@doi [\aap] {10.1051/0004-6361/201526949}, \href {https://ui.adsabs.harvard.edu/abs/2016A&A...588A.148H} {588, A148}

\bibitem[\protect\citeauthoryear{{Ivanova}, {Belczynski}, {Fregeau}  \& {Rasio}}{{Ivanova} et~al.}{2005}]{Ivanova_2005}
{Ivanova} N.,  {Belczynski} K.,  {Fregeau} J.~M.,   {Rasio} F.~A.,  2005, \mn@doi [\mnras] {10.1111/j.1365-2966.2005.08804.x}, \href {https://ui.adsabs.harvard.edu/abs/2005MNRAS.358..572I} {358, 572}

\bibitem[\protect\citeauthoryear{{Johnson}, {Dupree}, {Mateo}, {Bailey}, {Olszewski}  \& {Walker}}{{Johnson} et~al.}{2020}]{Johnson_2020}
{Johnson} C.~I.,  {Dupree} A.~K.,  {Mateo} M.,  {Bailey} John~I. I.,  {Olszewski} E.~W.,   {Walker} M.~G.,  2020, \mn@doi [\aj] {10.3847/1538-3881/ab8819}, \href {https://ui.adsabs.harvard.edu/abs/2020AJ....159..254J} {159, 254}

\bibitem[\protect\citeauthoryear{{Kamann}, {Wisotzki}  \& {Roth}}{{Kamann} et~al.}{2013}]{Kamann2013}
{Kamann} S.,  {Wisotzki} L.,   {Roth} M.~M.,  2013, \mn@doi [\aap] {10.1051/0004-6361/201220476}, \href {https://ui.adsabs.harvard.edu/abs/2013A&A...549A..71K} {549, A71}

\bibitem[\protect\citeauthoryear{{Kamann} et~al.,}{{Kamann} et~al.}{2016}]{Kamann_2016}
{Kamann} S.,  et~al., 2016, \mn@doi [\aap] {10.1051/0004-6361/201527065}, \href {https://ui.adsabs.harvard.edu/abs/2016A&A...588A.149K} {588, A149}

\bibitem[\protect\citeauthoryear{{Kamann} et~al.,}{{Kamann} et~al.}{2018}]{Kamann_2018}
{Kamann} S.,  et~al., 2018, \mn@doi [\mnras] {10.1093/mnras/stx2719}, \href {https://ui.adsabs.harvard.edu/abs/2018MNRAS.473.5591K} {473, 5591}

\bibitem[\protect\citeauthoryear{{K{\i}ro{\u{g}}lu}, {Weatherford}, {Kremer}, {Ye}, {Fragione}  \& {Rasio}}{{K{\i}ro{\u{g}}lu} et~al.}{2022}]{Kiroglu_2022}
{K{\i}ro{\u{g}}lu} F.,  {Weatherford} N.~C.,  {Kremer} K.,  {Ye} C.~S.,  {Fragione} G.,   {Rasio} F.~A.,  2022, \mn@doi [\apj] {10.3847/1538-4357/ac5895}, \href {https://ui.adsabs.harvard.edu/abs/2022ApJ...928..181K} {928, 181}

\bibitem[\protect\citeauthoryear{{Kremer} et~al.,}{{Kremer} et~al.}{2020}]{Kremer_2020}
{Kremer} K.,  et~al., 2020, \mn@doi [\apjs] {10.3847/1538-4365/ab7919}, \href {https://ui.adsabs.harvard.edu/abs/2020ApJS..247...48K} {247, 48}

\bibitem[\protect\citeauthoryear{{Latour}, {H{\"a}mmerich}, {Dorsch}, {Heber}, {Husser}, {Kamann}, {Dreizler}  \& {Brinchmann}}{{Latour} et~al.}{2024}]{Latour_2024}
{Latour} M.,  {H{\"a}mmerich} S.,  {Dorsch} M.,  {Heber} U.,  {Husser} T.~O.,  {Kamann} S.,  {Dreizler} S.,   {Brinchmann} J.,  2024, \mn@doi [\aap] {10.1051/0004-6361/202449709e}, \href {https://ui.adsabs.harvard.edu/abs/2024A&A...685C...2L} {685, C2}

\bibitem[\protect\citeauthoryear{{Lebzelter} \& {Wood}}{{Lebzelter} \& {Wood}}{2016}]{LebzelterWood2016}
{Lebzelter} T.,  {Wood} P.~R.,  2016, \mn@doi [\aap] {10.1051/0004-6361/201527315}, \href {https://ui.adsabs.harvard.edu/abs/2016A&A...585A.111L} {585, A111}

\bibitem[\protect\citeauthoryear{{Lee}, {Joo}, {Sohn}, {Rey}, {Lee}  \& {Walker}}{{Lee} et~al.}{1999}]{Lee1999}
{Lee} Y.~W.,  {Joo} J.~M.,  {Sohn} Y.~J.,  {Rey} S.~C.,  {Lee} H.~C.,   {Walker} A.~R.,  1999, \mn@doi [\nat] {10.1038/46985}, \href {https://ui.adsabs.harvard.edu/abs/1999Natur.402...55L} {402, 55}

\bibitem[\protect\citeauthoryear{{Lee}, {Kang}, {Lee}  \& {Lee}}{{Lee} et~al.}{2009}]{Lee2009}
{Lee} J.-W.,  {Kang} Y.-W.,  {Lee} J.,   {Lee} Y.-W.,  2009, \mn@doi [\nat] {10.1038/nature08565}, \href {https://ui.adsabs.harvard.edu/abs/2009Natur.462..480L} {462, 480}

\bibitem[\protect\citeauthoryear{{Leigh}, {L{\"u}tzgendorf}, {Geller}, {Maccarone}, {Heinke}  \& {Sesana}}{{Leigh} et~al.}{2014}]{Leigh2014}
{Leigh} N. W.~C.,  {L{\"u}tzgendorf} N.,  {Geller} A.~M.,  {Maccarone} T.~J.,  {Heinke} C.,   {Sesana} A.,  2014, \mn@doi [\mnras] {10.1093/mnras/stu1437}, \href {https://ui.adsabs.harvard.edu/abs/2014MNRAS.444...29L} {444, 29}

\bibitem[\protect\citeauthoryear{{Limberg}, {Souza}, {P{\'e}rez-Villegas}, {Rossi}, {Perottoni}  \& {Santucci}}{{Limberg} et~al.}{2022}]{Limberg2022}
{Limberg} G.,  {Souza} S.~O.,  {P{\'e}rez-Villegas} A.,  {Rossi} S.,  {Perottoni} H.~D.,   {Santucci} R.~M.,  2022, \mn@doi [\apj] {10.3847/1538-4357/ac8159}, \href {https://ui.adsabs.harvard.edu/abs/2022ApJ...935..109L} {935, 109}

\bibitem[\protect\citeauthoryear{{Marigo} et~al.,}{{Marigo} et~al.}{2017}]{Marigo_2017}
{Marigo} P.,  et~al., 2017, \mn@doi [\apj] {10.3847/1538-4357/835/1/77}, \href {https://ui.adsabs.harvard.edu/abs/2017ApJ...835...77M} {835, 77}

\bibitem[\protect\citeauthoryear{{Milone} et~al.,}{{Milone} et~al.}{2012}]{Milone_2012}
{Milone} A.~P.,  et~al., 2012, \mn@doi [\apjl] {10.1088/2041-8205/754/2/L34}, \href {https://ui.adsabs.harvard.edu/abs/2012ApJ...754L..34M} {754, L34}

\bibitem[\protect\citeauthoryear{{Moe} \& {Di Stefano}}{{Moe} \& {Di Stefano}}{2017}]{MoeDiStefano2017}
{Moe} M.,  {Di Stefano} R.,  2017, \mn@doi [\apjs] {10.3847/1538-4365/aa6fb6}, \href {https://ui.adsabs.harvard.edu/abs/2017ApJS..230...15M} {230, 15}

\bibitem[\protect\citeauthoryear{{M\"uller-Horn}, {G\"ottgens}, {Dreizler}, {Kamann}, {Martens}, {Saracino}  \& {Ye}}{{M\"uller-Horn} et~al.}{2024}]{MullerHorn2024}
{M\"uller-Horn} J.,  {G\"ottgens} F.,  {Dreizler} S.,  {Kamann} S.,  {Martens} S.,  {Saracino} S.,   {Ye} C.~S.,  2024, \mnras

\bibitem[\protect\citeauthoryear{{Neumayer}, {Seth}  \& {B{\"o}ker}}{{Neumayer} et~al.}{2020}]{Neumayer_2020}
{Neumayer} N.,  {Seth} A.,   {B{\"o}ker} T.,  2020, \mn@doi [\aapr] {10.1007/s00159-020-00125-0}, \href {https://ui.adsabs.harvard.edu/abs/2020A&ARv..28....4N} {28, 4}

\bibitem[\protect\citeauthoryear{{Nitschai} et~al.,}{{Nitschai} et~al.}{2023}]{Nitschai_2023}
{Nitschai} M.~S.,  et~al., 2023, \mn@doi [\apj] {10.3847/1538-4357/acf5db}, \href {https://ui.adsabs.harvard.edu/abs/2023ApJ...958....8N} {958, 8}

\bibitem[\protect\citeauthoryear{{Nitschai} et~al.,}{{Nitschai} et~al.}{2024}]{Nitschai2024}
{Nitschai} M.~S.,  et~al., 2024, \mn@doi [\apj] {10.3847/1538-4357/ad5289}, \href {https://ui.adsabs.harvard.edu/abs/2024ApJ...970..152N} {970, 152}

\bibitem[\protect\citeauthoryear{{Noyola}, {Gebhardt}  \& {Bergmann}}{{Noyola} et~al.}{2008}]{Noyola_2008}
{Noyola} E.,  {Gebhardt} K.,   {Bergmann} M.,  2008, \mn@doi [\apj] {10.1086/529002}, \href {https://ui.adsabs.harvard.edu/abs/2008ApJ...676.1008N} {676, 1008}

\bibitem[\protect\citeauthoryear{{Pfeffer}, {Lardo}, {Bastian}, {Saracino}  \& {Kamann}}{{Pfeffer} et~al.}{2021}]{Pfeffer_2021}
{Pfeffer} J.,  {Lardo} C.,  {Bastian} N.,  {Saracino} S.,   {Kamann} S.,  2021, \mn@doi [\mnras] {10.1093/mnras/staa3407}, \href {https://ui.adsabs.harvard.edu/abs/2021MNRAS.500.2514P} {500, 2514}

\bibitem[\protect\citeauthoryear{{Platais} et~al.,}{{Platais} et~al.}{2024}]{Platais_2024}
{Platais} I.,  et~al., 2024, \mn@doi [\apj] {10.3847/1538-4357/ad167c}, \href {https://ui.adsabs.harvard.edu/abs/2024ApJ...963...60P} {963, 60}

\bibitem[\protect\citeauthoryear{{Price-Whelan}, {Hogg}, {Foreman-Mackey}  \& {Rix}}{{Price-Whelan} et~al.}{2017}]{Price-Whelan2017}
{Price-Whelan} A.~M.,  {Hogg} D.~W.,  {Foreman-Mackey} D.,   {Rix} H.-W.,  2017, \mn@doi [\apj] {10.3847/1538-4357/aa5e50}, \href {https://ui.adsabs.harvard.edu/abs/2017ApJ...837...20P} {837, 20}

\bibitem[\protect\citeauthoryear{{Price-Whelan} et~al.,}{{Price-Whelan} et~al.}{2020}]{Price-Whelan2020}
{Price-Whelan} A.~M.,  et~al., 2020, \mn@doi [\apj] {10.3847/1538-4357/ab8acc}, \href {https://ui.adsabs.harvard.edu/abs/2020ApJ...895....2P} {895, 2}

\bibitem[\protect\citeauthoryear{{Sana} et~al.,}{{Sana} et~al.}{2012}]{Sana2012}
{Sana} H.,  et~al., 2012, \mn@doi [Science] {10.1126/science.1223344}, \href {https://ui.adsabs.harvard.edu/abs/2012Sci...337..444S} {337, 444}

\bibitem[\protect\citeauthoryear{{Saracino} et~al.,}{{Saracino} et~al.}{2023}]{Saracino_2023}
{Saracino} S.,  et~al., 2023, \mn@doi [\mnras] {10.1093/mnras/stad2706}, \href {https://ui.adsabs.harvard.edu/abs/2023MNRAS.526..299S} {526, 299}

\bibitem[\protect\citeauthoryear{{Sarajedini} et~al.,}{{Sarajedini} et~al.}{2007}]{Sarajedini_2007}
{Sarajedini} A.,  et~al., 2007, \mn@doi [\aj] {10.1086/511979}, \href {https://ui.adsabs.harvard.edu/abs/2007AJ....133.1658S} {133, 1658}

\bibitem[\protect\citeauthoryear{{Sharma} \& {Rodriguez}}{{Sharma} \& {Rodriguez}}{2024}]{Sharma_Rodriguez_2024}
{Sharma} K.,  {Rodriguez} C.~L.,  2024, \mn@doi [arXiv e-prints] {10.48550/arXiv.2405.05397}, \href {https://ui.adsabs.harvard.edu/abs/2024arXiv240505397S} {p. arXiv:2405.05397}

\bibitem[\protect\citeauthoryear{{Weilbacher} et~al.,}{{Weilbacher} et~al.}{2020}]{pipeline}
{Weilbacher} P.~M.,  et~al., 2020, \mn@doi [\aap] {10.1051/0004-6361/202037855}, \href {https://ui.adsabs.harvard.edu/abs/2020A&A...641A..28W} {641, A28}

\bibitem[\protect\citeauthoryear{{Wragg} et~al.,}{{Wragg} et~al.}{2024}]{Wragg2024}
{Wragg} F.,  et~al., 2024, \mnras

\bibitem[\protect\citeauthoryear{{Ye}, {Kremer}, {Rodriguez}, {Rui}, {Weatherford}, {Chatterjee}, {Fragione}  \& {Rasio}}{{Ye} et~al.}{2022}]{Ye_2022}
{Ye} C.~S.,  {Kremer} K.,  {Rodriguez} C.~L.,  {Rui} N.~Z.,  {Weatherford} N.~C.,  {Chatterjee} S.,  {Fragione} G.,   {Rasio} F.~A.,  2022, \mn@doi [\apj] {10.3847/1538-4357/ac5b0b}, \href {https://ui.adsabs.harvard.edu/abs/2022ApJ...931...84Y} {931, 84}

\bibitem[\protect\citeauthoryear{{Zahn}}{{Zahn}}{1975}]{Zahn1975}
{Zahn} J.~P.,  1975, \aap, \href {https://ui.adsabs.harvard.edu/abs/1975A&A....41..329Z} {41, 329}

\bibitem[\protect\citeauthoryear{{Zaris}, {Veske}, {Samsing}, {M{\'a}rka}, {Bartos}  \& {M{\'a}rka}}{{Zaris} et~al.}{2020}]{Zaris2020}
{Zaris} J.,  {Veske} D.,  {Samsing} J.,  {M{\'a}rka} Z.,  {Bartos} I.,   {M{\'a}rka} S.,  2020, \mn@doi [\apjl] {10.3847/2041-8213/ab89a3}, \href {https://ui.adsabs.harvard.edu/abs/2020ApJ...894L...9Z} {894, L9}

\bibitem[\protect\citeauthoryear{{Zechmeister} \& {K{\"u}rster}}{{Zechmeister} \& {K{\"u}rster}}{2009}]{Zechmeister2009}
{Zechmeister} M.,  {K{\"u}rster} M.,  2009, \mn@doi [\aap] {10.1051/0004-6361:200811296}, \href {https://ui.adsabs.harvard.edu/abs/2009A&A...496..577Z} {496, 577}

\bibitem[\protect\citeauthoryear{{Zocchi}, {Gieles}  \& {H{\'e}nault-Brunet}}{{Zocchi} et~al.}{2019}]{Zocchi_2019}
{Zocchi} A.,  {Gieles} M.,   {H{\'e}nault-Brunet} V.,  2019, \mn@doi [\mnras] {10.1093/mnras/sty1508}, \href {https://ui.adsabs.harvard.edu/abs/2019MNRAS.482.4713Z} {482, 4713}

\bibitem[\protect\citeauthoryear{{van de Ven}, {van den Bosch}, {Verolme}  \& {de Zeeuw}}{{van de Ven} et~al.}{2006}]{van_de_Ven_2006}
{van de Ven} G.,  {van den Bosch} R.~C.~E.,  {Verolme} E.~K.,   {de Zeeuw} P.~T.,  2006, \mn@doi [\aap] {10.1051/0004-6361:20053061}, \href {https://ui.adsabs.harvard.edu/abs/2006A&A...445..513V} {445, 513}

\bibitem[\protect\citeauthoryear{{van der Marel} \& {Anderson}}{{van der Marel} \& {Anderson}}{2010}]{VanDerMarel_2010}
{van der Marel} R.~P.,  {Anderson} J.,  2010, \mn@doi [\apj] {10.1088/0004-637X/710/2/1063}, \href {https://ui.adsabs.harvard.edu/abs/2010ApJ...710.1063V} {710, 1063}

\makeatother
\end{thebibliography}

\clearpage
\appendix
\section{Additional tests}
This Appendix presents two additional tests aimed at understanding the limitations of our dataset and exploring potential improvements. We include them here to avoid disrupting the main text.

In Section \ref{sec:testII}, we applied \textsc{Ultranest} to simulations of binaries with MS companions, fixing the eccentricity to e = 0 to assess whether short-period binaries (P $<$ 2d) could be better constrained. In Section \ref{sec:testIII}, we explored the impact of additional observational epochs for $\omega$ Cen, estimating how many more binaries could be constrained in this scenario.

\subsection{\textsc{Ultranest} with fixed eccentricity}
\label{sec:testII}
As done in Section \ref{sec:orbitalfitting}, we used the simulation of binaries with MS companions for further testing, focusing on systems with $P_{var}>0.8$ and orbital periods P $<$ 2d, which were assigned a fixed eccentricity of e = 0. Since eccentricity is one of the most uncertain parameters when using sparse RV data, we aimed to see if fixing the eccentricity could improve the results. Specifically, we examined whether constraining more binaries or better recovering their properties was possible when applying \textsc{Ultranest} with the same priors but fixing eccentricity to 0.

\begin{figure}
    \centering
	\includegraphics[width=0.46\textwidth]{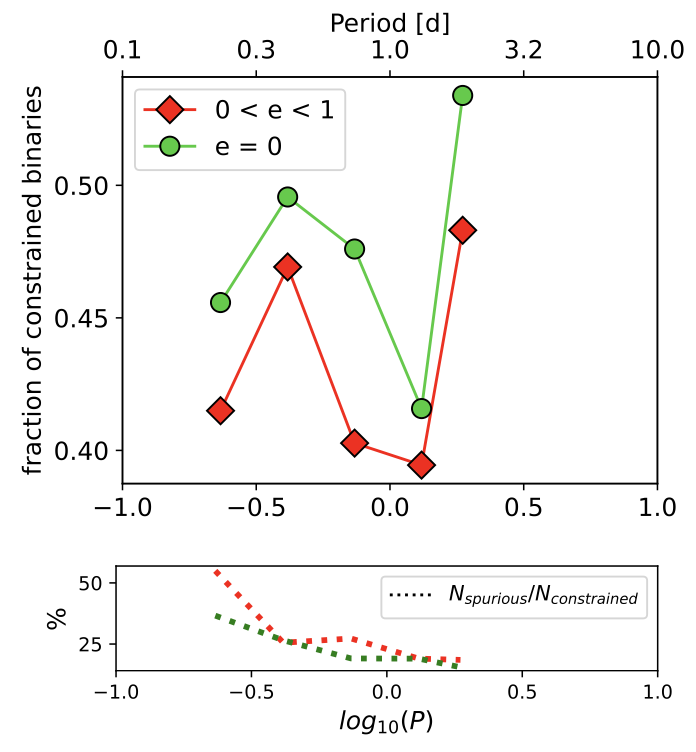}
    \caption{Comparison of the results of simulation I for binaries with P $<$ 2d under the assumptions of variable and fixed eccentricity, respectively. Large red diamonds refer to the former, while large green dots correspond to the latter. By fixing the eccentricity, \textsc{Ultranest} seems able to constrain a larger fraction of short-period binaries, of 2-3\% overall. Furthermore, panel (a) show the fraction of binaries with spurious solutions as a function of orbital period for the two tests, respectively. This plot highlight how the number of binaries with spurious solutions is significantly reduced in the short-period regime, going from a variable to a fixed eccentricity. This can be taken in consideration when we apply \textsc{Ultranest} to the observed sample of binary candidates in a cluster like $\omega$ Cen.}
    \label{fig:testII}
\end{figure}

Figure \ref{fig:testII} compares the fraction of constrained binaries using two methods: varying the eccentricity (red diamonds) and fixing it to 0 (green dots). While the overall fraction of constrained binaries increases slightly by 2-3\% across the period range of 0.1-2 d, this small improvement suggests that fixing eccentricity helps constrain binaries with short periods but does not substantially increase the sample size. However, the real impact is shown in the bottom panel of the Figure. It indeed presents the fraction of binaries with spurious solutions (those whose recovered period deviates by more than 10\% from the simulated period). The color scheme remains the same: red for variable eccentricity and green for fixed eccentricity. This plot highlights a significant decrease—by 15-20\%—in the number of binaries with spurious solutions with P $<$ 0.3d when fixing eccentricity.

These results suggest that while fixing eccentricity has a minimal impact on increasing the number of constrained binaries, it substantially improves the recovery of short-period binaries, hence the purity of the sample, which is critical for characterizing binary properties in clusters. This justifies our approach for the $\omega$ Cen binary candidates, where RV curves were processed using both fixed and variable eccentricity in \textsc{Ultranest} (see Section \ref{sec:orbitalfitting} for the results). This method could enhance future binary studies and complement standard techniques.

\subsection{Simulation with additional epochs}
\label{sec:testIII}
Although our simulations are idealized, they provide valuable insights into the limitations of our current data and suggest ways to enhance our understanding with additional observations. In Section \ref{sec:orbitalfitting}, we constrained the orbital properties of 19 candidate binary systems in $\omega$ Cen. We now explore how many more binaries we could constrain with additional observations spread over time. To address this, we simulated a scenario where each star in our observed sample received five additional RV measurements between January and July 2025, spaced 45 days apart. We aimed to replicate the conditions of the previous simulation (Section \ref{sec:testI}), hereafter referred to as simulation III, for comparison with simulation I.

\begin{figure}
    \centering
	\includegraphics[width=0.47\textwidth]{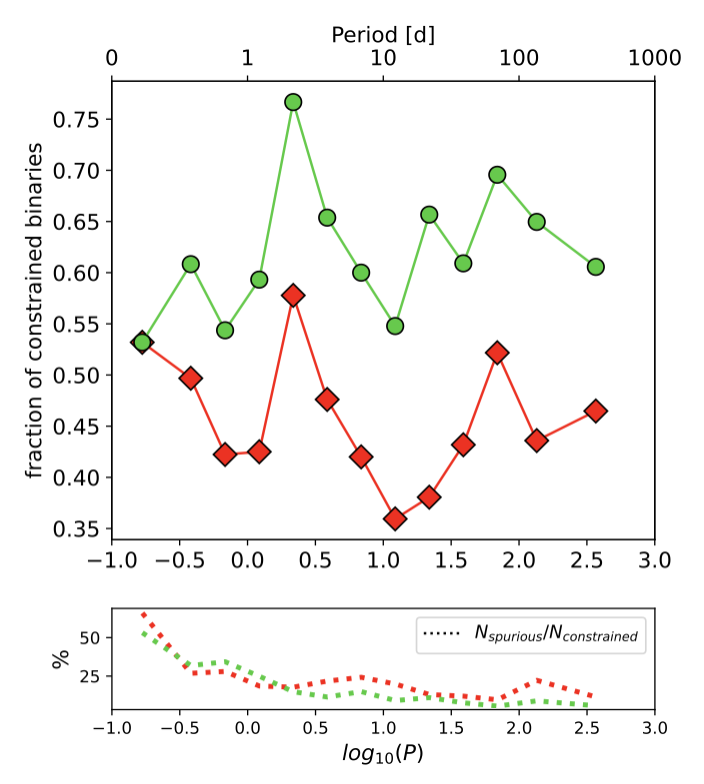}
    \caption{Completeness curves for the simulation using the current observational setup (red diamonds) and for the simulation with five additional observations taken in 2025, spaced by 45 d (green dots). We observe an overall increase of the completeness in orbital periods by 15-20\%, with a major contribution from binaries with long periods (P $>$ 10d). Panel (a) instead presents the fraction of binaries with spurious solutions in simulations I and III, in red and green respectively. These plots show that a larger number of observations also implies a slightly improved ability to recover binary properties well.}
    \label{fig:testIII}
\end{figure}

We simulated a sample of 38 249 sources, with 33.5\% identified as genuine binaries. The sample processed by \textsc{Ultranest} included 6 428 binaries with $P_{var}>0.8$, excluding less than 1\% that were false binaries. Figure \ref{fig:testIII} compares completeness curves for simulations I (red) and III (green), focusing on binaries common to both simulations—those meeting the criteria of being true binaries with $P_{var}>0.8$. Of 3 731 binaries shared between simulations, only 58\% of the total number in simulation III, this subset indicates that many previously discarded stars were reclassified as binaries with the additional epochs.

Simulation III constrained 785 new binaries, a 50\% increase compared to simulation I. Figure \ref{fig:testIII} shows the fraction of constrained binaries as a function of the orbital period. Green dots represent simulation III results, while red diamonds denote simulation I. The completeness increased consistently across all periods, from a fraction of a day to several hundred days. Specifically, simulation III recovers 68\% (compared to 53\% in simulation I) of binaries with P $<$ 1d, 62\% (45\% in simulation I) with 1 $\leq$ P $<$ 10 d, 58\% (39\% in simulation I) with 10 $\leq$ P $<$ 100 d, and 59\% (42\% in simulation I) with 100 $\leq$ P $<$ 500 d. Binaries with longer periods benefit most from the additional observations. The bottom panel in the figure presents the fraction of binaries with spurious solutions in simulation I and III, using the same color code. The purity of the sample seems almost the same for both simulations, with the tendency of an improvement for binaries with periods longer than a few days. This suggests that additional observational epochs not only increase the sample size but also enhance our ability to accurately determine orbital parameters. This finding underscores the importance of follow-up proposals to improve both the quantity and quality of constrained binaries.

The observed binary sample in $\omega$ Cen exhibits significant differences from the simulated sample. The actual number of constrained binaries is notably lower than what simulations suggest. Section \ref{sec:LPB} provides a detailed analysis of this discrepancy. Despite this, the comparison is valuable for understanding the potential impact of additional observations. Adding five more epochs spaced 45 days apart in 2025 could increase the number of constrained binaries by approximately 50\%, potentially raising the sample from 19 to 30. This would significantly enhance our ability to characterize the binary population in $\omega$ Cen. However, exploring alternative observational strategies to further maximize the sample size remains necessary and is beyond the scope of this paper.

\section{Additional material}
\label{app:extra}
\begin{figure*}
\begin{subfigure}
    \centering
    \includegraphics[width=0.49\textwidth]{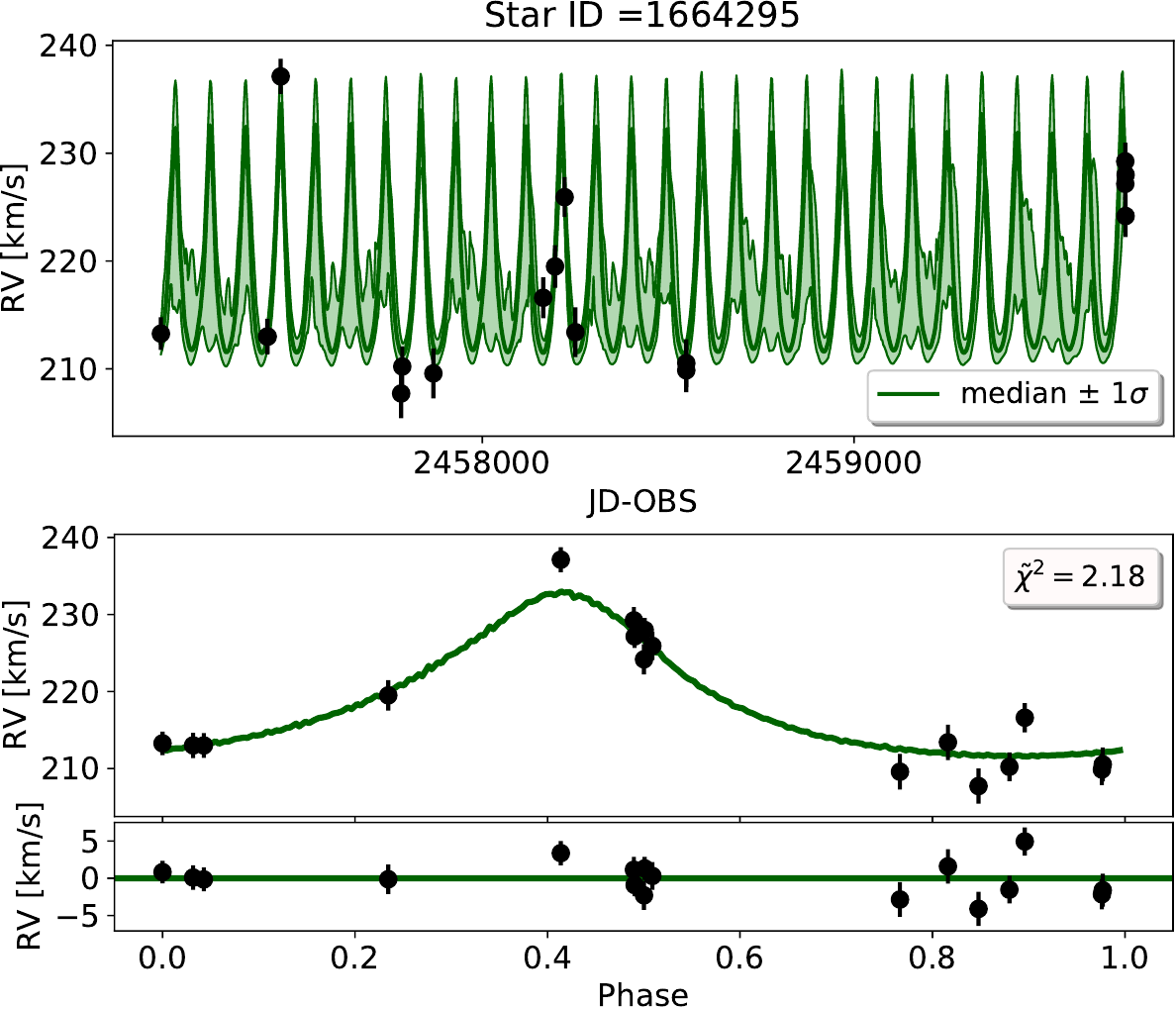}
\end{subfigure}
\begin{subfigure}
    \centering
    \includegraphics[width=0.49\textwidth]{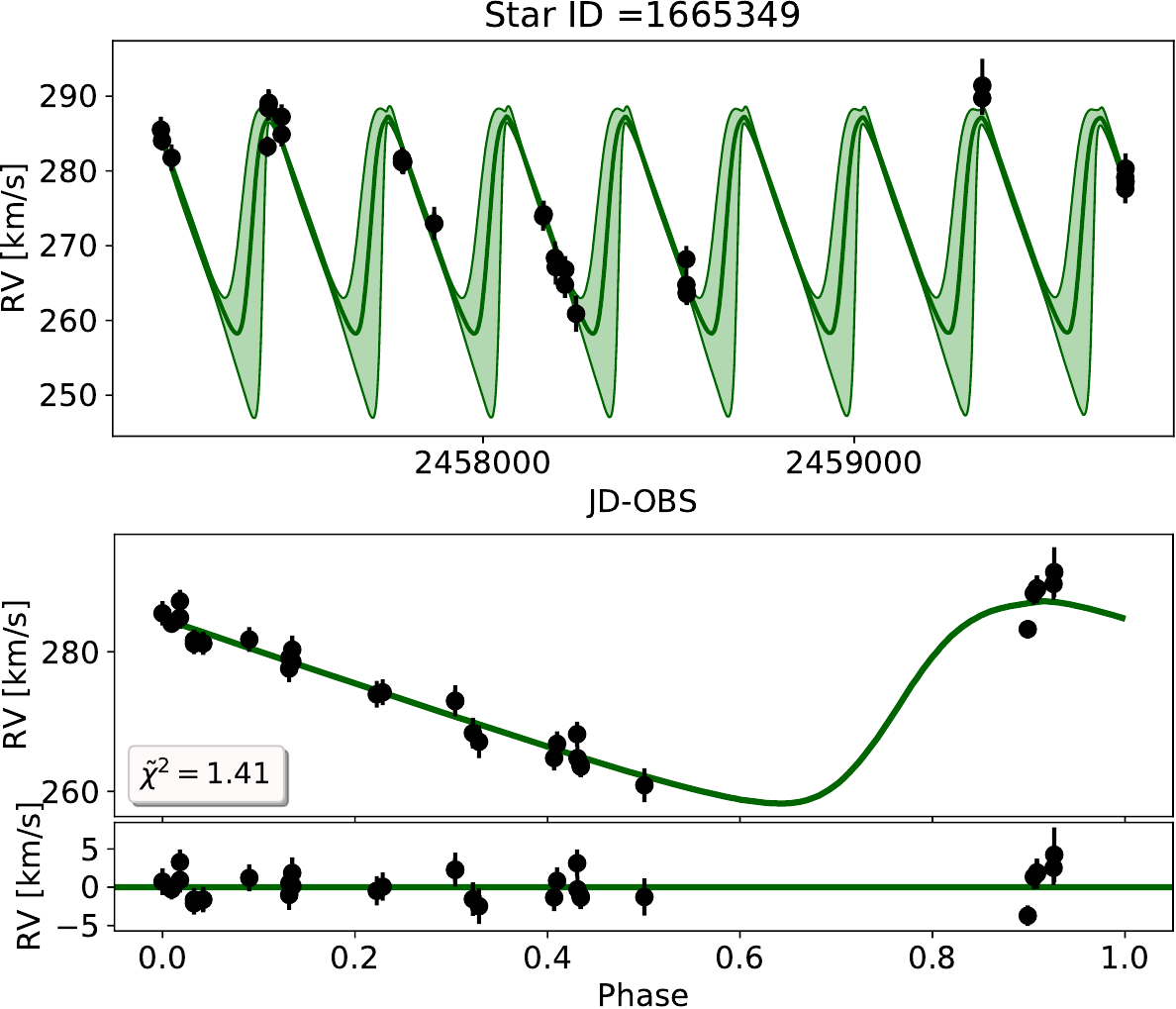}
\end{subfigure}
 \vfill
\begin{subfigure}
    \centering
    \includegraphics[width=0.49\textwidth]{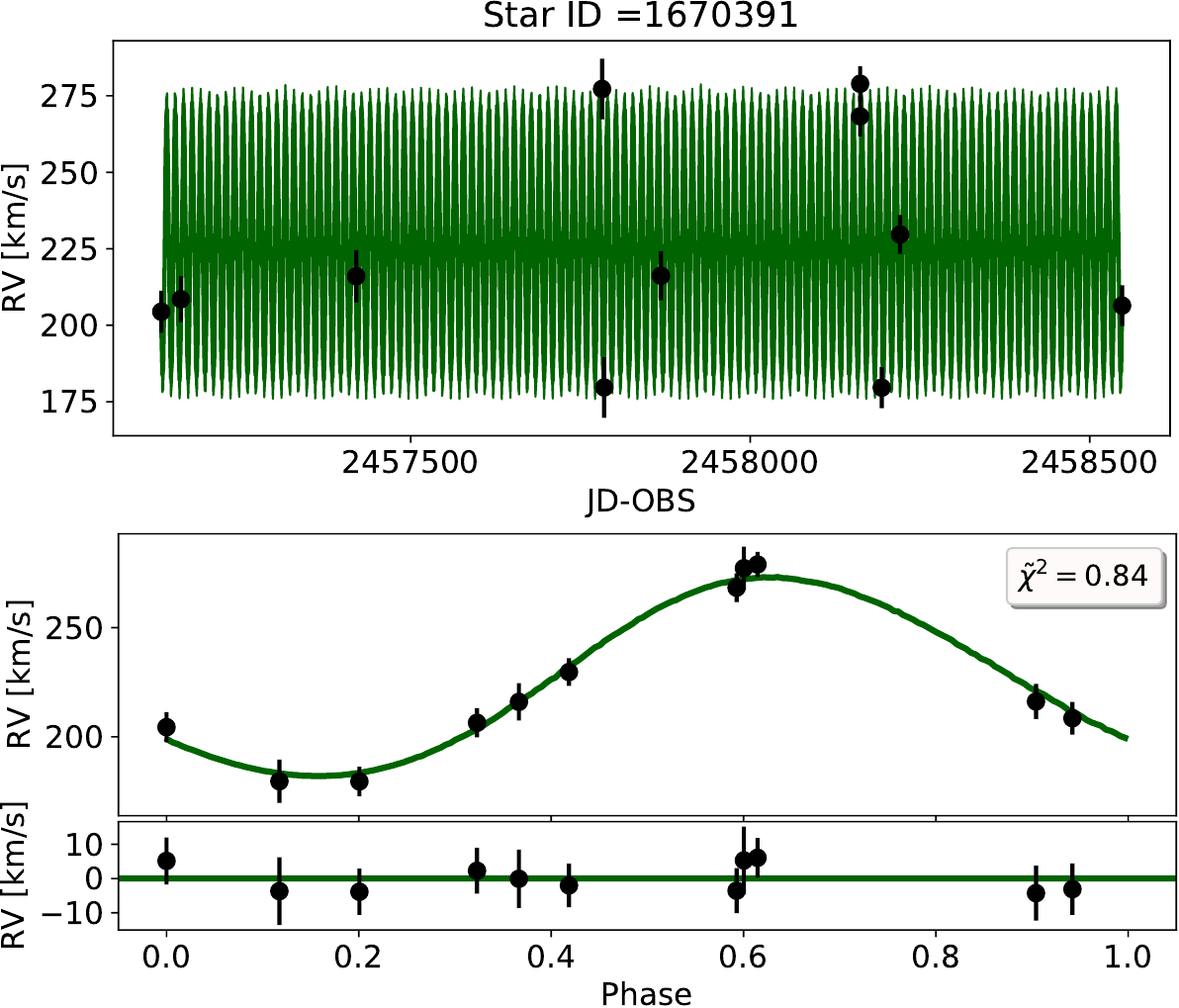}
\end{subfigure}
\begin{subfigure}
    \centering
    \includegraphics[width=0.49\textwidth]{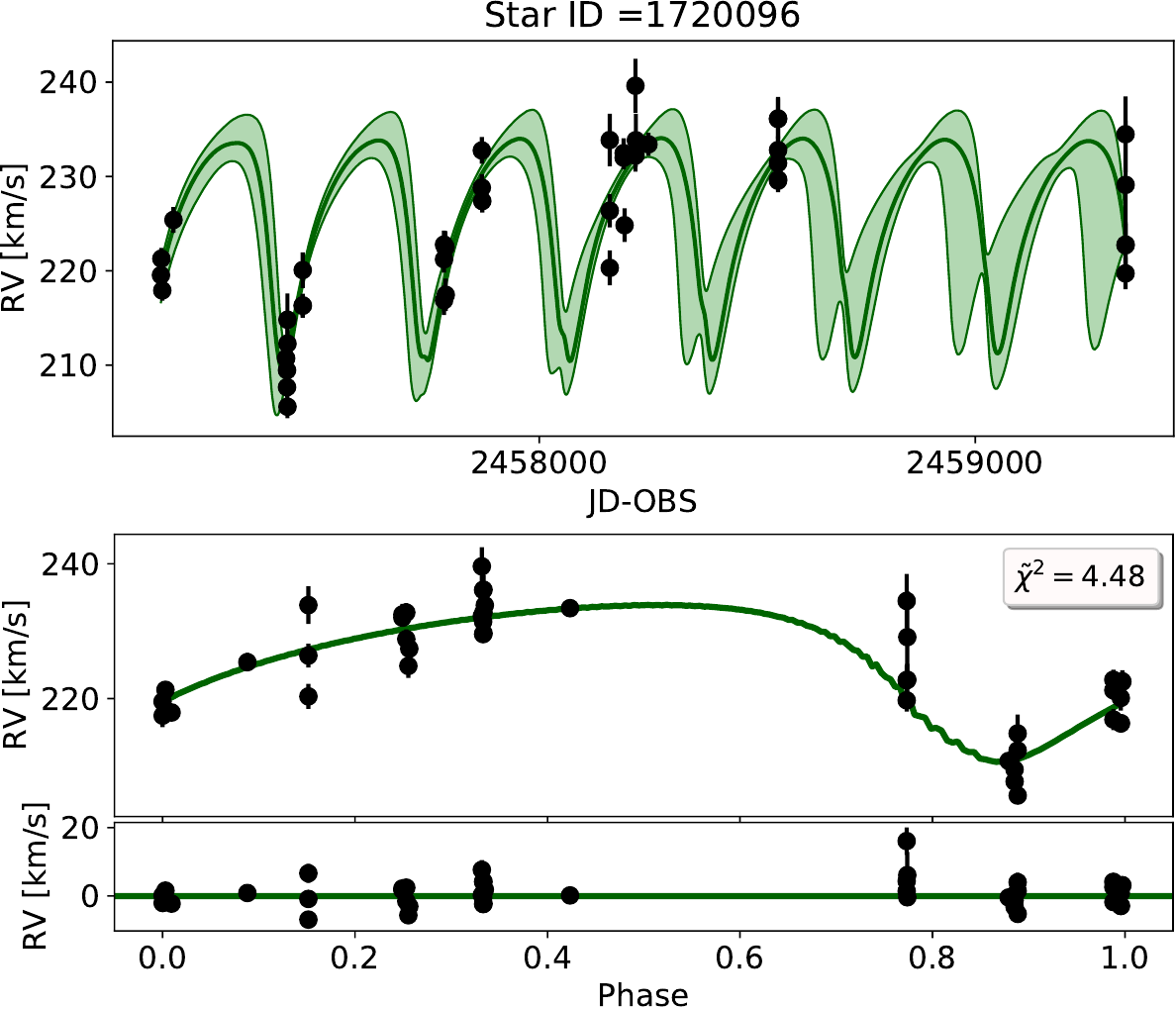}
\end{subfigure}
 \vfill
\begin{subfigure}
    \centering
    \includegraphics[width=0.49\textwidth]{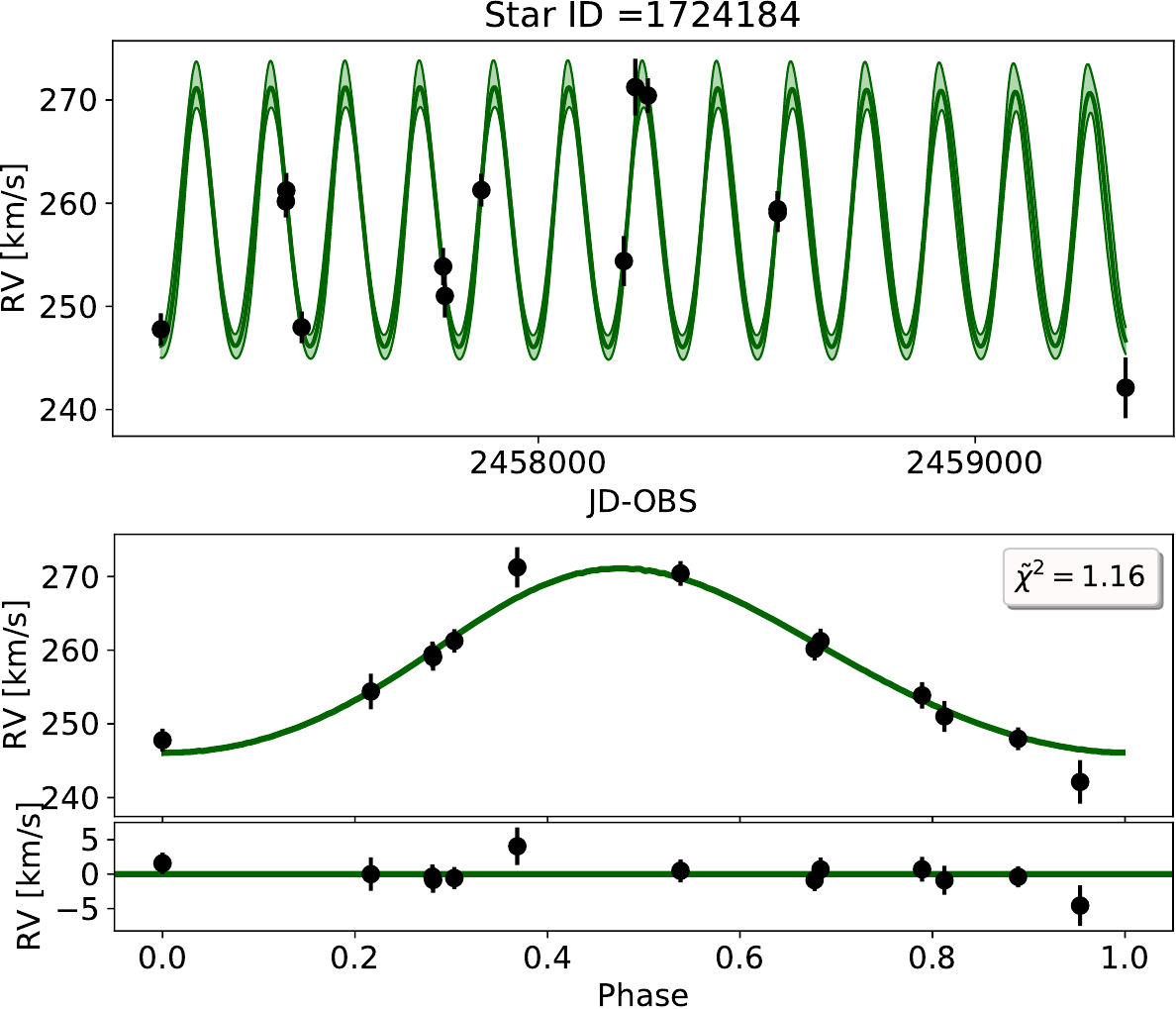}
\end{subfigure}
\begin{subfigure}
    \centering
    \includegraphics[width=0.49\textwidth]{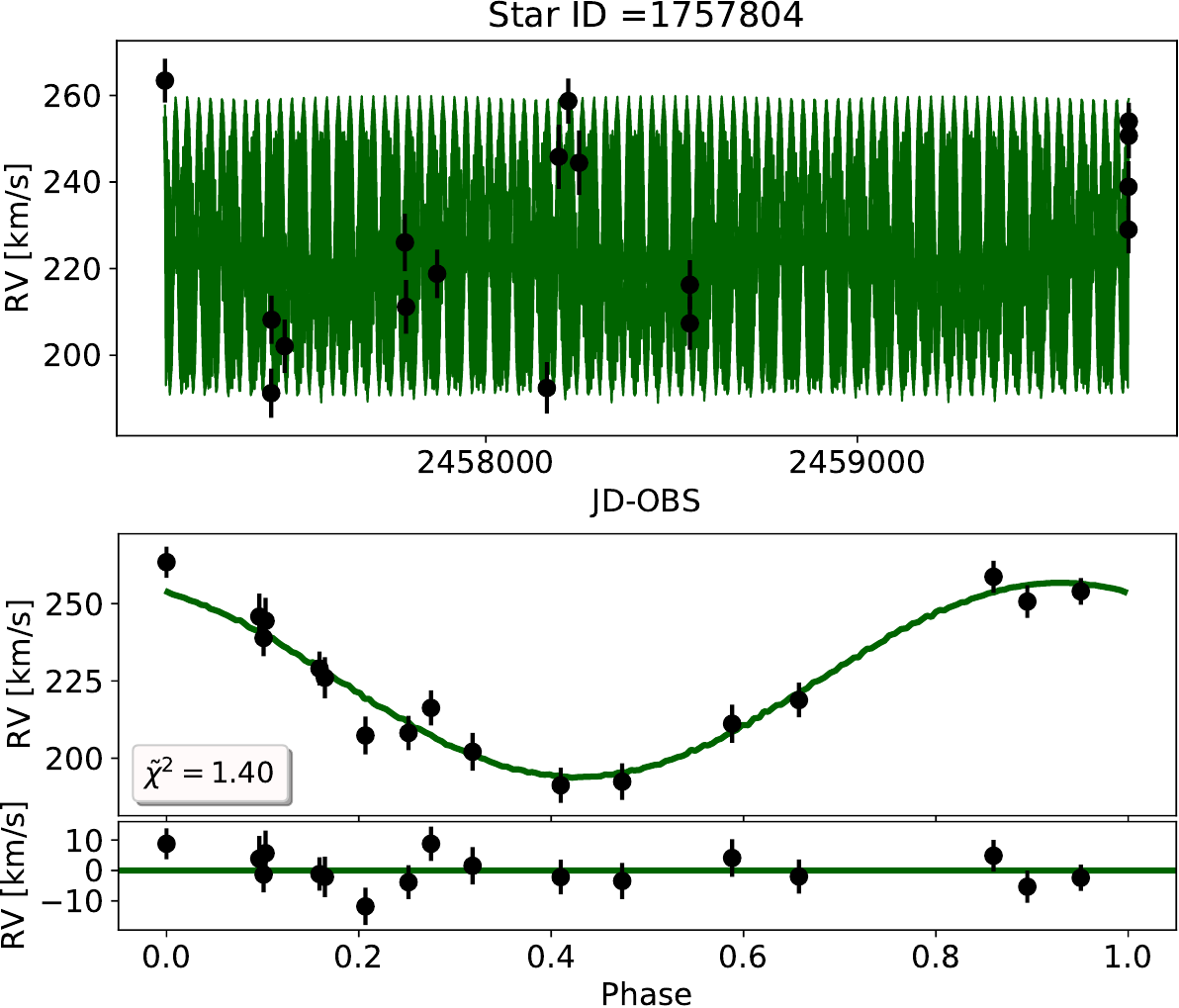}
\end{subfigure}
\caption{Constrained binaries in $\omega$ Cen sorted by Star ID. The upper panel of every plot shows the observed RV curve (black points), the best-fit median model and the $\pm$ 1 $\sigma$ models (green continues lines). The green shaded area is the allowed region by propagating the uncertainties on the parameters. The lower panel shows the same RVs phase folded with the period from the median model. The colour code is the same as in upper panel. Moreover, it also contains the residuals after subtracting this model from the data. The reduced $\chi^2$ of the best-fit median model is also mentioned.}
\label{fig:plot}
\end{figure*}

\begin{figure*}
\begin{subfigure}
    \centering
    \includegraphics[width=0.49\textwidth]{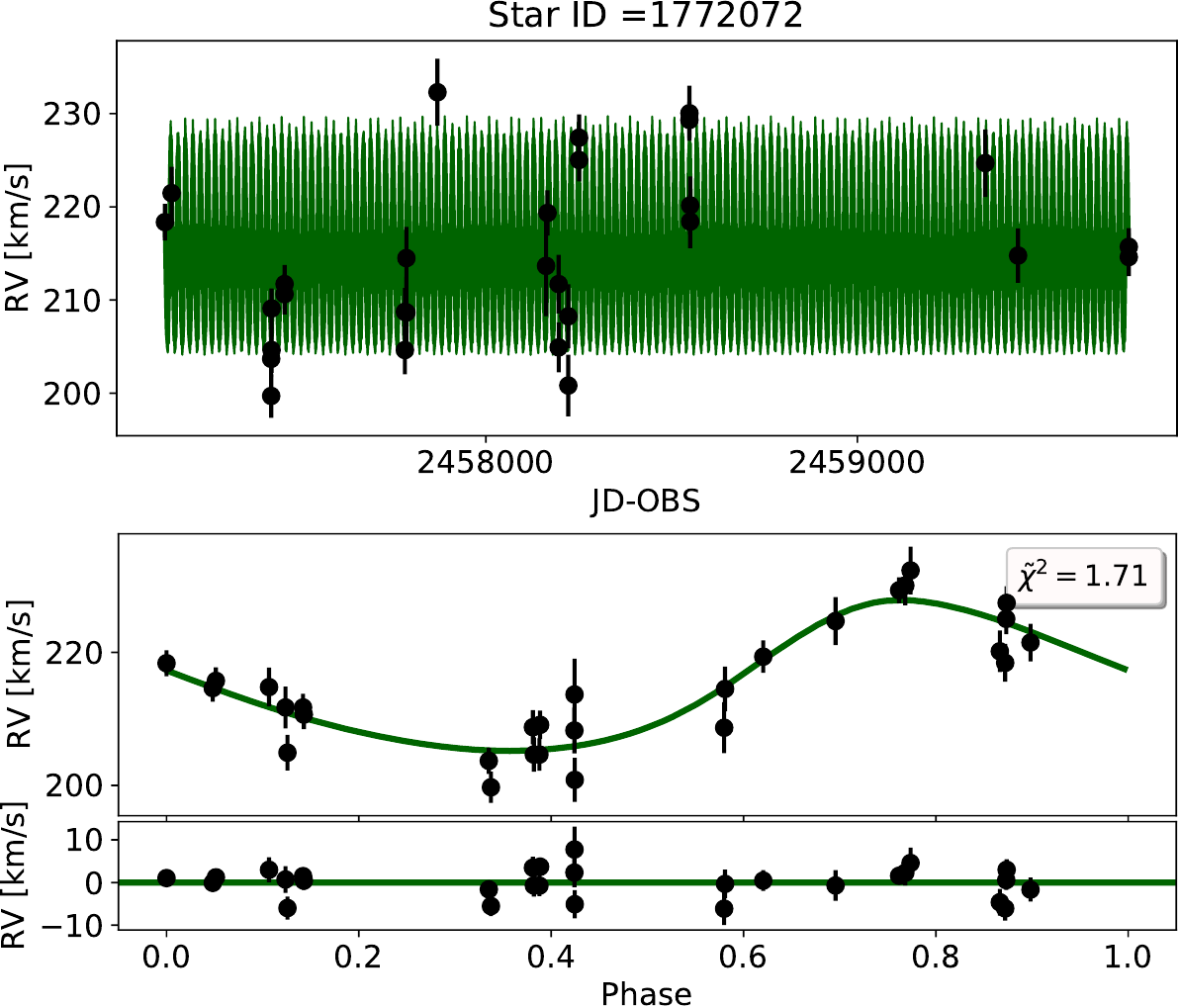}
\end{subfigure}
\begin{subfigure}
    \centering
    \includegraphics[width=0.49\textwidth]{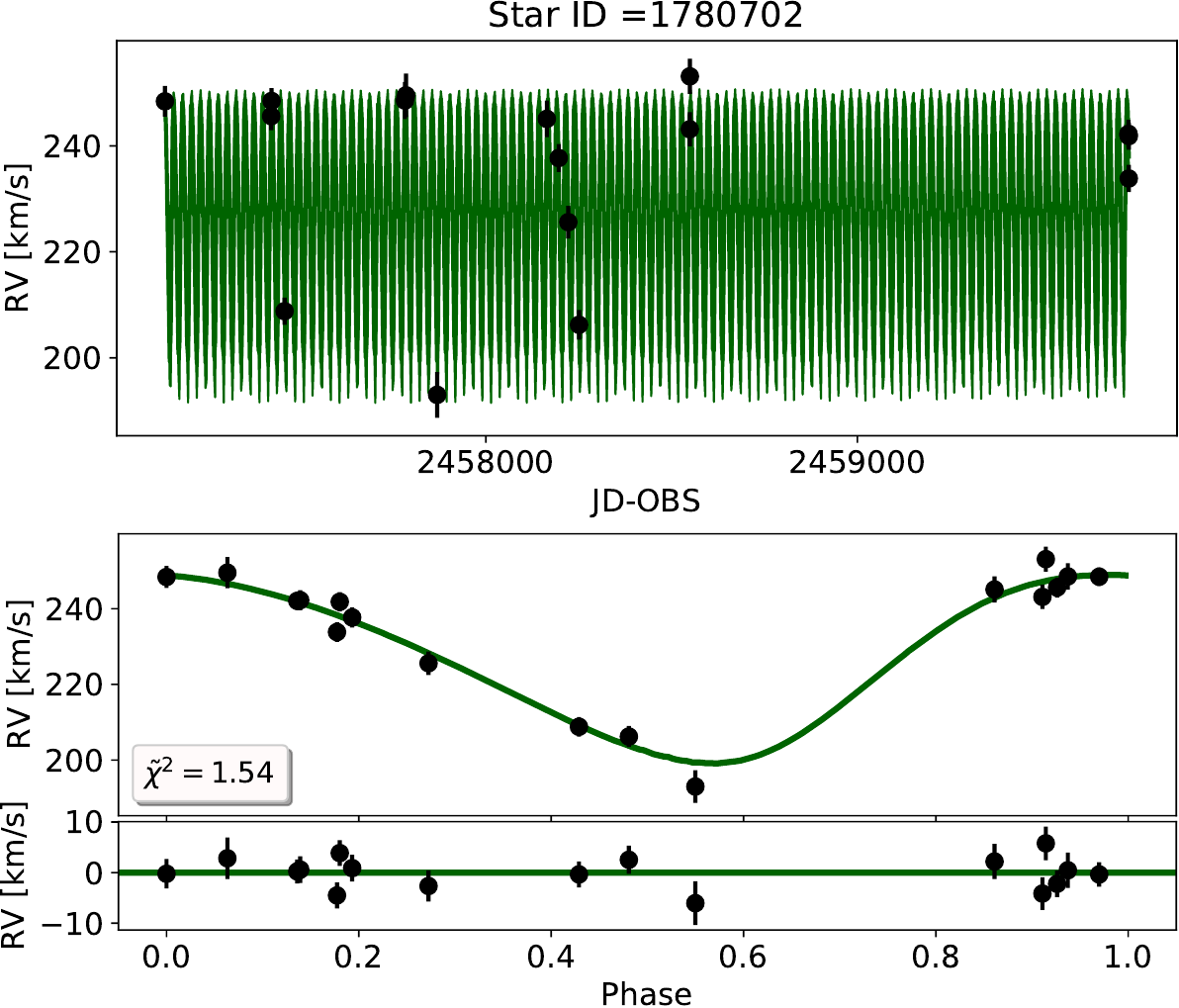}
\end{subfigure}
 \vfill
\begin{subfigure}
    \centering
    \includegraphics[width=0.49\textwidth]{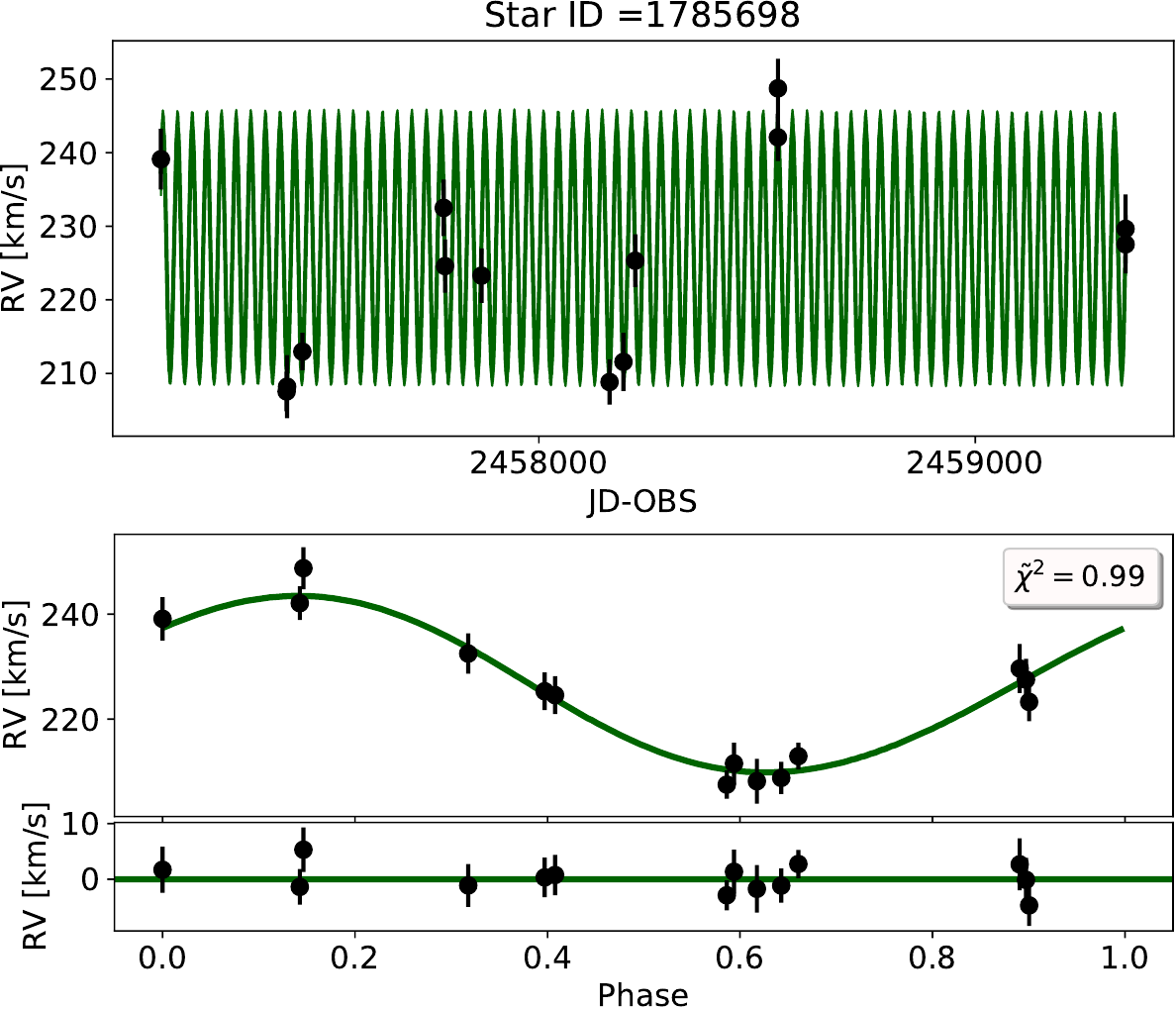}
\end{subfigure}
\begin{subfigure}
    \centering
    \includegraphics[width=0.49\textwidth]{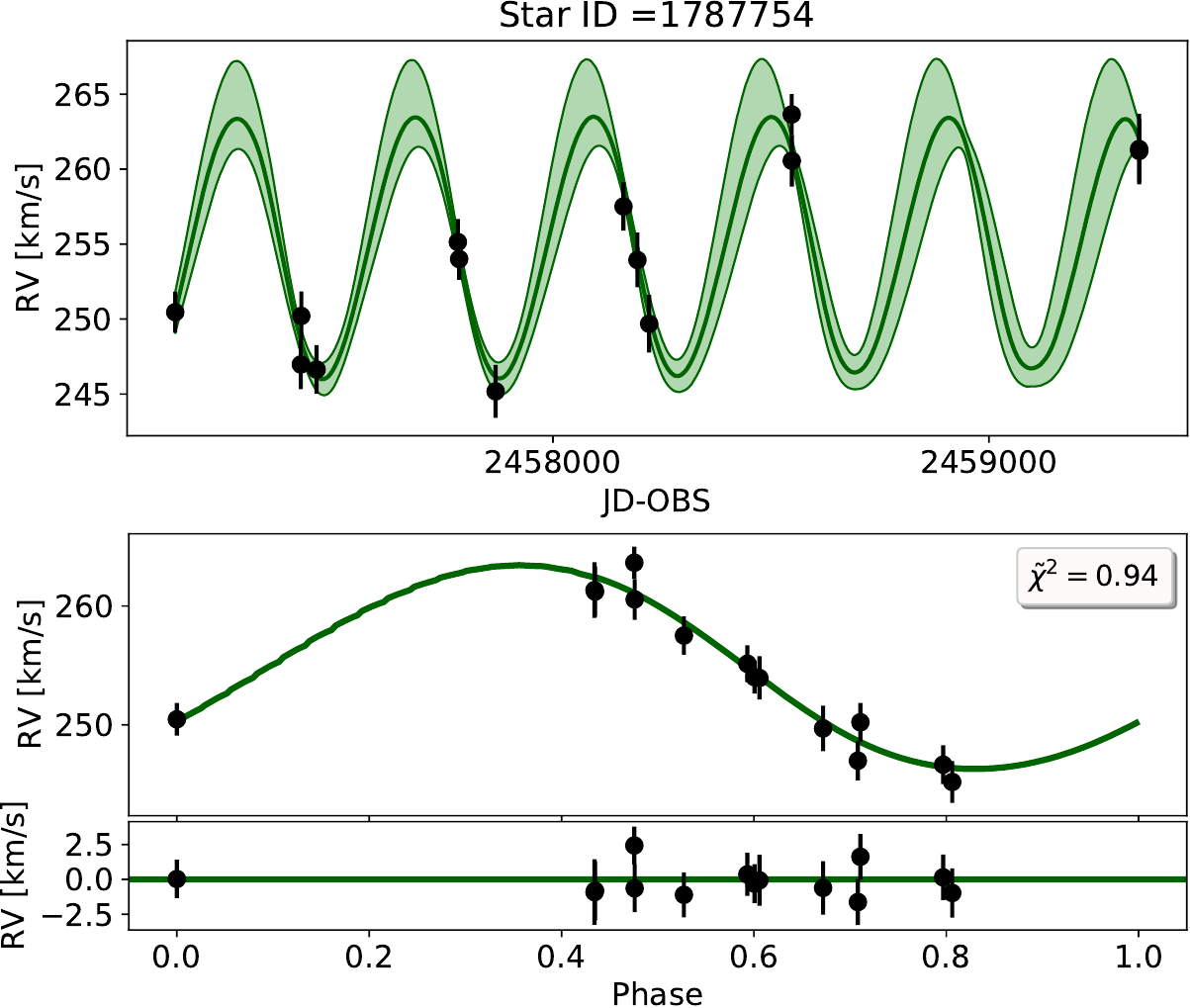}
\end{subfigure}
 \vfill
\begin{subfigure}
    \centering
    \includegraphics[width=0.49\textwidth]{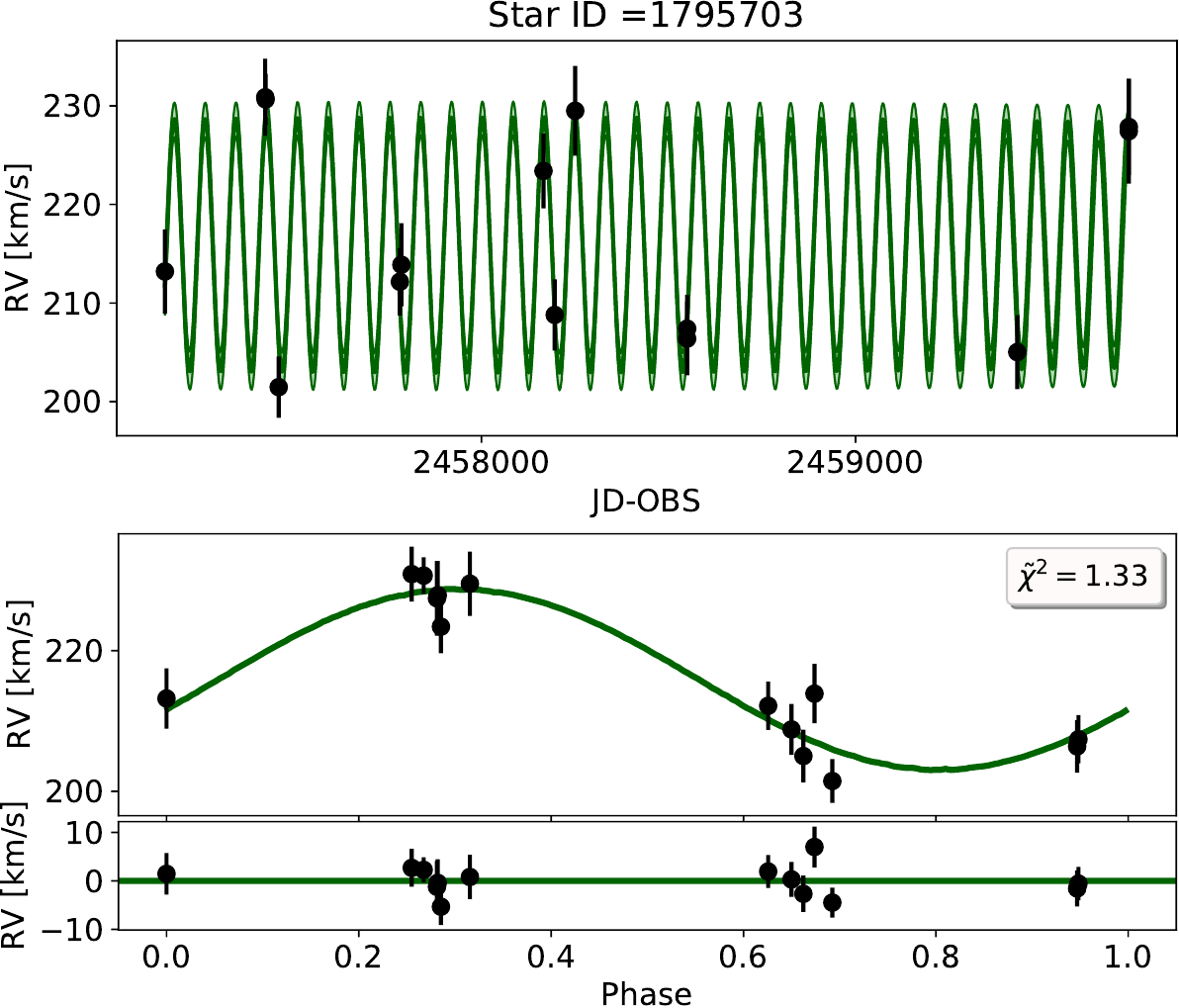}
\end{subfigure}
\begin{subfigure}
    \centering
    \includegraphics[width=0.49\textwidth]{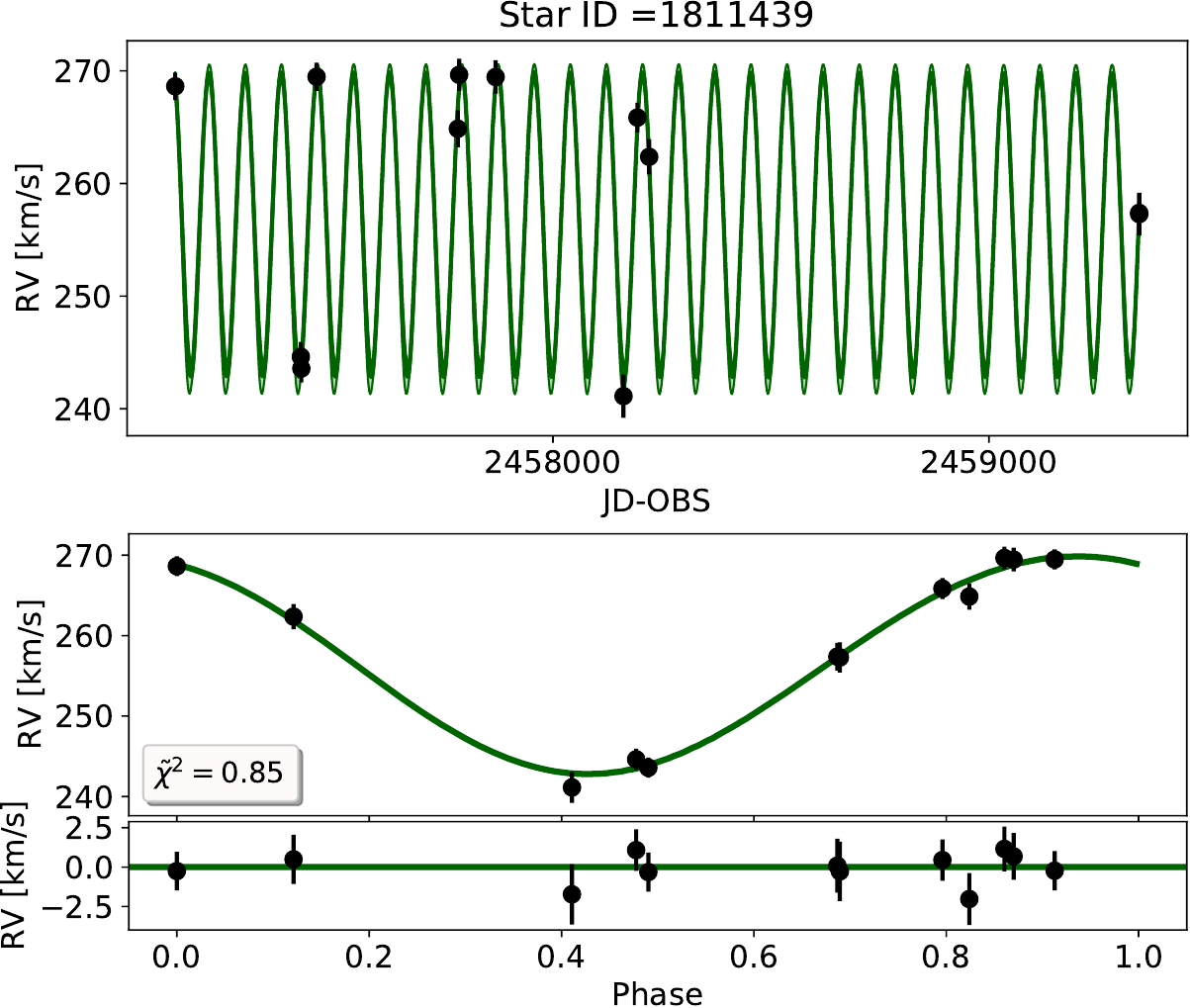}
\end{subfigure}
\caption{Same as in Figure \ref{fig:plot}.}
\label{fig:plot1}
\end{figure*}

\begin{figure*}
\begin{subfigure}
    \centering
    \includegraphics[width=0.49\textwidth]{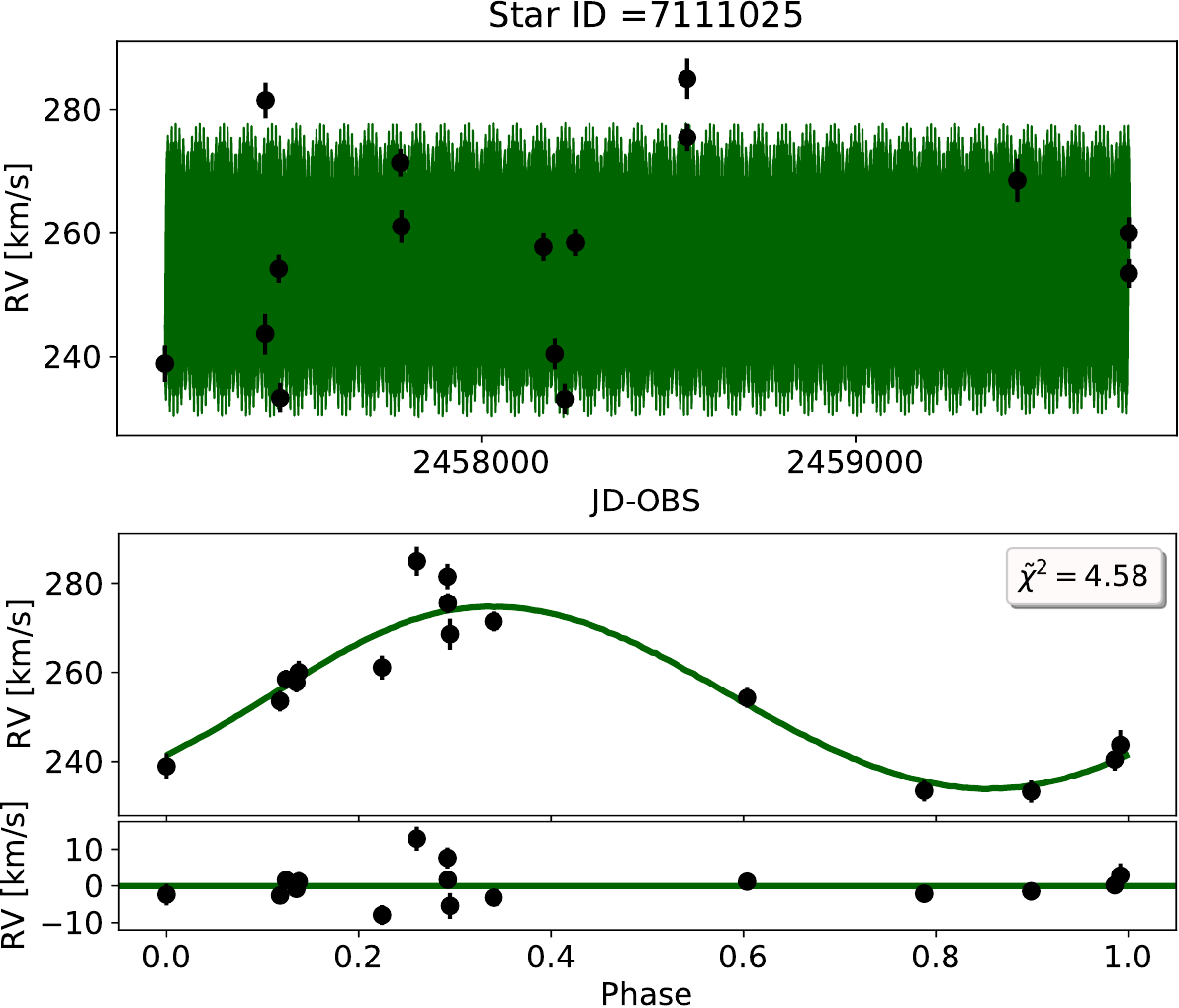}
\end{subfigure}
\begin{subfigure}
    \centering
    \includegraphics[width=0.49\textwidth]{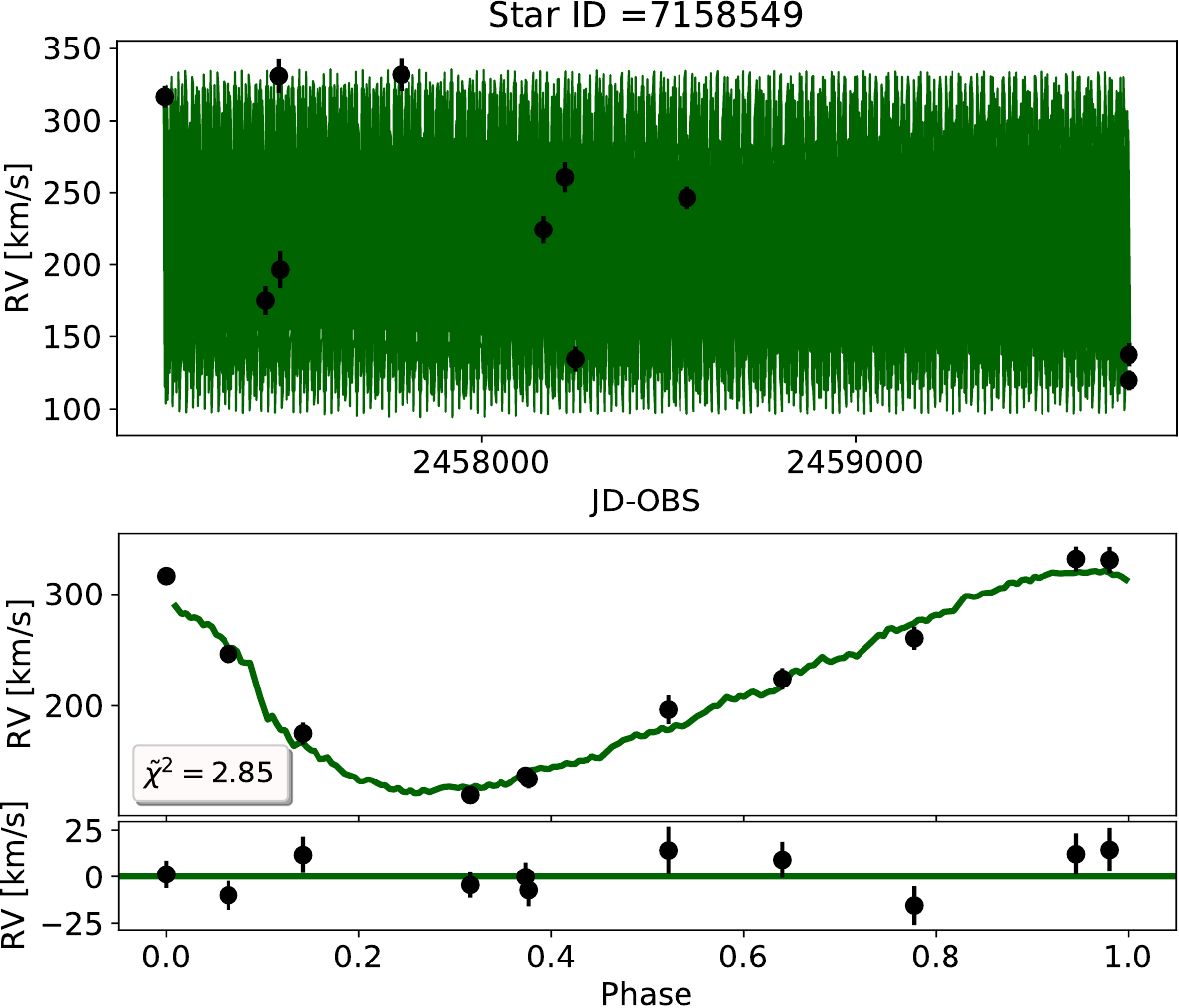}
\end{subfigure}
 \vfill
\begin{subfigure}
    \centering
    \includegraphics[width=0.49\textwidth]{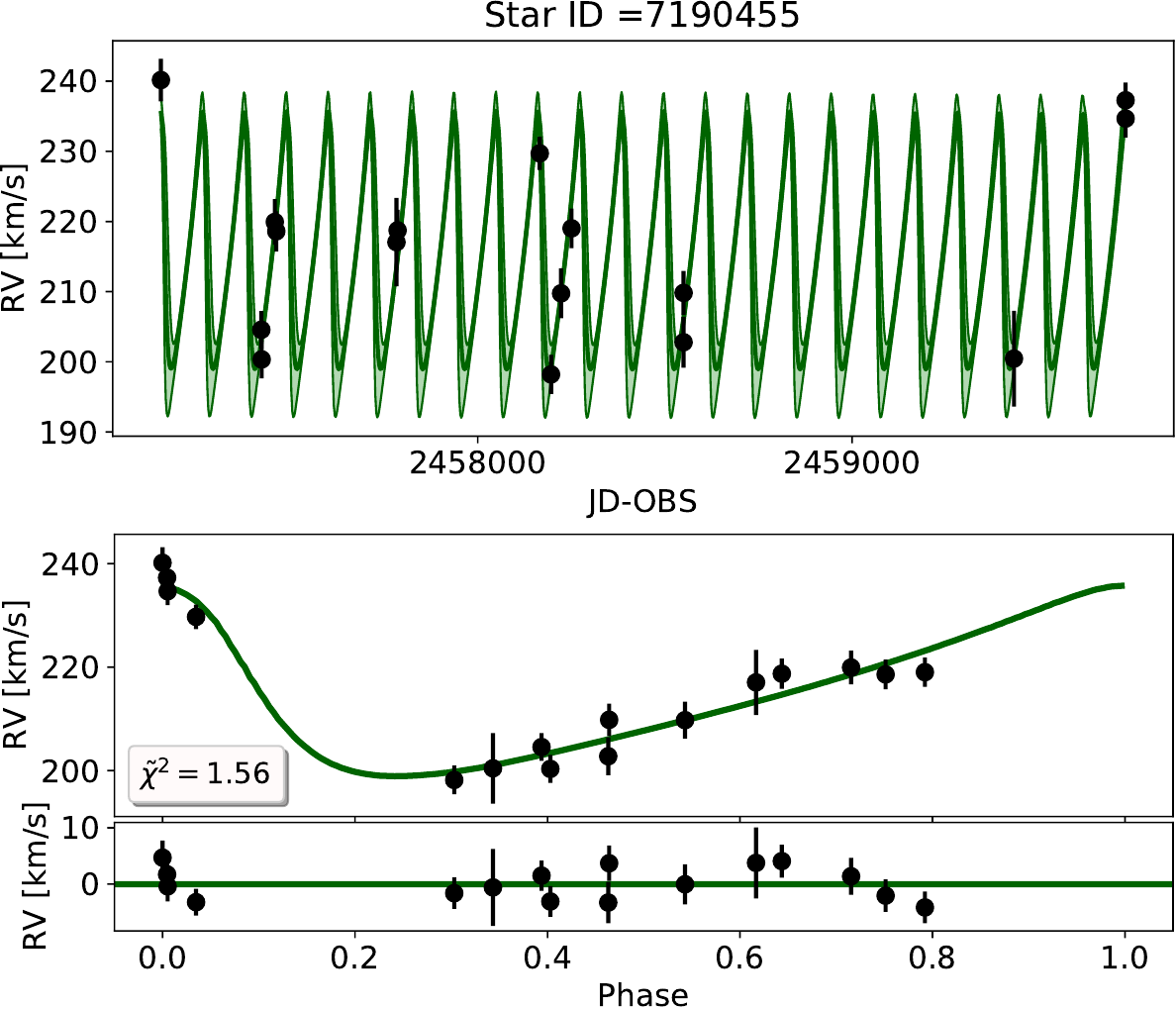}
\end{subfigure}
\begin{subfigure}
    \centering
    \includegraphics[width=0.49\textwidth]{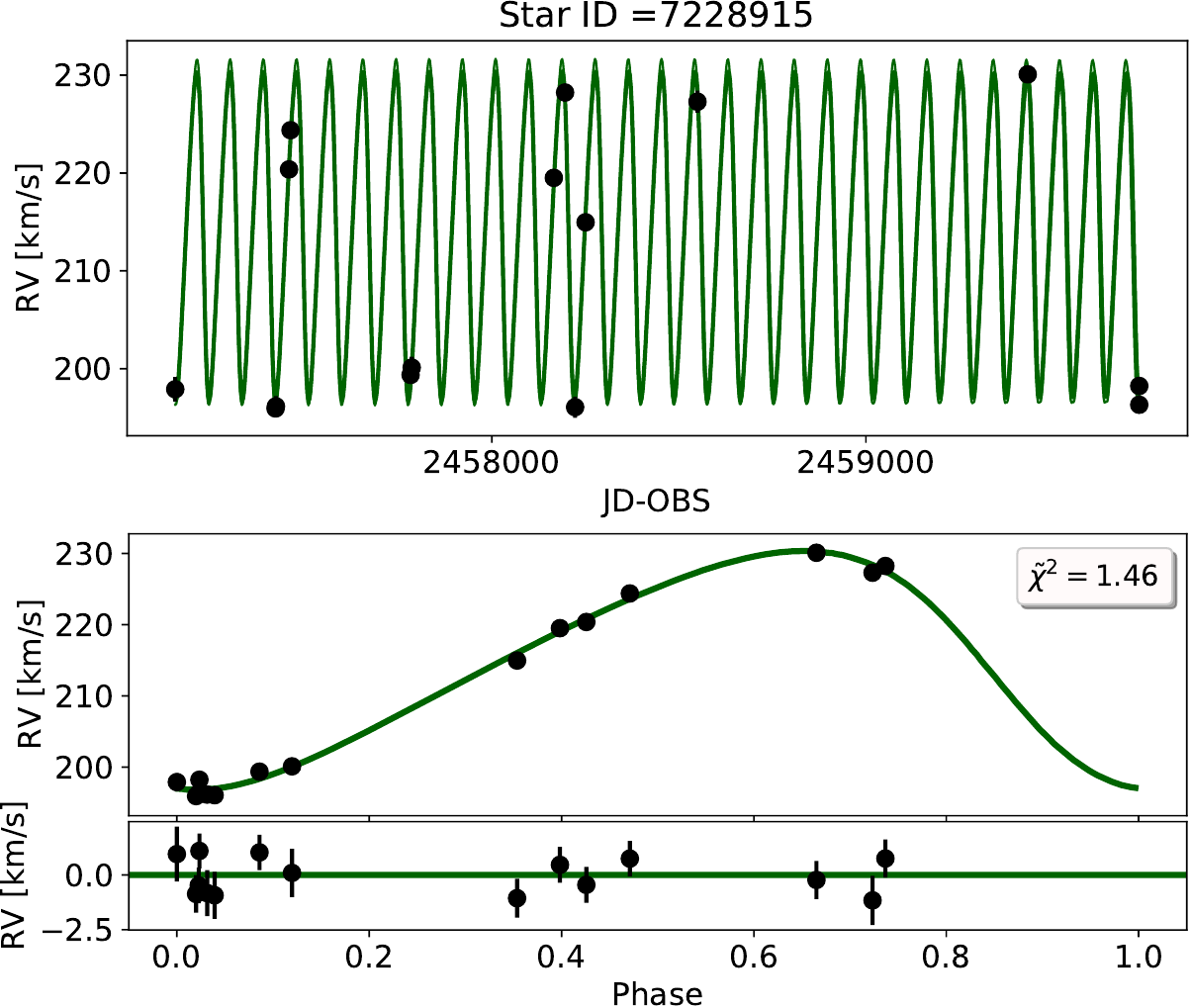}
\end{subfigure}
 \vfill
\begin{subfigure}
    \centering
    \includegraphics[width=0.49\textwidth]{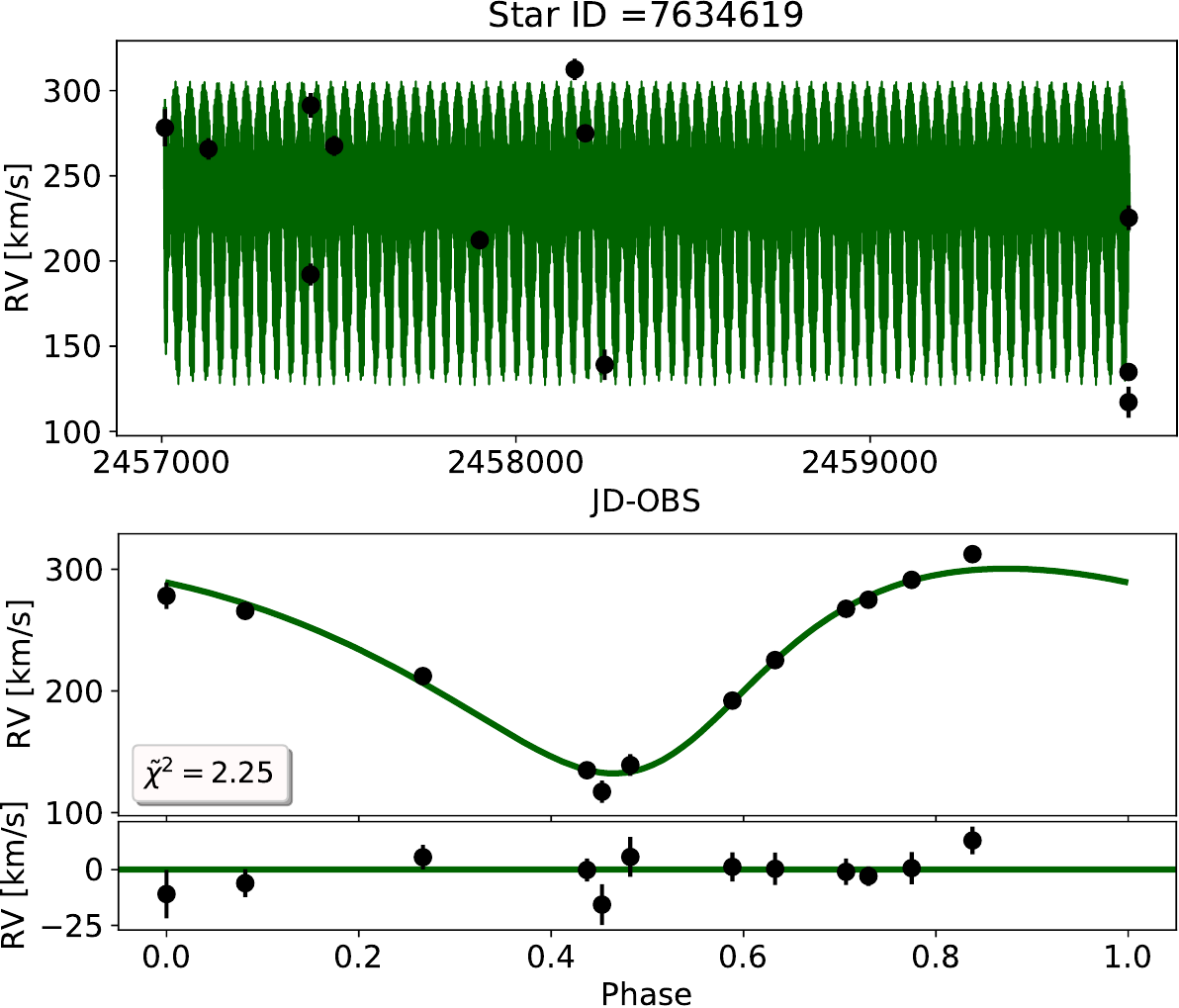}
\end{subfigure}
\begin{subfigure}
    \centering
    \includegraphics[width=0.49\textwidth]{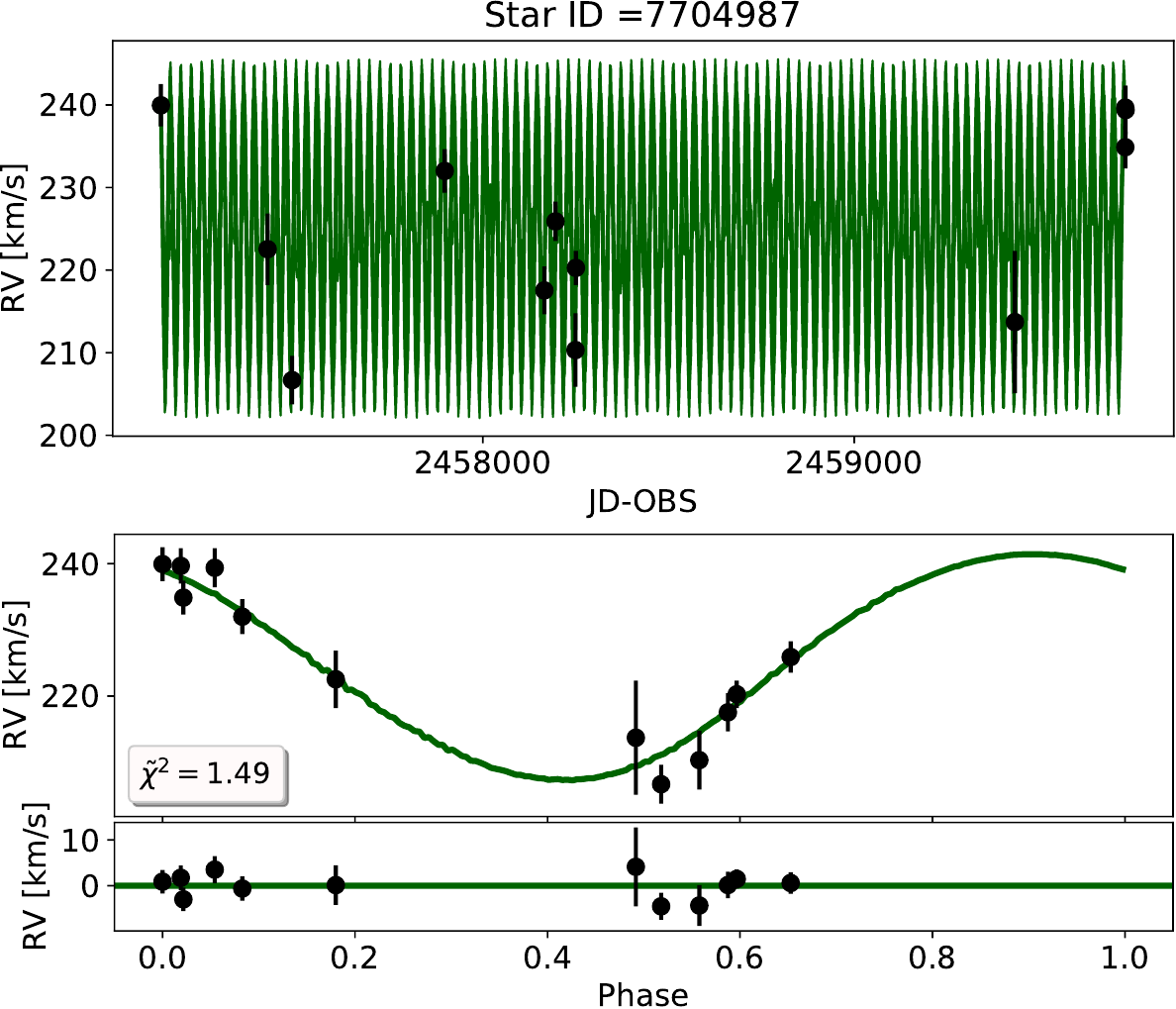}
\end{subfigure}
\caption{Same as in \ref{fig:plot}.}
\label{fig:plot2}
\end{figure*}

\begin{figure*}
\begin{subfigure}
    \centering
    \includegraphics[width=0.49\textwidth]{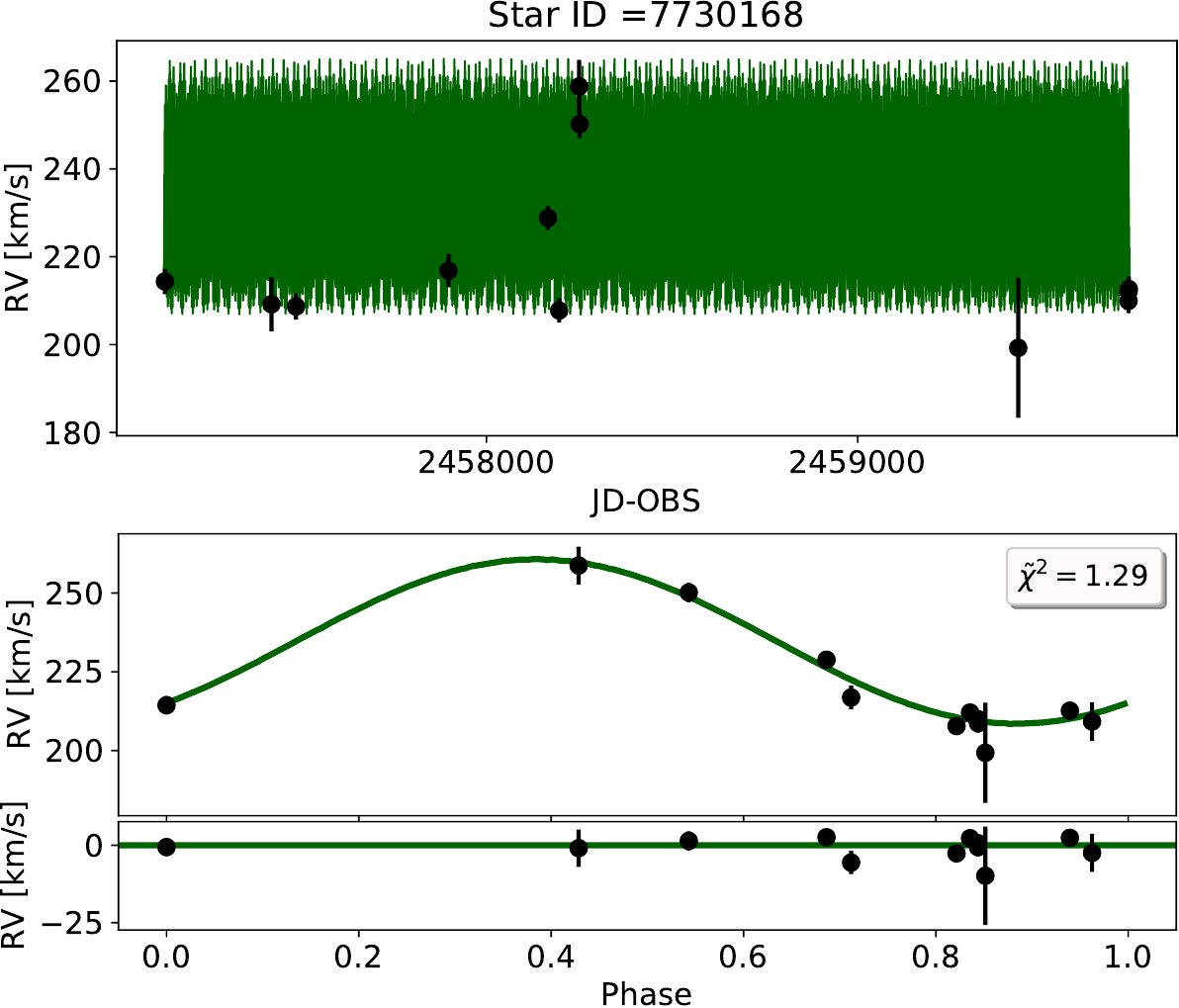}
\end{subfigure}
\caption{Same as in \ref{fig:plot}.}
\label{fig:plot3}
\end{figure*}

This Appendix contains supplementary material (Tables \ref{tab:astrophoto} and \ref{tab:constrained} and Figures \ref{fig:plot}, \ref{fig:plot1}, \ref{fig:plot2} and \ref{fig:plot3}), helping to better understand the content of the paper, without interrupting its flow.

\begin{table*}
    \centering
    \caption{Astrometric and photometric properties of constrained binaries in $\omega$ Cen, from the HST catalog and the comparison with PARSEC stellar models.}\label{tab:astrophoto}
    \begin{tabular}{|l|llllll|}
    \hline
        \textbf{Star Id} & \textbf{RA} & \textbf{Dec} & \textbf{F435W} & \textbf{F625W} & \textbf{$M_{1}$} & \textbf{$\sigma_{M1}$}\\ \hline
        \textbf{} & [$^\circ$] & [$^\circ$] & [mag] & [mag] & [$M_{\odot}$] & [$M_{\odot}$] \\             
        1664295 & 201.702819 & -47.4851368 & 16.2923 & 14.8896 & 0.77 & 0.03 \\ 
        1665349 & 201.697407 & -47.4859776 & 15.6385 & 14.2728 & 0.79 & 0.04 \\ 
        1670391 & 201.6799852 & -47.485817 & 18.2523 & 17.3338 & 0.74 & 0.05 \\ 
        1720096 & 201.6756022 & -47.4793909 & 15.3557 & 13.9222 & 0.79 & 0.04 \\ 
        1724184 & 201.6586312 & -47.4788943 & 16.1161 & 14.8248 & 0.78 & 0.04 \\ 
        1757804 & 201.7102218 & -47.4719659 & 16.6021 & 16.2086 & 1.20 & 0.06 \\ 
        1772072 & 201.6599485 & -47.470203 & 17.775 & 16.5706 & 0.74 & 0.03 \\ 
        1780702 & 201.7161967 & -47.4675106 & 17.7569 & 16.5187 & 0.73 & 0.03 \\ 
        1785698 & 201.6926747 & -47.4694025 & 17.6695 & 16.4858 & 0.74 & 0.04 \\ 
        1787754 & 201.6883243 & -47.4669057 & 16.6207 & 15.246 & 0.76 & 0.03 \\ 
        1795703	& 201.6529693 & -47.4687259	& 17.6513 &	16.4829	& 0.74 & 0.04 \\
        1811439 & 201.6841355 & -47.4652308 & 15.5123 & 14.196 & 0.80 & 0.04 \\ 
        7111025 & 201.64031819 & -47.50575637 & 15.8992 & 15.4701 & 1.20 & 0.05 \\ 
        7158549 & 201.61994011 & -47.49779472 & 20.4877 & 19.3614 & 0.67 & 0.05 \\ 
        7190455 & 201.6332041 & -47.49772488 & 18.0588 & 17.0723 & 0.78 & 0.05 \\ 
        7228915 & 201.62204674 & -47.49048452 & 15.0125 & 13.457 & 0.82 & 0.05 \\ 
        7634619 & 201.66616557 & -47.42096561 & 20.3122 & 19.3045 & 0.69 & 0.05 \\ 
        7704987 & 201.66867026 & -47.40084593 & 19.1629 & 18.1656 & 0.73 & 0.05 \\ 
        7730168	& 201.67537888 & -47.40043293 & 18.1273	& 17.2267 &	0.78 & 0.05 \\ \hline
    \end{tabular}
\end{table*}

\begin{table*}
  \begin{adjustbox}{addcode={\begin{minipage}{\width}}{\caption{Orbital properties of the 19 well-constrained binaries in $\omega$ Cen.}\label{tab:constrained}\end{minipage}},rotate=90,center}
    \begin{tabular}
    {p{0.7cm}p{0.8cm}p{0.8cm}p{0.8cm}p{0.6cm}p{0.6cm}p{0.6cm}p{0.8cm}p{0.8cm}p{0.8cm}p{0.8cm}p{0.8cm}p{0.8cm}p{0.6cm}p{0.6cm}p{0.6cm}p{0.6cm}p{0.6cm}p{0.6cm}p{0.9cm}p{1.cm}}%
    \hline \hline
            \textbf{ID} & \textbf{P\textsubscript{min}} & \textbf{P} & \textbf{P\textsubscript{max}} & \textbf{e\textsubscript{min}} & \textbf{e} & \textbf{e\textsubscript{max}} & \textbf{K\textsubscript{min}} & \textbf{K} & \textbf{K\textsubscript{max}} & \textbf{$V_{\rm sys,min}$} & \textbf{$V_{\rm sys}$} & \textbf{$V_{\rm sys,max}$} & \textbf{M\textsubscript{2,min}} & \textbf{M\textsubscript{2}} & \textbf{M\textsubscript{2,max}} & \textbf{MF\textsubscript{min}} & \textbf{MF} & \textbf{MF\textsubscript{max}} & \textbf{Fit} & \textbf{Comment} \\ \hline
        1664295 & 94.0838 & 94.356 & 95.2832 & 0.2242 & 0.3961 & 0.4939 & 10.2635 & 12.3047 & 13.8994 & 217.8609 & 218.8233 & 219.7413 & 0.1879 & 0.2175 & 0.237 & 0.0098 & 0.0141 & 0.0175 & Ultranest & Unimodal\\ 
        - & 94.2662 & 94.3562 & 94.4842 & 0.3891 & 0.4618 & 0.5223 & 12.0444 & 12.308 & 15.6641 & 217.7951 & 218.4488 & 219.5363 & 0.2249 & 0.2448 & 0.2984 & 0.0134 & 0.0127 & 0.0234 & TheJoker & Unimodal\\ 
        1665349 & 318.2753 & 319.352 & 320.2915 & 0.2822 & 0.4446 & 0.6571 & 12.7594 & 16.1767 & 23.9086 & 267.3921 & 271.9012 & 275.524 & 0.4029 & 0.5096 & 0.7045 & 0.0606 & 0.1009 & 0.1947 & Ultranest & Unimodal \\ 
        - & 318.4176 & 319.3404 & 319.9048 & 0.4267 & 0.5414 & 0.6501 & 15.4944 & 18.6721 & 23.0524 & 267.2429 & 269.0389 & 271.3752 & 0.5454 & 0.6832 & 0.7522 & 0.091 & 0.1283 & 0.1786 & TheJoker & Unimodal\\ 
        1670391 & 0.6702 & 0.6702 & 0.6702 & 0.0146 & 0.0614 & 0.1365 & 43.4492 & 46.8791 & 50.2869 & 223.3559 & 225.8229 & 228.3687 & 0.1525 & 0.1662 & 0.1788 & 0.0057 & 0.0071 & 0.0086 & Ultranest & Bimodal \\ 
        - & ~ & ~ & ~ & ~ & ~ & ~ & ~ & ~ & ~ & ~ & ~ & ~ & ~ & ~ & ~ & ~ & ~ & ~ & ~ \\ 
        1720096 & 311.4699 & 326.5524 & 330.7393 & 0.3891 & 0.4925 & 0.6043 & 11.469 & 13.1836 & 16.4143 & 225.143 & 226.4939 & 227.9395 & 0.3286 & 0.3737 & 0.4487 & 0.0381 & 0.0512 & 0.0768 & Ultranest & Unimodal \\ 
        - & ~ & ~ & ~ & ~ & ~ & ~ & ~ & ~ & ~ & ~ & ~ & ~ & ~ & ~ & ~ & ~ & ~ & ~ & ~ \\ 
        1724184 & 169.7355 & 170.5128 & 171.3736 & 0.0362 & 0.1142 & 0.2009 & 11.6473 & 12.7804 & 14.1266 & 256.9338 & 257.5549 & 258.1631 & 0.2872 & 0.3214 & 0.3604 & 0.0278 & 0.0362 & 0.0472 & Ultranest & Unimodal \\ 
        - & 169.8026 & 170.5977 & 171.4474 & 0.0318 & 0.1058 & 0.1924 & 11.4976 & 12.5745 & 14.0043 & 257.0199 & 257.6229 & 258.2192 & 0.317 & 0.3541 & 0.402 & 0.0268 & 0.0346 & 0.0462 & TheJoker & Unimodal \\ 
        1757804 & 1.2469 & 1.247 & 1.247 & 0.00 & 0.00 & 0.00 & 30.3962 & 32.5646 & 34.5938 & 223.7491 & 225.4005 & 226.9811 & 0.1286 & 0.139 & 0.1489 & 0.0036 & 0.0045 & 0.0054 & Ultranest & Bimodal \\ 
        - & 1.2469 & 1.247 & 1.247 & 0.0039 & 0.0301 & 0.0912 & 30.8145 & 32.345 & 33.8827 & 223.6468 & 224.6465 & 226.8755 & 0.1937 & 0.2051 & 0.2159 & 0.0038 & 0.0044 & 0.0050 & TheJoker & Unimodal \\ 
        1772072 & 19.9467 & 19.9531 & 19.9592 & 0.0694 & 0.2085 & 0.3092 & 10.7477 & 11.6198 & 12.5267 & 214.4552 & 215.01 & 215.5626 & 0.1127 & 0.1203 & 0.1268 & 0.0026 & 0.0030 & 0.0035 & Ultranest & Unimodal \\ 
        - & 19.9472 & 19.9528 & 19.9582 & 0.0449 & 0.2229 & 0.3092 & 11.0568 & 11.8008 & 12.4188 & 214.4404 & 215.0007 & 215.431 & 0.124 & 0.1327 & 0.1421 & 0.0028 & 0.0032 & 0.0034 & TheJoker & Unimodal \\ 
        1780702 & 23.9773 & 23.9867 & 23.9969 & 0.0752 & 0.1894 & 0.3014 & 23.2801 & 25.4377 & 28.912 & 224.0965 & 226.2394 & 227.9513 & 0.3013 & 0.3311 & 0.3769 & 0.0312 & 0.0388 & 0.0522 & Ultranest & Unimodal \\ 
        - & 23.9773 & 23.9882 & 23.9963 & 0.0735 & 0.198 & 0.3404 & 23.8666 & 25.7079 & 31.885 & 222.7668 & 225.9248 & 227.1983 & 0.3333 & 0.3635 & 0.4527 & 0.0336 & 0.0399 & 0.0672 & TheJoker & Unimodal \\ 
        1785698 & 33.5265 & 33.5524 & 33.5807 & 0.0145 & 0.056 & 0.132 & 15.8288 & 17.264 & 18.7102 & 225.5169 & 226.5668 & 227.5749 & 0.2155 & 0.2391 & 0.2617 & 0.0138 & 0.0178 & 0.0222 & Ultranest & Unimodal \\ 
        - & 33.5294 & 33.5517 & 33.5765 & 0.0141 & 0.061 & 0.1304 & 15.9215 & 17.3375 & 18.7382 & 225.6386 & 226.6048 & 227.5985 & 0.2372 & 0.2623 & 0.2878 & 0.014 & 0.0181 & 0.0224 & TheJoker & Unimodal \\ 
        1787754 & 400.3544 & 406.8219 & 415.6467 & 0.0159 & 0.0686 & 0.1678 & 7.8049 & 8.947 & 10.4741 & 253.7788 & 254.8174 & 256.1226 & 0.2492 & 0.2967 & 0.3616 & 0.0198 & 0.03 & 0.0475 & Ultranest & Unimodal \\ 
        - & 398.8283 & 404.4212 & 414.7457 & 0.0166 & 0.0743 & 0.183 & 8.3437 & 9.9001 & 12.7982 & 254.1858 & 255.4306 & 257.5533 & 0.3006 & 0.3736 & 0.5172 & 0.024 & 0.0404 & 0.0858 & TheJoker & Unimodal \\ 
        1795703	& 82.1658 & 82.3796	& 82.5614 & 0.00 & 0.00 & 0.00 & 11.7882 & 13.1441 & 14.4823 &	214.8484 & 215.857 & 216.872 & 0.2166 & 0.2475 & 0.2794 & 0.014 & 0.0194 & 0.026 & Ultranest & Unimodal \\
        - & ~ & ~ & ~ & ~ & ~ & ~ & ~ & ~ & ~ & ~ & ~ & ~ & ~ & ~ & ~ & ~ & ~ & ~ & ~ \\
        1811439 & 82.7262 & 82.8422 & 82.9493 & 0.0102 & 0.0438 & 0.1034 & 13.0113 & 13.6618 & 14.3998 & 255.9338 & 256.5433 & 257.1112 & 0.2448 & 0.2599 & 0.2761 & 0.0189 & 0.0219 & 0.0253 & Ultranest & Unimodal \\ 
        - & 82.7249 & 82.8456 & 82.9527 & 0.0103 & 0.0439 & 0.1046 & 13.0197 & 13.6904 & 14.4185 & 255.9244 & 256.5876 & 257.1659 & 0.28 & 0.297 & 0.316 & 0.019 & 0.022 & 0.0254 & TheJoker & Unimodal \\ 
        7111025 & 3.3944 & 3.3946 & 3.3949 & 0.0151 & 0.0604 & 0.1456 & 19.3438 & 21.4075 & 23.6122 & 252.0958 & 253.8919 & 255.6672 & 0.1127 & 0.1259 & 0.1393 & 0.0026 & 0.0034 & 0.0045 & Ultranest & Unimodal \\ 
        - & 3.3946 & 3.3947 & 3.3948 & 0.0031 & 0.0179 & 0.0666 & 20.5962 & 21.6083 & 22.7232 & 252.9971 & 253.6171 & 254.519 & 0.1797 & 0.1897 & 0.2006 & 0.0031 & 0.0036 & 0.0041 & TheJoker & Unimodal \\ 
        7158549 & 1.135 & 1.1351 & 1.8116 & 0.1038 & 0.2743 & 0.3263 & 101.2181 & 106.4847 & 119.0781 & 208.9761 & 213.9498 & 218.1579 & 0.5542 & 0.5681 & 0.8333 & 0.1203 & 0.1266 & 0.2683 & Ultranest & Unimodal \\ 
        - & 1.8116 & 1.8116 & 1.8116 & 0.0843 & 0.1339 & 0.1468 & 109.4052 & 112.0801 & 120.973 & 210.8851 & 213.3213 & 217.2259 & 0.8084 & 0.8349 & 0.9493 & 0.2437 & 0.2578 & 0.3224 & TheJoker & Unimodal \\ 
        7190455 & 111.8467 & 112.0176 & 112.2327 & 0.3595 & 0.4813 & 0.6064 & 17.4326 & 19.9294 & 23.808 & 212.0477 & 214.325 & 216.7104 & 0.3698 & 0.407 & 0.4547 & 0.05 & 0.062 & 0.0791 & Ultranest & Unimodal \\ 
        - & 111.8716 & 112.0487 & 112.1831 & 0.3822 & 0.5256 & 0.6203 & 17.3308 & 21.0076 & 23.9362 & 211.3317 & 214.1284 & 215.8687 & 0.4052 & 0.4553 & 0.5304 & 0.0477 & 0.0664 & 0.0771 & TheJoker & Unimodal \\ 
        7228915 & 88.6444 & 88.7903 & 88.9353 & 0.1851 & 0.2203 & 0.2567 & 16.3481 & 16.8171 & 17.3682 & 213.851 & 214.2773 & 214.7269 & 0.3287 & 0.3379 & 0.3485 & 0.0382 & 0.0407 & 0.0437 & Ultranest & Unimodal \\ 
        - & 88.6396 & 88.7696 & 88.9024 & 0.1971 & 0.2181 & 0.2574 & 16.3167 & 16.8045 & 17.5147 & 213.9669 & 214.2557 & 214.7679 & 0.3737 & 0.3916 & 0.4045 & 0.0377 & 0.0407 & 0.0448 & TheJoker & Unimodal \\ 
        7634619 & 5.1 & 5.1002 & 5.1003 & 0.2051 & 0.2432 & 0.2816 & 81.7585 & 84.7245 & 87.6786 & 230.281 & 232.6315 & 234.919 & 0.8384 & 0.8754 & 0.9096 & 0.2714 & 0.294 & 0.3154 & Ultranest & Unimodal \\ 
        - & ~ & ~ & ~ & ~ & ~ & ~ & ~ & ~ & ~ & ~ & ~ & ~ & ~ & ~ & ~ & ~ & ~ & ~ & ~ \\ 
        7704987 & 28.1905 & 28.2091 & 28.2312 & 0.0178 & 0.0752 & 0.1844 & 16.8152 & 19.3499 & 22.1505 & 223.3895 & 224.8799 & 226.2564 & 0.2162 & 0.2558 & 0.2976 & 0.0139 & 0.021 & 0.0303 & Ultranest & Bimodal \\ 
        - & 28.1968 & 28.2092 & 28.2218 & 0.0279 & 0.107 & 0.2314 & 17.3987 & 20.2065 & 23.935 & 222.3423 & 224.4642 & 225.9762 & 0.2433 & 0.291 & 0.3588 & 0.0154 & 0.0238 & 0.037 & TheJoker & Bimodal \\
        7730168	& 9.5825 & 9.5846 & 9.5866 & 0.00 & 0.00 & 0.00 & 24.2821 & 26.4106 & 28.6492 &	232.8145 & 234.8553	& 236.9172 & 0.2182	& 0.2417 & 0.2673 & 0.0142	& 0.0183 & 0.0234 & Ultranest & Unimodal \\
        - & ~ & ~ & ~ & ~ & ~ & ~ & ~ & ~ & ~ & ~ & ~ & ~ & ~ & ~ & ~ & ~ & ~ & ~ & ~ \\
        
\hline \hline
    \end{tabular}
\end{adjustbox}
\end{table*}

\bsp	
\label{lastpage}
\end{document}